\documentclass[twocolumn, 10pt, aps, superscriptaddress, showpacs, prb, citeautoscript, longbibliography]{revtex4-2}

\usepackage[ngerman, english]{babel}   
\usepackage[utf8]{inputenc}
\usepackage{lmodern}
\usepackage{amssymb}
\usepackage{amsmath}
\usepackage{graphicx}
\usepackage{diagbox} 
\usepackage{hhline}  
\usepackage{float} 

\usepackage{xr} 
\usepackage{xcite} 
  
\usepackage[usenames,dvipsnames]{xcolor} 
\usepackage[colorlinks, citecolor={blue!50!black}, urlcolor={blue!50!black}, linkcolor={red!50!black},hypertexnames=false]{hyperref}

\usepackage[position=top,caption=false]{subfig}  
\graphicspath{{figures/}} 
\usepackage{tikz}             

\usepackage{tabularx}                    
\usepackage{multirow, bigdelim}  
\usepackage{longtable}
\usepackage{makecell}                   
\usepackage{cellspace}

\usepackage{mdwlist}      

\usepackage{verbatim}       
\usepackage{bbm}              
\usepackage{bm} 
\usepackage{dsfont}			
\usepackage{nicefrac}        
\usepackage[vcentermath]{youngtab} 
\usepackage{cancel}

\usepackage{etoolbox} 

\usepackage[normalem]{ulem}

\newcommand{\K}[2]{\mathcal{K}_{#1}^{#2}}

\newcommand{\Kb}[2]{\mathcal{K}_{#1^\prime}^{#2}}

\newcommand{\doubleI}{\mathds{1}}     

\renewcommand{\Re}{\mathrm{Re}}
\renewcommand{\Im}{\mathrm{Im}}
\newcommand{\sgn}{\mathrm{sgn}}

\newcommand{\Sec}[1]{Sec.~\ref{#1}}
\newcommand{\App}[1]{App.~\ref{#1}}
\newcommand{\Eq}[1]{Eq.~\eqref{#1}}
\newcommand{\Eqs}[1]{Eqs.~\eqref{#1}}
\newcommand{\Fig}[1]{Fig.~\ref{#1}}
\newcommand{\Figs}[1]{Figs.~\ref{#1}}

\definecolor{darkgreen}{rgb}{0,0.5,0}
\definecolor{orange}{rgb}{1,0.5,.3}
\definecolor{darkred}{rgb}{.7,0,0}
\definecolor{purple}{rgb}{0.6,0,0.5}
\definecolor{darkpetrol}{RGB}{0,73,76}

\usepackage[normalem]{ulem} 
\definecolor{darkgreen}{rgb}{0,0.5,0}
\definecolor{purple}{rgb}{0.6,0,0.5}
\definecolor{orange}{rgb}{1,0.5,0}
\definecolor{darkred}{rgb}{.7,0,0}
\definecolor{darkblue}{rgb}{0,0,.6}
\definecolor{grey}{rgb}{.6,.6,.6}
\definecolor{dimgreen}{rgb}{0.2,0.6,0.1}

\newcommand{\dd}{\mathrm{d}}
\newcommand{\ee}{\mathrm{e}}
\newcommand{\ii}{\mathrm{i}}

\begin{document}

\title{Subleading logarithmic behavior in the parquet formalism}

\author{Marcel Gievers}
\affiliation{Arnold Sommerfeld Center for Theoretical Physics, 
	Center for NanoScience,\looseness=-1\,  and Munich 
	Center for Quantum Science and Technology,\looseness=-2\, Ludwig-Maximilians-Universität München, 80333 Munich, Germany}
\affiliation{Max Planck Institute  of  Quantum  Optics,  Hans-Kopfermann-Straße  1,  85748  Garching,  Germany}

\author{Richard Schmidt}
\affiliation{Institut für Theoretische Physik, Universität Heidelberg, 69120 Heidelberg, Germany}

\author{Jan von Delft}
\affiliation{Arnold Sommerfeld Center for Theoretical Physics, 
	Center for NanoScience,\looseness=-1\,  and Munich 
	Center for Quantum Science and Technology,\looseness=-2\, Ludwig-Maximilians-Universität München, 80333 Munich, Germany}

\author{Fabian B. Kugler}
\affiliation{Center for Computational Quantum Physics, Flatiron Institute, 162 5th Avenue, New York, New York 10010, USA}

\date{\today}
\pacs{}

\begin{abstract}
The Fermi-edge singularity in x-ray absorption spectra of metals is a paradigmatic case of a logarithmically divergent perturbation series. Prior work has thoroughly analyzed the leading logarithmic terms. Here, we investigate the perturbation theory beyond leading logarithms and formulate self-consistent equations to incorporate all leading and next-to-leading logarithmic terms. This parquet solution of the Fermi-edge singularity goes beyond the previous first-order parquet solution and sheds new light on the parquet formalism regarding logarithmic behavior. We present numerical results in the Matsubara formalism and discuss the characteristic power laws. We also show that, within the single-boson exchange framework, multi-boson exchange diagrams are needed already at the leading logarithmic level.
\end{abstract}

\maketitle

\section{Introduction}

Perturbative expansions are ubiquitous in theoretical physics, and logarithmic divergences therein often lead to power laws in observables. In the 1960s, several works developed self-consistent methods to sum up the leading logarithmic terms from Feynman diagrams of all orders. Among other systems, these techniques were successfully applied to meson scattering~\cite{Diatlov1957Asymptotic}, the Kondo model~\cite{Abrikosov1965}, the one-dimensional interacting Fermi gas~\cite{Bychkov1966,Anderson1970}, and the Fermi-edge singularity in x-ray absorption in metals~\cite{roulet1969singularities,nozieres1969singularities2}. These self-consistent summations take into account two diagrammatic channels, but exclude self-energy corrections. Following Ref.~\cite{roulet1969singularities}, we refer to them as the \emph{first-order parquet} approach.

Presently, the Hubbard model is one of the most studied many-body problems of condensed matter physics~\cite{Schaefer2021,Qin2022Hubbard}. To capture the interplay of its competing fluctuations, another type of self-consistent summation of diagrams was developed, which we here call \emph{full parquet} approach. It involves the self-energy and the effective interactions in all three channels of two-particle reducibility~\cite{Bickers1991Conserving,Bickers1992,Chen1992Numerical,Bickers2004}. As the perturbative series of the Hubbard model does not exhibit logarithmic divergences, such a treatment was not motivated by logarithmic terms but by the fulfillment of crossing symmetry and self-consistency on the one- and two-particle level~\cite{Bickers2004, Held2008}. With increasing computational power, the numerical solution of the parquet equations has nowadays become a viable tool~\cite{Tam2013Solvings,Valli2015Dynamical,Li2016,Li2019The,Kauch2020Generic,Astretsov2020Dual,Eckhardt2020Truncated}.

In this work, we examine what the transition from the first-order to the full parquet approach entails for logarithmically divergent problems. To this end, we revisit the Fermi-edge singularity, which was recently revived as an inspiring workhorse to better understand diagrammatic techniques~\cite{Lange_2015,kugler2018fermi,KuglerPRL2018b,Diekmann2021,Diekmann2024Leading}. Although it can be solved in a one-particle scheme~\cite{nozieres1969singularities3,Schotte1969}, a precise analysis in perturbation theory remains challenging. It was shown that  the leading logarithmic behavior can be obtained using the one-loop functional renormalization group~\cite{Diekmann2021,Diekmann2024Leading}, while the full summation of all parquet diagrams is only recovered in a multiloop expansion~\cite{KuglerPRL2018b,Kugler2018a,KuglerNJP2018e}. 

Remarkably, we find that a well-defined subset of diagrams from the full parquet solution offers a convenient way to not only capture the leading logarithmic singularity, but also the next-to-leading contributions. Hence, we close the gap between the traditional summation of leading logarithms (first-order parquet) and the one- and two-particle self-consistent summation (full parquet), giving new insights into the structure of logarithmically divergent perturbation theories. The power-law exponent of the particle-hole susceptibility obtained from our approach is closer to the exact result than that obtained from  previous diagrammatic analyses respecting only the leading logarithmic contributions. Our analysis beyond the leading logarithms is possible (despite the concerns of Ref.~\cite{nozieres1969singularities2}) as our numerical results include the full dependence of individual diagrams beyond logarithmic accuracy. Moreover, we briefly explain that the self-energy diagrams needed to capture Anderson's orthogonality catastrophe~\cite{Anderson1967Infrared} go beyond the present approach.

Treating the full frequency dependence of the effective interaction requires huge numerical effort. In recent years, there were several attempts to make use of frequency asymptotics to lower the numerical costs~\cite{Wentzell2020}. One of them is the decomposition of the full interaction vertex into bosonic exchange processes, the so-called \emph{single-boson exchange}~\cite{Krien2019,Krien2019a,Krien2019b,Krien2020a,Krien2020b,Krien2021,Harkov2021,Harkov2021a,krien2022plain,Bonetti2022SBE,fraboulet2022single,gievers2022multiloop} as well as the remaining and numerically most expensive \emph{multi-boson exchange} terms. We show that multi-boson exchange terms are essential already at the leading logarithmic level; neglecting them is thus not justified in the present case.

The rest of our paper is organized as follows. Section~\ref{sec:Model} serves as a reminder of the model and the diagrammatic approach. In \Sec{sec:Perturbation-theory}, we discuss the lowest terms in perturbation theory. In \Sec{sec:Parquet}, we explain our self-consistent summation scheme and show the corresponding numerical results. We conclude in \Sec{sec:Conclusion}.

\section{Model and method}
\label{sec:Model}

\begin{figure}
    \centering
    \includegraphics[width=0.6\linewidth]{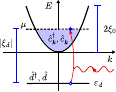}
    \caption{Illustration of the model defined by \Eq{eq:Hamiltonian}. The conduction band with quadratic dispersion relation and finite bandwidth $2\xi_0$ hosts electrons $\hat{c}^\dagger_{\bm{k}}, \hat{c}^{\phantom{\dagger}}_{\bm{k}}$. The deep core level at energy $\varepsilon_d$ hosts a single electron $\hat{d}^\dagger, \hat d$, which gets excited to the Fermi level by absorbing a photon.
    }
    \label{fig:Model}
\end{figure}

The model for the Fermi-edge singularity of x-ray absorption in metals was introduced in the late 1960s~\cite{Mahan1967Excitons,roulet1969singularities,nozieres1969singularities2,nozieres1969singularities3,nozieres1994effect}. It is visualized in \Fig{fig:Model}. A conduction band with quadratic dispersion relation $\varepsilon_{\bm{k}}=\bm{k}^2/(2m)$ hosts electrons represented by creation and annihilation operators $\hat c_{\bm{k}}^\dagger, \hat c^{\phantom{\dagger}}_{\bm{k}}$. In addition, there is a deep, localized core level
at energy $\varepsilon_d<0$, which hosts a single $\hat d^\dagger, \hat d$ electron. Spin indices for the electrons are omitted since the spin degeneracy only results in a doubled density of states~\footnote{This is justified since, as in the original model~\cite{nozieres1969singularities3}, we ignore exchange processes, in which the deep hole and one conduction electron reverse their spins.}. An empty core level
corresponds to the presence of a core hole, with an effective attraction of strength $U>0$ to the conduction electrons. The Hamiltonian reads
\begin{align}
    \hat H
    &=
    \sum_{\bm{k}}\varepsilon_{\bm{k}}\hat c_{\bm{k}}^\dagger\hat c_{\bm{k}}+\varepsilon_d\hat d^\dagger \hat d -\frac{U}{V}\sum_{\bm{k},\bm{k}'}\hat c^\dagger_{\bm{k}}\hat c^{\phantom{\dagger}}_{\bm{k}'}\hat d\hat d^\dagger
    .
    \label{eq:Hamiltonian}
\end{align}
We assume a finite bandwidth $2\xi_0$ of the conduction band and set the chemical potential to half the bandwidth, i.e., $\mu = \xi_0$. Absorption of a photon of frequency $\omega\simeq -\xi_d = \mu - \varepsilon_d$ excites the core electron to a state near the Fermi level. Due to the sharp Fermi edge, a singularity arises in absorption and emission spectra.

We are interested in two quantities. First, we analyze the particle-hole susceptibility,
\begin{align}
    \label{eq:Susceptibility-definition}
    X(t)
    &=
    -\ii\frac{1}{V}\sum_{\bm{k},\bm{k}'}\bigl\langle\,\mathcal{T}\hat d^\dagger(t)\hat c^{\phantom{\dagger}}_{\bm{k}}(t)\hat c_{\bm{k}'}^\dagger \hat d\,\bigr\rangle
    ,
\end{align}
where $\mathcal{T}$ is the time-ordering operator. $X(t)$ is the response function to the photo-excitation of a conduction-particle and core-hole pair. Its imaginary part in frequency space corresponds to the x-ray transition rate~\cite{roulet1969singularities,nozieres1969singularities2,Diekmann2024Leading}. Second, we investigate the propagator of the $d$ electron,
\begin{align}
    \label{eq:Green_d-definition}
    G(t) 
    &= 
    -\ii\,\bigl\langle\,\mathcal{T}\hat d(t)\hat d^\dagger\,\bigr\rangle
    ,
\end{align}
which encodes single-particle excitations. The expectation values in \Eqs{eq:Susceptibility-definition} and \eqref{eq:Green_d-definition} are taken with respect to the ground state $|\Psi_0\rangle$. For large enough $|\xi_d|$, the ground state is given by the occupied core level and the Fermi sea of conduction electrons in the sense of Fermi-liquid theory. Consequently, at zero temperature $T=0$ (and more generally  for $T \ll |\xi_d|$), $X(t)$ is purely retarded while $G(t)$ is purely advanced, as follows from the effect of the time-ordering operators $\mathcal{T}$ in \Eqs{eq:Susceptibility-definition}--\eqref{eq:Green_d-definition}. In \App{sec:FDA}, we provide numerically exact solutions of the two quantities using the functional determinant approach~\cite{levitov1996electron,klich2003elementary,schmidt2018universal}.

The fact that $G(t)$ is purely advanced has important consequences. It implies that there are no self-energy contributions to the $c$ electrons beyond the Hartree term. Since there is precisely one local $d$ level, this Hartree term reads $U/V f(\xi_d)$, with the Fermi--Dirac distribution function $f(\varepsilon)=1/(1+\ee^{\beta\varepsilon})$, where $\beta = 1/T$. Now, bringing the Hamiltonian, \Eq{eq:Hamiltonian}, into normal order, we get an additional term $-U/V$. This term and the $c$-electron Hartree self-energy cancel exactly at $T=0$. We may thus suppress these $\sim\!U$ single-particle terms altogether, thereby effectively working with Hartree-dressed propagators for the $c$ electrons~\cite{Diekmann2021,Diekmann2024Leading}.

In Matsubara field theory, the expectation value $\langle ...\rangle = \frac{1}{Z}\mathrm{tr}(\ee^{-\beta(\hat H-\mu\hat N)}...)$ (with $Z= \mathrm{tr}\,\ee^{-\beta(\hat H-\mu\hat N)}$) can be written as a functional integral involving the action
\begin{align}
    \nonumber 
    S
    &= 
    -\frac{1}{\beta}\sum_{\nu,\bm{k}} \bar c_{\bm{k},\nu} (\ii\nu-\xi_{\bm{k}}) c_{\bm{k},\nu} -\frac{1}{\beta}\sum_{\nu} \bar d_\nu (\ii\nu-\xi_d) d_\nu
    \\
    &\phantom{=} 
    + \frac{U}{V} \frac{1}{\beta^3} \sum_{\omega,\nu,\nu',\bm{k},\bm{k}'} \bar d_\nu \bar c_{\bm{k},\nu'+\omega} c_{\bm{k}',\nu+\omega} d_{\nu'}
    .
    \label{eq:Action_cd}
\end{align}
Here, $\xi_{\bm{k}}=\varepsilon_{\bm{k}}-\mu$, $\xi_d=\varepsilon_d-\mu$, and the fields depend on the fermionic (bosonic) Matsubara frequencies $\nu$ ($\omega$). Evidently, only the $c$ electron at the core hole, $\tfrac{1}{\sqrt{V}} \sum_{\bm{k}}c_{\bm{k},\nu}$, referred to as the \emph{local} $c$ electron, interacts with the $d$ hole. We may thus integrate out all $c$ electrons that are not located at the core hole.

In the following, we assume a constant local density of states of the conduction electrons
\begin{align}
    \rho(\varepsilon)
    =
    \frac{1}{V} \sum_{\bm{k}} \delta(\varepsilon - \varepsilon_{\bm{k}})
    =
    \rho \, \Theta(2\xi_0 - \varepsilon)\Theta(\varepsilon)
    .
    \label{eq:local_dos}
\end{align}
This holds in two dimension, and in three dimensions, it is an approximation motivated by the dominance of effects near the Fermi level.
We can hence rescale the local $c$ electrons by $\sqrt{\rho}$ to avoid trivial factors of $\rho$.
Thereby, we obtain the action:
\begin{align}
    \nonumber 
    S
    &= 
    -\frac{1}{\beta}\sum_{\nu} \bar c_\nu g_\nu^{-1} c_\nu -\frac{1}{\beta}\sum_{\nu} \bar d_\nu (\ii\nu-\xi_d) d_\nu
    \\
    &\phantom{=} 
    + u \frac{1}{\beta^3} \sum_{\omega,\nu,\nu'} \bar c_{\nu'+\omega} c_{\nu+\omega} \bar d_\nu  d_{\nu'}
    \label{eq:action_rescaled}
\end{align}
with the dimensionless interaction $u = \rho \, U$ and the dimensionless local $c$ propagator $g_\nu = -\tfrac{1}{\rho V}\sum_{\bm{k}}\langle c_{\bm{k},\nu}\bar c_{\bm{k},\nu}\rangle$. At half filling, $\mu = \xi_0$, the latter is given by
\begin{align}
    g_\nu 
    &=
    \int_{0}^{2\xi_0} \frac{\dd \varepsilon}{\ii\nu-\varepsilon+\xi_0}
    =
    -2\ii\,\arctan\frac{\xi_0}{\nu}
    =
    g^\mathrm{sm}_\nu
    .
    \label{eq:Gc_smooth}
\end{align}
We refer to this as the \emph{smooth} propagator. Previous diagrammatic works~\cite{roulet1969singularities,nozieres1969singularities2,Lange_2015,Kugler2018a,KuglerPRL2018b,Diekmann2021,Diekmann2024Leading} approximated it by 
\begin{align}
    g^\mathrm{sh}_\nu
    =
    -\ii\pi\,\sgn(\nu)\Theta(\xi_0-|\nu|)
    ,
    \label{eq:Gc_sharp}
\end{align}
its \emph{sharp} form. This expression is convenient for analytical calculations of the power law around $\omega\simeq|\xi_d|$. However, it is problematic for self-consistent numerical computations, as it violates basic properties, such as the $\sim\! 1/(\ii\nu)$ decay for large $|\nu|$. Indeed, for our calculations beyond logarithmic accuracy, it is crucial to use $g^\mathrm{sm}$ instead of $g^\mathrm{sh}$. Note that we also obtain a dimensionless susceptibility in terms of the rescaled local $c$ fields,
\begin{align}
    \chi(\omega) = X(\omega)/\rho
    .
\end{align}

Seminal works from the 1960s showed that $\chi$ and $G$ exhibit characteristic power laws close to the threshold $\omega_0$ (cf.\ Eq.~(66) in Ref.~\cite{nozieres1969singularities3} and Refs.~\cite{Mahan1967Excitons,Anderson1967Infrared,roulet1969singularities,nozieres1969singularities2}):
\begin{subequations}
    \label{eq:exact_power-laws}
    \begin{align}
        \chi(\omega + \ii 0^+)
        &\simeq
        \frac{1}{\alpha_X}\left[1-\left(\frac{\omega + \ii 0^+-\omega_0}{-\xi_0}\right)^{-\alpha_X}\right]
        ,
        \label{eq:exact_power-law_chi}
        \\
        G(\nu - \ii 0^+)
        &\simeq
        \frac{1}{\nu - \ii 0^+ +\omega_0}\left(\frac{\nu - \ii 0^+ +\omega_0}{\xi_0}\right)^{\alpha_G}
        .
        \label{eq:exact_power-law_G}
    \end{align}
\end{subequations}
Here, the power law of $\chi$ characterizes the x-ray edge singularity, while that of $G$ is related to Anderson's orthogonality catastrophe. Note that \Eqs{eq:exact_power-laws} are given in real frequencies, in contrast to all the other expressions in this paper. The power-law exponents $\alpha_X = 2\delta/\pi-(\delta/\pi)^2$ and $\alpha_G = (\delta/\pi)^2$ depend on the $s$-wave scattering phase shift $\delta$ evaluated at the Fermi surface. For the present model [cf.\ \Eq{eq:local_dos}], this is related to the interaction strength via $\delta=\arctan(\pi u)$. We have verified the power laws in \Eqs{eq:exact_power-laws} with our numerically exact data using the functional determinant approach (cf.\ \App{sec:FDA}).

The threshold frequency $\omega_0$ depends sensitively on how the UV cutoff $\xi_0$ is implemented in the model~\cite{roulet1969singularities,nozieres1969singularities2}. As discussed later, our diagrammatic analysis allows for computing $\omega_0$ independently from all other quantities (cf.\ \Sec{sec:Threshold} and \App{sec:FDA_Threshold}). For now, we set $\omega_0$ to its bare value $\omega_0\to -\xi_d$. In the final results, $-\xi_d$ can be replaced by $\omega_0$.  

Expanding the power-law expressions, \Eqs{eq:exact_power-laws}, in $u$ 
reveals the logarithmic divergences. For small $u$, we can approximate $\delta/\pi = u + \mathcal{O}(u^3)$ (cf.\ \App{sec:Taylor-expansions_supplement}). The resulting form for $\chi$ in imaginary frequencies is:
\begin{align}
    \nonumber\chi(\ii\omega)
    &\simeq
    \frac{1}{2u-u^2}\left[ 1-\left(\frac{\ii\omega+\xi_d}{-\xi_0}\right)^{-2u+u^2}\right]
    \\
    \nonumber &\equiv
    \frac{1}{2u-u^2}\left[ 1-\ee^{(-2u+u^2)L}\right]
    \\
    \nonumber &=
    L - u L^2 + u^2 [\tfrac{2}{3} L^3 + \tfrac{1}{2} L^2] - u^3  [\tfrac{1}{3}L^4+\tfrac{2}{3}L^3]
    \\
    &\phantom{=}
    + u^4 [\tfrac{2}{15}L^5+\tfrac{1}{2}L^4+\tfrac{1}{6}L^3] + \mathcal{O}(u^5)
    .
    \label{eq:Taylor-expansion}
\end{align}
Here, we introduced the logarithmic factor
\begin{align}
    L(\omega)=\ln\frac{\ii\omega+\xi_d}{-\xi_0}.
    \label{eq:definition_L}
\end{align}
Taking only the highest power of $L$ in each order of $u$ yields the leading logarithmic result, where $2u-u^2$ is replaced by $2u$ (cf.\ Eq.~(42) in Ref.~\cite{roulet1969singularities} and Eq.~(57) in Ref.~\cite{nozieres1969singularities2}).

Analogously, $G$ in imaginary frequencies has an expansion in terms of $\bar{L}(\nu) = L(-\nu) = \ln[(\ii\nu-\xi_d)/\xi_0]$:
\begin{align}
    \nonumber G(\ii\nu)
    &\simeq
    \frac{1}{\ii\nu-\xi_d}\left(\frac{\ii\nu-\xi_d}{\xi_0}\right)^{u^2}
    \equiv \frac{1}{\ii\nu-\xi_d}\ee^{u^2\bar{L}}
    \\
    &= \frac{1}{\ii\nu-\xi_d}\left[1+u^2 \bar{L} + \tfrac{1}{2}u^4 \bar{L}^2 + \mathcal{O}(u^6)\right].
    \label{eq:Taylor-expansion_G}
\end{align}

In this work, we will show that a suitable summation of parquet diagrams not only contains the leading logarithmic result of $\chi$, but also the second-highest power of $L$ at each order of $u$ in \Eq{eq:Taylor-expansion}. Taking into account even lower powers of $L$ would require diagrams beyond the parquet approximation. Differently from $\chi$, the expansion of $G$, \Eq{eq:Taylor-expansion_G}, is in terms of $u^2\bar{L}$. So, with higher orders of $u$,
the difference in the powers of $u$ and of $\bar{L}$ increases.
Hence, a perturbative analysis of $G$ (and thus the overlap related to Anderson's orthogonality catastrophe~\cite{Anderson1967Infrared}) would require going beyond leading/subleading logarithms and beyond the parquet approximation (cf.\ \App{sec:AOC_Self-energy}).

\subsection{Numerical parameters}
\label{sec:numerical-parameters}

For all our plots, we fix the dimensionless interaction strength to $u=0.28$, if not stated otherwise. The analytical results are presented for $T=0$, where $\frac{1}{\beta}\sum_\nu \to \int_{-\infty}^\infty\frac{\dd\nu}{2\pi}\equiv\int_\nu$. There, we mostly use $g^\mathrm{sh}_\nu$, \Eq{eq:Gc_sharp}, as we focus on the behavior near the threshold, $|\ii\omega+\xi_d| \ll \xi_0$. The numerical results are obtained for a finite temperature $T/\xi_0=0.002$ and a discrete grid of Matsubara frequencies. For numerically determined perturbative results, we compare both propagator choices $g_\nu$ in \Eqs{eq:Gc_smooth} and \eqref{eq:Gc_sharp}. Details of the implementation are given in \App{sec:Numerics}.

The remaining parameter is the excitation energy $\xi_d$. Physically, one imagines $\varepsilon_d < 0$ and $\xi_d \ll -T$, so that $f(\xi_d) \simeq 1$, corresponding to an occupied core level. In our diagrammatic approach, we have already used $f(\xi_d) \simeq 1$ by canceling the term $U/V$ from normal-ordering with the $c$-electron Hartree self-energy equal to $U/V f(\xi_d)$. Consequently, there are no more $c$ Hartree diagrams (involving a closed $d$ line) in the expansion, and the $d$-level occupation is never actually probed. Instead, there is a single $d$ line threading through all diagrams of $\chi(\ii\omega)$, and one may choose to keep the external frequency argument paired with $\xi_d$, e.g., in the form $\ii\omega + \xi_d$ (cf.\ \Sec{sec:Threshold}). As a result, $\xi_d$ can be shifted to any (negative) value~\cite{Diekmann2021,Diekmann2024Leading}.

Here, we use $\xi_d/\xi_0=-0.01$. The reason is that, in the Matsubara formalism, $\xi_d$ broadens the characteristic features of the correlation functions $\chi$ and $G$, and, to reduce the effects of such a broadening, we use small $|\xi_d|\ll\xi_0$. Additionally, this corresponds to larger values of $\ln(-\xi_d/\xi_0)$, which is beneficial to clearly separate logarithmic terms of different powers at small frequencies in the perturbative expansions, \Eqs{eq:Taylor-expansion}--\eqref{eq:Taylor-expansion_G}. However, we keep $|\xi_d|>\pi T$
so that features below $\xi_d$ (the lowest non-thermal energy scale) are resolved by the Matsubara grid. After analytical continuation, the parameter $\xi_d$ eventually only shifts the threshold frequency $\omega_0$ and is irrelevant for the analysis of the power-law behavior.

\section{Perturbation theory}
\label{sec:Perturbation-theory}

To get an intuition about typical diagrammatic contributions to the infrared divergence to subleading accuracy, we analyze Feynman diagrams at low orders. We start with the well-known leading logarithmic terms in the particle-hole susceptibility. Next, we discuss subleading terms in the self-energy and the vertex. Finally, we consider a multi-boson exchange diagram and present a rule to generally assess the logarithmic behavior in the present model. The logarithmic behavior of the diagrams in Secs.~\ref{sec:leading-logarithms}--\ref{sec:gamma_t} was already discussed in Refs.~\cite{roulet1969singularities,nozieres1969singularities2}. We here extend their analysis by giving numerical results along with some exact analytical results [\Eqs{eq:simple-bubble} and \eqref{eq:Sigma_2_exact_main_text_full}] as well as a general rule for extracting the logarithmic behavior.

\subsection{Leading logarithmic diagrams}
\label{sec:leading-logarithms}

\begin{figure}
    \centering
    \includegraphics[width=0.7\linewidth]{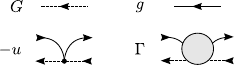}
    \caption{Diagrammatic representation of the $d$ propagator $G$ and the $c$ propagator $g$ as well as the bare vertex $\Gamma^{(1)}=-u$ and the full vertex $\Gamma$.
    }
    \label{fig:definitions-diagrams}
\end{figure}

Utilizing a similar notation and diagrammatic representation as in Refs.~\cite{KuglerPRL2018b,Kugler2018a,KuglerNJP2018e,gievers2022multiloop}, the particle-hole susceptibility is given by
\begin{align}
    \nonumber \chi(\omega)
    &= \quad
    \begin{gathered}
        \includegraphics[width=110pt]{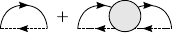}
    \end{gathered}\\
    &= \int_\nu\Pi^a_{\omega,\nu} + \int_{\nu,\nu'}\Pi^a_{\omega,\nu}\Gamma^a_{\omega,\nu,\nu'}\Pi^a_{\omega,\nu'}.
    \label{eq:chi_diagram}
\end{align}
Here, $\Gamma$ refers to the full four-point vertex. By energy conversation, it depends on only three frequencies (cf.~\App{sec:Vertex-conventions}). The index $a$ in $\Gamma^a$ signifies that its frequencies are parametrized with respect to the $a$ channel ($a$ stands for \emph{antiparallel} and $p$, used below, for \emph{parallel}). The bubbles $\Pi^r_{\omega,\nu}$ are products of Green's functions,
\begin{align}
    \Pi^{a}_{\omega,\nu} 
    = 
    G_{\nu-\omega} g_{\nu}, 
    \quad 
    \Pi^p_{\omega,\nu} 
    = 
    G_{\nu-\omega} g_{-\nu}
    =
    - \Pi^{a}_{\omega,\nu} 
    ,
    \label{eq:bubbles-definition}
\end{align}
having used $g_{-\nu}=-g_{\nu}$ in the last step. Diagrammatically, $G$ is represented by a dashed line and $g$ by a solid line. The vertex $\Gamma$ is denoted by a gray circle, its lowest-order contribution $\Gamma^{(1)}=-u$ by a black dot (cf.\ \Fig{fig:definitions-diagrams}).

The lowest-order term of the susceptibility $\chi^{(0)}$ is an integrated $a$ bubble:
\begin{align}
    \nonumber \chi^{(0)}(\omega) &= ~
    \begin{gathered}
        \includegraphics[width=30pt]{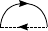}
    \end{gathered}
    ~ = ~ - ~
    \begin{gathered}
        \includegraphics[width=30pt]{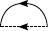}
    \end{gathered}\\
    &= \int_\nu\Pi^{a}_{\omega,\nu} =
    \ln\frac{\ii\omega+\xi_d}{\ii\omega+\xi_d-\xi_0}
    \simeq L(\omega).
    \label{eq:simple-bubble}
\end{align}
The first expression is the exact result [cf.\ \Eq{eq:simple-bubble_exact}] with $g^\mathrm{sm}$, \Eq{eq:Gc_smooth}. The second gives the logarithmic behavior, \Eq{eq:definition_L}, near the threshold (indicated by the symbol ``$\simeq$''), which is also obtained with $g^\mathrm{sh}$ (cf.~\Eqs{eq:integral_bubble-f}--\eqref{eq:simple-bubble_details2} and Refs.~\cite{Mahan1967Excitons,roulet1969singularities,kugler2018fermi,KuglerPRL2018b,Diekmann2021,Diekmann2024Leading}).
Indeed, the approximation is justified for $|\ii\omega+\xi_d|\ll\xi_0$, which, after analytic continuation $\ii\omega\to\omega+\ii 0^+$, corresponds to frequencies close to the absorption threshold $-\xi_d = |\xi_d|$. 

\begin{figure}
    \centering
    \includegraphics[width=1.0\linewidth]{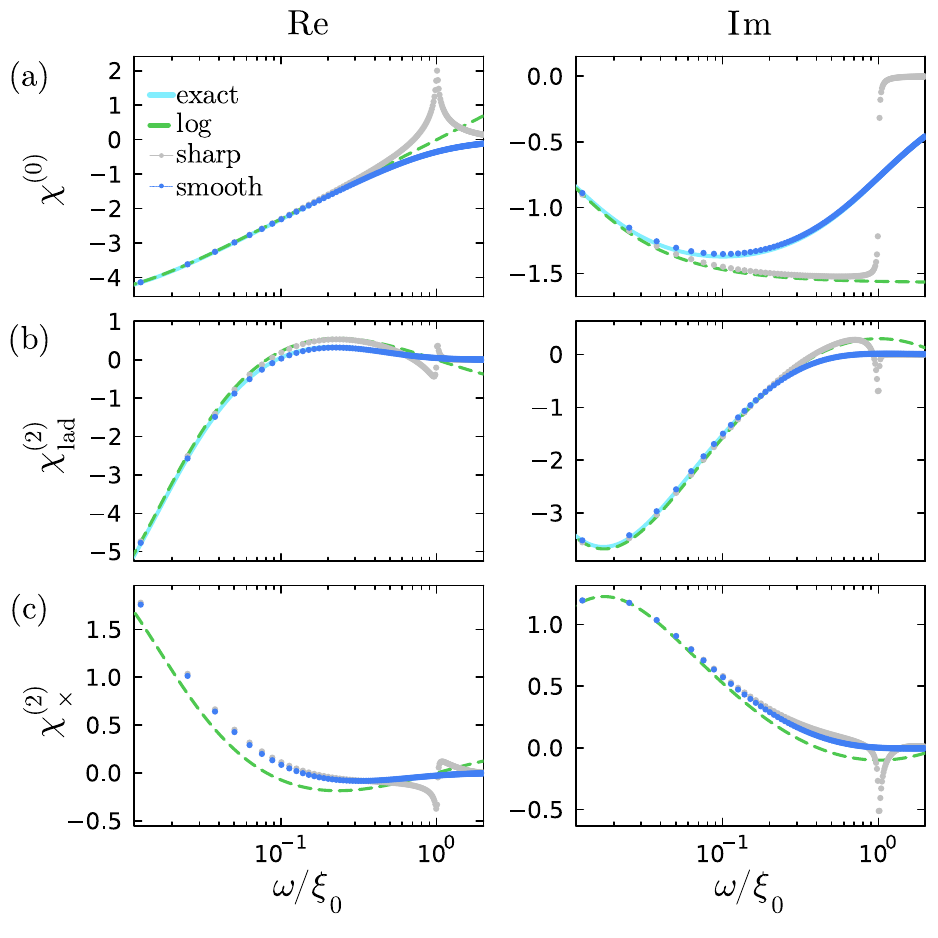}
    \caption{
    (a) Zeroth-order diagram $\chi^{(0)} \!\simeq\! L$, \Eq{eq:simple-bubble},
    (b) second-order ladder diagram $\chi_\mathrm{lad}^{(2)} \!\simeq\! u^2 L^3$, \Eq{eq:ladder-diagrams}, and
    (c) second-order cross diagram $\chi_\times^{(2)} \!\simeq\! -\tfrac{1}{3}u^2 L^3$, \Eq{eq:cross-diagram}, comparing numerical results with smooth $g^\mathrm{sm}$ (blue dots) and sharp $g^\mathrm{sh}$ (gray dots) to the analytical logarithmic form (green, dashed) and the exact result (light blue, solid).
    }
    \label{fig:ladder-diagrams}
\end{figure}

Figure~\ref{fig:ladder-diagrams}(a) shows the frequency dependence of $\chi^{(0)}$, \Eq{eq:simple-bubble}. The result from the smooth propagator $g^\mathrm{sm}$ is the exact result at $u = 0$ and is seen to obey $\mathrm{Re}\,\chi^{(0)}<0$. As expected, the numerical data obtained from $g^\mathrm{sm}$ at finite temperature (blue dots) lie on top of the analytical exact result at $T=0$ (light blue line). Furthermore, we note that  (i) the numerical result with the sharp propagator $g^\mathrm{sh}$, \Eq{eq:Gc_sharp}, (gray dots) yields artifacts around $\omega\simeq\pm\xi_0$ and violates $\Re\,\chi^{(0)}<0$; (ii) the (approximate) analytical result $L(\omega)$ violates $\mathrm{Re}\,\chi^{(0)}<0$ as well as $\lim_{|\omega|\to 0} \chi^{(0)} \to 0$.

The simplest diagrams of the vertex $\Gamma$ are ladder diagrams, which are products of $\chi^{(0)}$. We consider ladder diagrams in the antiparallel ($\gamma^a$) and parallel ($\gamma^p$) channels, built from the antiparallel ($\Pi^a$) and parallel ($\Pi^p$) bubble, respectively. Their $n$th-order contributions are (the external legs are amputated): 
\begin{subequations}
    \label{eq:ladder-diagrams_gamma}
    \begin{align}
        [\gamma^{a}_\mathrm{lad}]^{(n\geq 2)}_\omega
        &= \quad
        \begin{gathered}
            \includegraphics[width=110pt]{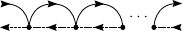}
        \end{gathered}
        \\
        \nonumber &=
        (-u)^n\left[\int_\nu\Pi^{a}_{\omega,\nu}\right]^{n-1} \simeq (-u)^n [L(\omega)]^{n-1},
        \\
        [\gamma^{p}_\mathrm{lad}]^{(n\geq 2)}_\omega
        &= \quad
        \begin{gathered}
            \includegraphics[width=110pt]{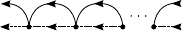}
        \end{gathered}
        \\
        \nonumber &=
        (-u)^n\left[\int_\nu \Pi^{p}_{\omega,\nu}\right]^{n-1} \simeq (-u)^n[-L(\omega)]^{n-1}.
    \end{align}
\end{subequations}

The ladder diagrams of $\chi$ have two more integrated bubbles, and their logarithmic behavior is thus:
\begin{align}
    \chi^{(n)}_\mathrm{lad}(\omega)
    &= \quad
    \begin{gathered}
        \includegraphics[width=110pt]{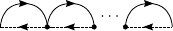}
    \end{gathered}
     \label{eq:ladder-diagrams}
    \\
    \nonumber &=
    \int_{\nu,\nu'}\Pi^a_{\omega,\nu}[\gamma^{a}_\mathrm{lad}]^{(n)}_\omega\Pi^a_{\omega,\nu'}
    \simeq
    (-u)^n[L(\omega)]^{n+1}.
\end{align}
Figure~\ref{fig:ladder-diagrams}(b) shows the frequency dependence of the second-order ladder diagram $\chi^{(2)}_\mathrm{lad}$, \Eq{eq:ladder-diagrams}. Similarly as in Fig.~\ref{fig:ladder-diagrams}(a), one notices artifacts at $\omega\simeq\xi_0$ in the numerical solution with $g^\mathrm{sh}$ as well as spurious high-frequency behavior in this solution and the analytical result. While the full susceptibility obeys $\Re\,\chi < 0$, for a single diagrammatic contribution, $\Re\,\chi^{(2)}_\mathrm{lad}<0$ need not hold. 

The so-called \emph{crossed diagram} is the first non-ladder diagram which contributes to the leading logarithm of the susceptibility. It is obtained by integrating $[\gamma^p]^{(2)}$ with two $a$ bubbles (see \App{sec:leading-logarithms_details} for details):
\begin{align}
    \chi_{\times}^{(2)}(\omega)
    &= \quad
    \begin{gathered}
        \includegraphics[width=90pt]{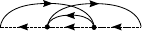}
    \end{gathered}
    \label{eq:cross-diagram}
    \\
    \nonumber &=
    \int_{\nu,\nu'}\Pi^a_{\omega,\nu}[\gamma^p]^{(2)}_{\omega-\nu-\nu'}\Pi^a_{\omega,\nu'}
    \simeq
    -\tfrac{1}{3}u^2 L^3(\omega).
\end{align}
Figure~\ref{fig:ladder-diagrams}(c) shows the numerical result for $\chi_{\times}^{(2)}(\omega)$. The opposite sign compared to $\chi^{(0)}$ and $\chi^{(2)}_\mathrm{lad}$ shows that the crossed diagram counteracts the growth of the full result.

Generally, susceptibility diagrams proportional to $u^n L^{n+1}$ are referred to as \emph{leading log}. Summing up only ladder diagrams yields the random phase approximation (RPA), resulting in an unphysical bound state~\cite{Mahan1967Excitons}. It was shown that, for a complete summation of leading-log diagrams, one has to take into account the interplay between the $a$ and $p$ channels. This is referred to as the first-order parquet approach~\cite{roulet1969singularities,nozieres1969singularities2} because it takes only the highest power of logarithms in each order of the power-law expansion, \Eq{eq:Taylor-expansion}. In a more general perturbative treatment, diagrams proportional to lower orders of $L$ show up, i.e., $u^n L^{n+1-p}$ with $n+1>p>0$. In this work, we will go beyond Ref.~\cite{roulet1969singularities} by including all diagrams with $p=1$, which we denote as \emph{subleading log}.

\subsection{Self-energy}
\label{sec:Self-energy}

\begin{figure}
    \label{fig:Sigma-2}
    \centering
    \includegraphics[width=1.0\linewidth]{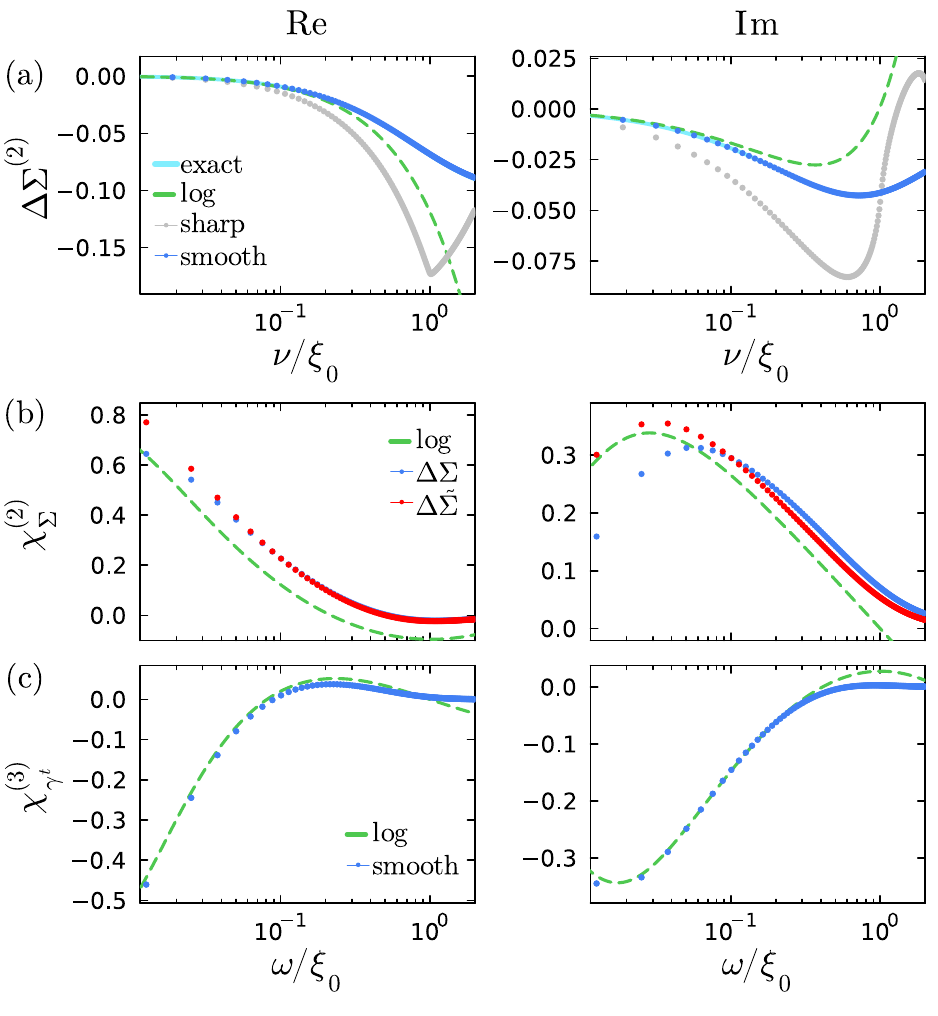}
    \caption{
    (a) Second-order self-energy $\Sigma^{(2)} \simeq u^2 \ii\tilde\nu\bar L$, \Eq{eq:Sigma-2}, where the zero-frequency value is subtracted, i.e., $\Delta\Sigma_\nu^{(2)} = \Sigma_\nu^{(2)}-\Re\,\Sigma_{\nu=0}^{(2)}$ [cf.\ \Eq{eq:Self-energy_subtraction}].
    (b) The corresponding susceptibility $\chi^{(2)}_\Sigma \simeq \tfrac{1}{2} u^2 L^2$, \Eq{eq:chi_2_Sigma}. Here, we only use the numerical results computed with the smooth propagator $g^\mathrm{sm}$, first with the self-energy difference $\Delta\Sigma^{(2)}_\nu$ (blue dots) and then with $\Delta\tilde\Sigma^{(2)}$, \Eq{eq:Self-energy_subtraction_remedy}, including the logarithmic part at $\nu=0$ (red dots).
    (c) Third-order diagram $\chi_{\gamma^t}^{(3)}\simeq\tfrac{1}{3}u^3L^3$, \Eq{eq:gamma_t_3}, originating from the $t$-reducible vertex $[\gamma^t]^{(3)}$.
    }
    \label{fig:selfenergy-2}
\end{figure}

To include subleading-log contributions, we next consider the impact of the $d$-electron self-energy~\cite{nozieres1969singularities2}, which was neglected in most previous diagrammatic analyses~\cite{Mahan1967Excitons,roulet1969singularities,Lange_2015,kugler2018fermi,KuglerPRL2018b,Diekmann2021,Diekmann2024Leading}. The $d$ Hartree self-energy shifts the threshold frequency by $u\xi_0$ [cf.\ \Eq{eq:Hartree}]. The second-order diagram $\Sigma^{(2)}$ is the first to exhibit (subleading) logarithmic behavior. Its general expression,
\begin{align}
    \nonumber\Sigma^{(2)}_\nu
    &= \quad 
    \begin{gathered}
        \includegraphics[width=84pt]{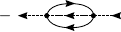}
    \end{gathered}
    \\
    &=
    -u^2\int_{\omega,\nu'}G_{\nu-\omega}g_{\nu'-\omega}g_{\nu'},
    \label{eq:Sigma-2}
\end{align}
is evaluated to (cf.\ \App{sec:Self-energy_details}):
\begin{subequations}
    \label{eq:Sigma_2_exact_main_text_full}
    \begin{align}
        \frac{1}{u^2} \Sigma^{(2)}_\nu
        & = \ii\tilde{\nu}\ln\frac{\ii\tilde{\nu}}{\ii\tilde{\nu}+\xi_0}
        +
        (\ii\tilde{\nu}+2\xi_0)\ln\frac{\ii\tilde{\nu}+2\xi_0}{\ii\tilde{\nu}+\xi_0}
        \label{eq:Sigma_2_exact_main_text}
        \\
        & \simeq
        \ii\tilde{\nu} \ln\frac{\ii\tilde{\nu}}{\xi_0} +  2 \xi_0 \ln 2
        ,
        \label{eq:Sigma_2_approximated}
    \end{align}
\end{subequations}
where $\ii\tilde{\nu} = \ii\nu - \xi_d$. The first summand gives the logarithmic behavior $(\ii\nu-\xi_d) L(-\nu)$
that contributes to the shape of the x-ray edge singularity and was given in previous studies (cf.\ line after Eq.~(13) in Ref.~\cite{nozieres1969singularities2}). The second, proportional to $\xi_0$ is constant and thus merely shifts the threshold frequency, just like the Hartree term $u\xi_0$ mentioned before.

As discussed in \Sec{sec:numerical-parameters}, we want to exclude shifts of the threshold frequency since they blur the singular frequency dependence. The easiest way to do so is to generally subtract the zero-frequency part \cite{nozieres1969singularities2}. In the Matsubara formalism, this amounts to replacing $\Sigma_\nu$ by 
\begin{align}
    \Delta\Sigma_\nu = \Sigma_\nu - \mathrm{Re}\,\Sigma_{\nu=0},
    \label{eq:Self-energy_subtraction}
\end{align}
as $\mathrm{Im}\,\Sigma_\nu$ is antisymmetric and thus vanishes at zero frequency. (In practice, we approximate $\Re\,\Sigma_{\nu=0}$ by $\Re\,\Sigma_{\nu=\pi T}$.) However, the (imaginary-frequency) logarithmic terms of the self-energy also have a finite contribution at $\nu=0$, which we do not want to subtract. Indeed, the first term in \Eq{eq:Sigma_2_approximated} evaluates to $-\xi_d L(0)$ at zero frequency. If the latter contribution to the shift is added back, we obtain
\begin{align}
    \Delta\tilde\Sigma_\nu = \Delta\Sigma_\nu - u^2 \xi_d L(0).
    \label{eq:Self-energy_subtraction_remedy}
\end{align}

Figure~\ref{fig:selfenergy-2}(a) shows our numerical results for the self-energy. The finite-$T$ results using $g^\mathrm{sm}$ [blue dots in \Fig{fig:selfenergy-2}(a)] lie on top of the $T=0$ exact result, \Eq{eq:Sigma_2_exact_main_text} [light blue line in \Fig{fig:selfenergy-2}(a)]. We see that using $g^\mathrm{sh}$, \Eq{eq:Gc_sharp}, violates causality, i.e., leads to $\Im\, \Sigma^{(2)} > 0$ for $\nu > 0$ [gray dots in \Fig{fig:selfenergy-2}(a)], whereas the result from $g^\mathrm{sm}$, \Eq{eq:Gc_smooth}, naturally obeys this property [blue dots in \Fig{fig:selfenergy-2}(a)]. At low frequencies, the results from $g^\mathrm{sm}$ agree well with the analytic logarithmic behavior $u^2(\ii\nu-\xi_d) L(-\nu)$ (green), which however bends over to unphysical results with $\Im\, \Sigma^{(2)} > 0$ already for $\nu/\xi_0 \gtrsim 0.5$.

One directly sees that the corresponding second-order term of the $d$ propagator $G^{(2)} = G^{(0)}\Sigma^{(2)}G^{(0)} = u^2 L(-\nu) / (\ii\nu-\xi_d)$ matches the perturbative series, \Eq{eq:Taylor-expansion_G}. To find the corresponding subleading-log term for $\chi$, we insert $G^{(2)}$ into the integrated bubble, \Eq{eq:simple-bubble}:
\begin{align}
    \nonumber \chi^{(2)}_{\Sigma}(\omega)
    &= \quad 
    \begin{gathered}
        \includegraphics[width=105pt]{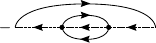}
    \end{gathered}
    \\
    &=
    \int_\nu [\Pi^a]^{(0)}_{\omega,\nu}\Sigma^{(2)}_{\nu-\omega} G^{(0)}_{\nu-\omega} \simeq \tfrac{1}{2}u^2 L^2(\omega).
    \label{eq:chi_2_Sigma}
\end{align}  
The same logarithmic term appears in the perturbative series of the power law, \Eq{eq:Taylor-expansion}, which was evaluated at the bare threshold frequency $\omega_0=-\xi_d$. Hence, the logarithmic term $u^2(\ii\nu-\xi_d)L(-\nu)$ of $\Sigma^{(2)}$, \Eq{eq:Sigma-2}, does indeed not change the threshold frequency (while the full expression, \Eq{eq:Sigma_2_exact_main_text_full}, does).

The numerical results for $\chi^{(2)}_\Sigma$ are first computed with $\Delta\Sigma^{(2)}$, \Eq{eq:Self-energy_subtraction} [blue dots in \Fig{fig:selfenergy-2}(b)]. To minimize the effect of the threshold shift when computing $\chi^{(2)}_\Sigma$, we use $\Delta\tilde\Sigma^{(2)}$, \Eq{eq:Self-energy_subtraction_remedy}, which adds back the logarithmic contribution $-u^2\xi_d L(0)$. Primarily, this brings the imaginary part of the data closer to the analytical result [red dots in \Fig{fig:selfenergy-2}(b)]. This strategy to compensate shifts of the threshold is further discussed in \Sec{sec:Threshold}.

At third order, there are two diagrams contributing to the self-energy, which cancel each other as $[\gamma^{a}_\mathrm{lad}]^{(3)}_\omega=[\gamma^{p}_\mathrm{lad}]^{(3)}_{-\omega}$ [cf.\ \Eq{eq:Sigma_3}]. This observation matches with the exact power law of the $d$ propagator, which only scales with $u^2$ [cf.\ \Eq{eq:Taylor-expansion_G}].

\subsection{$t$-reducible diagram}
\label{sec:gamma_t}

Generally, there is a third type of two-particle reducibility, namely in the \emph{transversal} channel (or short $t$ channel). Its first contribution occurs at third order and reads (cf.\ Eq.~(15) in Ref.~\cite{nozieres1969singularities2} and \App{sec:gamma_t_details} for details)
\begin{align}
    \nonumber [\gamma^t]^{(3)}_{\omega,\nu'}
    &= \quad 
    \begin{gathered}
        \includegraphics[width=105pt]{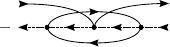}
    \end{gathered}
    \\
    \nonumber &=
    u^3\int_{\nu_1,\nu_2}G_{\nu_1}G_{\nu_1-\omega}g_{\nu_2}g_{\nu_1+\nu_2-\nu'}
    \\
    &=
    \frac{u}{\ii\omega}\left(\Sigma^{(2)}_{\nu'}-\Sigma^{(2)}_{\nu'-\omega}\right).
    \label{eq:gamma_t_3}
\end{align}
The expression in terms of the self-energy $\Sigma^{(2)}$ is exact and does not depend on the form of $g$. Indeed, it is a perturbative implementation of the $U(1)$ Ward identity (cf.\ \App{sec:Ward-identity}). Evidently, the logarithmic behavior of the self-energy and the $t$-reducible diagram are related~\cite{nozieres1969singularities2}. Note that, to save space and to emphasize the advanced property of the $d$ electron by a straight dashed line, we refrain from drawing the bubble of two $d$ propagators vertically, which originally motivates the term \emph{transversal}~\cite{KuglerPRL2018b,KuglerNJP2018e,gievers2022multiloop}.

The resulting third-order term for $\chi$ is (see \App{sec:gamma_t_details} for details):
\begin{align}
    \nonumber\chi^{(3)}_{\gamma^t}(\omega)
    &= \quad
    \begin{gathered}
        \includegraphics[width=135pt]{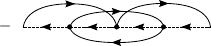}
    \end{gathered}
    \label{eq:chi_gamma_t_3}
    \\
    &=
    \int_{\nu,\nu'}\!\!\Pi^a_{\omega,\nu}[\gamma^t]^{(3)}_{\nu-\nu',\nu-\omega}\Pi^a_{\omega,\nu'}\simeq \tfrac{1}{3}u^3 L^3(\omega).
\end{align}
Figure~\ref{fig:selfenergy-2}(c) shows that our numerical data are close to this subleading-log behavior. Together with the two terms of third order coming from $\Sigma^{(2)}$,
\begin{align}
    \nonumber &
    \chi^{(2)}_{\Sigma}(-u)\chi^{(0)} + \chi^{(0)}(-u)\chi^{(2)}_{\Sigma}
    \\
    \nonumber &= ~\, 
    \begin{gathered}
        \includegraphics[width=130pt]{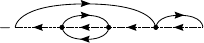}\phantom{~,}
    \end{gathered}
    \\
    \nonumber &\phantom{=}\quad
    \begin{gathered}
        \includegraphics[width=130pt]{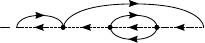}
    \end{gathered}
    \\
    &\simeq
    2\times \tfrac{1}{2} u^2 L^2 \times (-u) \times  L = -u^3 L^3,
\end{align}
we hence arrive at the correct subleading-log contribution to third order, namely $-\tfrac{2}{3}u^3 L^3$ [cf.\ \Eq{eq:Taylor-expansion}].

\subsection{Multi-boson exchange diagram}
\label{sec:MBE}

Multi-boson exchange (MBE) diagrams are two-particle reducible in a specific channel, i.e., their diagrams fall apart when cutting two propagator lines. However, they are not $U$ reducible, which means that their diagrams do not fall apart by removing one dot of a bare vertex~\cite{gievers2022multiloop}. Numerically, they are the most expensive objects to compute as they inherently depend on three frequency arguments. Thus computational resources can be saved if their impact on physical quantities is low compared to other diagrams.

\begin{figure}
    \centering
    \includegraphics[width=1.0\linewidth]{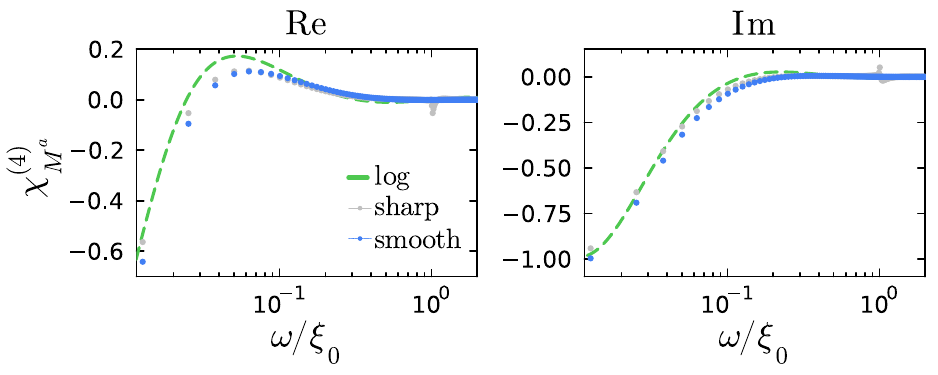}
    \caption{Susceptibility contribution from the MBE diagram $[M^a]^{(4)}$. We compare the results computed with $g^\mathrm{sm}$ (blue dots) and $g^\mathrm{sh}$ (gray dots) to the analytically determined logarithmic behavior $\tfrac{2}{15}u^4 L^5$, \Eq{eq:MBE-diagram} (green, dashed).}
    \label{fig:MBE-diagram}
\end{figure}

To analyze the relevance of MBE diagrams, we analytically check how the first multi-boson diagram $[M^a]^{(4)}$, occurring at fourth order,
\begin{align}
    \nonumber [M^a]^{(4)}_{\omega,\nu,\nu'}
    &= \quad
    \begin{gathered}
        \includegraphics[width=120pt]{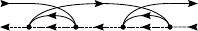}
    \end{gathered}
    \\
    &=
    \int_{\nu''}[\gamma^p]^{(2)}_{\omega-\nu-\nu''}\Pi^a_{\omega,\nu''}[\gamma^p]^{(2)}_{\omega-\nu''-\nu'},
    \label{eq:MBE4}
\end{align}
affects $\chi$. We find (see \App{sec:MBE4_details} for details):
\begin{align}
    \nonumber\chi^{(4)}_{M^a}(\omega)
    &= \quad
    \begin{gathered}
        \includegraphics[width=150pt]{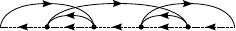}
    \end{gathered}
    \\
    &=
    \int_{\nu,\nu'}\Pi^a_{\omega,\nu}[M^a]^{(4)}_{\omega,\nu,\nu'}\Pi^a_{\omega,\nu'}\simeq \tfrac{2}{15}u^4 L^5(\omega),
    \label{eq:MBE-diagram}
\end{align}
which is leading log. Figure~\ref{fig:MBE-diagram} confirms this behavior in finite-$T$ numerical results. Hence, omitting MBE diagrams leads to an incomplete summation of diagrams already at leading-log order. We thus include MBE diagrams in the self-consist schemes presented in \Sec{sec:Parquet}. As seen there, neglecting them changes the results drastically at intermediate values of $u$. 

\subsection{Comment on the threshold frequency $\omega_0$}
\label{sec:Threshold}

Since we have numerical access to the full frequency dependence of individual diagrams, we can make a statement about the position of the threshold frequency, in contrast to previous diagrammatic analyses~\cite{roulet1969singularities,nozieres1969singularities2,Lange_2015,kugler2018fermi,KuglerPRL2018b,Diekmann2021,Diekmann2024Leading}. Generally, the threshold frequency $-\xi_d$ is shifted to a value $\omega_0$ which depends on the interaction $u$. A large value of $\omega_0$ blurs the characteristic features of imaginary-frequency data near $\omega=0$. For our numerical results presented in \Sec{sec:numerical-results}, we thus needed a strategy to compensate this effect, which is discussed in the following.

\begin{figure}
    \centering
    \includegraphics[width=0.82\linewidth]{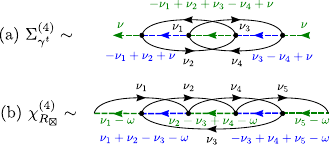}
    \caption{Demonstration that the single dashed line of $d$ propagators in diagrams implies that $\Sigma = \Sigma(\ii\nu-\xi_d)$ and $\chi = \chi(\ii\omega+\xi_d)$. (a) Fourth-order diagram to $\Sigma$ containing the $t$-reducible diagram $[\gamma^t]^{(3)}$. (b) Fourth-order diagram to $\chi$ containing the envelope diagram $R^{(4)}_\boxtimes$. The external frequencies $\nu$ and $\omega$ are contained in $d$ propagators, the frequencies $\nu_1,...,\nu_5$ are integrated over.}
    \label{fig:dashed-line}
\end{figure}

As mentioned in \Sec{sec:numerical-parameters}, our treatment of the $c$ Hartree self-energy and the advanced property of the $d$ propagator $G$ imply that there are no closed $d$ loops in the diagrammatic expansion. In other words, all non-vanishing diagrams must contain a single line of $d$ propagators. Hereby, it is always possible to find a frequency parametrization such that the external frequencies $\omega$ in $\chi$ and $\nu$ in $\Sigma$ are only contained in the $d$ propagators. (For four-point vertices, the according external frequencies are discussed in \App{sec:Vertex-conventions}.)

To demonstrate that this argument also holds for diagrams which are neither $a$- nor $p$-reducible, we show a self-energy diagram containing the third-order vertex $[\gamma^t]^{(3)}$ and a susceptibility diagram containing the fourth-order two-particle irreducible vertex $R^{(4)}_\boxtimes$ in \Fig{fig:dashed-line}. Here, we explicitly write the frequency arguments on every single propagator and conclude that the susceptibility $\chi$ is a function of $\ii\omega+\xi_d$ and the self-energy $\Sigma$ of $\ii\nu-\xi_d$. Thus, after analytical continuation to real frequencies, the effect of different values of $\xi_d$ can be recovered by corresponding shifts of the external frequencies~\footnote{In shifting $\xi_d$, we assume that no nonanalyticities of $\chi$ in $\ii\omega + \xi_d$ or of $\Sigma$ in $\ii\nu - \xi_d$ are crossed. Indeed, we have not detected such nonanalyticities in the analytical expressions from perturbation theory and our numerical data.}.

The threshold frequency $\omega_0$ appears likewise in the power laws of $\chi$ and $G$ [cf.\ \Eqs{eq:exact_power-laws}]. These forms demonstrate that, near the threshold, $\chi$ is described by a function of $\ii\omega-\omega_0$ and $\Sigma$ by a function of $\ii\nu+\omega_0$, suggesting that $\omega_0$ merely renormalizes its non-interacting correspondent $-\xi_d$. The threshold is fully determined by the self-energy as, after analytical continuation, i.e., $\ii\nu\to\nu-\ii 0^+$, the expression from the Dyson equation $1/(\ii\nu-\xi_d-\Sigma_\nu)$ becomes singular at $\nu = -\omega_0$.

Our perturbative analysis shows that the self-energy consists both of terms which do not affect the threshold-frequency and terms which renormalize it. In second order [cf.\ \Eq{eq:Sigma_2_exact_main_text_full}], the logarithmic term $u^2(\ii\nu-\xi_d)L(-\nu)$ does not change the threshold, since it appears in the power law expansion with an unrenormalized threshold $\omega_0=-\xi_d$ [cf.\ \Eq{eq:Taylor-expansion_G}], while the term $u^2\, 2\xi_0 \ln 2$, proportional to the UV cutoff $\xi_0$, does [cf.\ \Eqs{eq:Threshold_second-order} and \eqref{eq:Taylor-expansion_chi_w_threshold}]. Without knowing the analytical results, this separation of terms in the self-energy cannot be extended straightforwardly to higher orders in perturbation theory.

Similar to the second-order term, we suspect that more generally the dependence on the threshold frequency is mainly governed by a constant shift in the self-energy. So, to leave the threshold frequency unrenormalized in numerical computations of more general diagrams, we subtract the value $\mathrm{Re}\,\Sigma_{\nu=0}$ from the numerically computed self-energy $\Sigma_\nu$ [cf.\ \Eq{eq:Self-energy_subtraction}]. (The smallest imaginary Matsubara frequency $\ii\pi T$ is the closest value to the real threshold frequency $-\omega_0$.) Still, we have to recover those terms at $\nu=0$ that do not renormalize the threshold frequency. The only expression we know analytically is that of second order, so we add $-u^2\xi_d L(0)$ [cf.\ \Eq{eq:Self-energy_subtraction_remedy}], as done for $\chi^{(2)}_\Sigma$ [cf.\ red dots in \Fig{fig:selfenergy-2}(b)].

Although this strategy does not guarantee a full elimination of the threshold renormalization in our numerical computations, it does allow us to deduce reasonable values for the threshold frequency from the numerically determined subleading-log self-energy (cf.\ \App{sec:FDA_Threshold}, in particular \Fig{fig:FDA_threshold}).

\subsection{Logarithmic behavior in general diagrams}
\label{sec:logarithmic_behavior}

We show how to quickly deduce the leading power of the logarithm $L$ in the singular behavior of any diagram involving $d$ and $c$ propagators (see also App.~E of Ref.~\cite{nozieres1969singularities2}). Close to the threshold, the bare $d$ propagator behaves as $G_\nu \sim 1/(\ii\nu)$, while the local $c$ propagator obeys $g_\nu \sim\sgn (\nu)$ at small frequencies [cf.\ \Eq{eq:Gc_sharp}]. We estimate the leading logarithm by successively integrating over closed loops.

\begin{figure}
    \centering
    \includegraphics[width=0.95\linewidth]{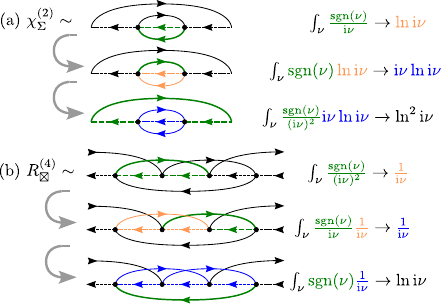}
    \caption{
    Step-wise procedure to deduce the logarithmic behavior of diagrams, exemplified for (a) $\chi_\Sigma^{(2)}\sim u^2 L^2$, \Eq{eq:chi_2_Sigma}, and (b) a fourth-order two-particle irreducible diagram, $R^{(4)}_\boxtimes \sim u^4 L$. We successively integrate loops involving $d$ and $c$ lines; the lines being integrated at a given step are marked green.
    }
    \label{fig:logarithmic-order}
\end{figure}

Take, e.g., the susceptibility diagram $\chi^{(2)}_\Sigma$, \Eq{eq:chi_2_Sigma}, shown in \Fig{fig:logarithmic-order}(a). An integration over the inner $dc$ bubble yields a logarithm $\int_\nu \sgn(\nu)/(\ii\nu)\sim \ln\ii\nu$. The subsequent integral, involving the second $c$ propagator, does not raise the power of the logarithm, $\int_\nu \sgn(\nu)\ln \ii\nu\sim \ii\nu\ln\ii\nu$. The final integral, involving two $d$ propagators and one $c$ propagator, in turn raises its power, $\int_\nu \sgn(\nu) \ln \ii\nu / (\ii\nu)\sim \ln^2 \ii\nu$. This matches \Eq{eq:chi_2_Sigma}.

More generally, any integrated $dc$ bubble yields a logarithm, $\int_\nu  \sgn(\nu)/(\ii\nu)\sim \ln \ii\nu$ [cf.\ \Eq{eq:simple-bubble}]. If every subsequent loops contains another $d$ line, 
each integral increases the power of the logarithm according to $\int_\nu \sgn(\nu) \ln^n\ii\nu/(\ii\nu) \sim \ln^{n+1}\ii\nu$ [cf.\ \Eq{eq:integral-identity}]. This is precisely what happens for the leading-log diagrams. By contrast, if a loop at a later stage does not contain a further $d$ line, then the power of the logarithm is not raised, $\int_\nu \ln^n\ii\nu \sim \ii\nu\ln^n\ii\nu$ [cf.\ \Eqs{eq:integral_log}--\eqref{eq:integral_log^n}]. Now, if the first loop in a vertex diagram has more $d$ lines than $c$ lines, this does not generate logarithmic behavior, as $\int_\nu \sgn(\nu)/(\ii\nu)^m \sim 1/(\ii\nu)^{m-1}$. Such an imbalance of dashed and solid lines within integration loops yields subleading diagrams. Indeed, many more loops with $c$ lines are required until the power of the logarithm is raised. Since, $\int_\nu \sgn(\nu) \ln^n\ii\nu/(\ii\nu)^m \sim \ln^n\ii\nu/(\ii\nu)^{m-1}$ [cf.\ \Eqs{eq:integral_log/iv^n}--\eqref{eq:integral_log^m/iv^n}], the first $(m-1)$ subsequent integrals with $c$ lines reduce the power $m$ that originates from the $d$ propagators, before further integrals can eventually raise the power $n$ of the logarithm.

This reasoning also applies to diagrams which are two-particle irreducible in all three channel (\emph{totally irreducible} diagrams). These go beyond the full parquet approach. The lowest-order vertex diagram of that type occurs at fourth order and is often called \emph{envelope} diagram $R^{(4)}_\boxtimes$ (cf.\ Fig.~5 in Ref.~\cite{nozieres1969singularities2}). In \Fig{fig:logarithmic-order}(b), we show that its logarithmic behavior can be estimated as $u^4 L$. With two more $dc$ bubbles, the corresponding susceptibility is $\chi^{(4)}_{R_\boxtimes}=\int\Pi^aR^{(4)}_\boxtimes\Pi^a\sim u^4 L^3$, which has two powers of $L$ less compared to the leading logarithm $\chi_\mathrm{lead}^{(n)}\sim u^n L^{n+1}$ and is thus beyond our subleading approximation $\chi_\mathrm{sub}^{(n)}\sim u^n L^n$. (Actually, there are two envelope diagrams at fourth order, whose leading contributions cancel by symmetry (cf.\ footnote 10 in Ref.~\cite{nozieres1969singularities2}).) Further totally irreducible diagrams like the fifth-order ``sealed'' envelope diagram $R^{(5)}_\boxtimes \sim u^5 L \Rightarrow \chi^{(5)}_\boxtimes \sim u^5 L^3$ have even fewer powers of $L$.

The strategy presented above allows us to estimate the logarithmic behavior, but of course does not yield the correct prefactor and does not account for possible cancellations of diagrams. Nevertheless, it is essential for the next step. To obtain the power-law behavior of $\chi$ up to a certain accuracy, one has to perform a self-consistent summation, which takes into account all diagrams with the corresponding power of logarithms. With the presented strategy, we can classify all parts of the parquet formalism by their dominant logarithmic behavior.

\section{Self-consistent summation}
\label{sec:Parquet}

In this section, we extend the self-consistent summation of all leading-log diagrams of Ref.~\cite{roulet1969singularities} (\emph{first-order parquet} approach) toward the \emph{full parquet} approach, widely used to describe physics related to the Hubbard model~\cite{Bickers1991Conserving,Bickers1992,Chen1992Numerical,Bickers2004}. We show that, in this way, we can additionally include all subleading-log diagrams in a systematic manner.

\subsection{Parquet approach}
\label{sec:Parquet-schemes}

\begin{figure*}
    \centering
    \includegraphics[width=1.0\linewidth]{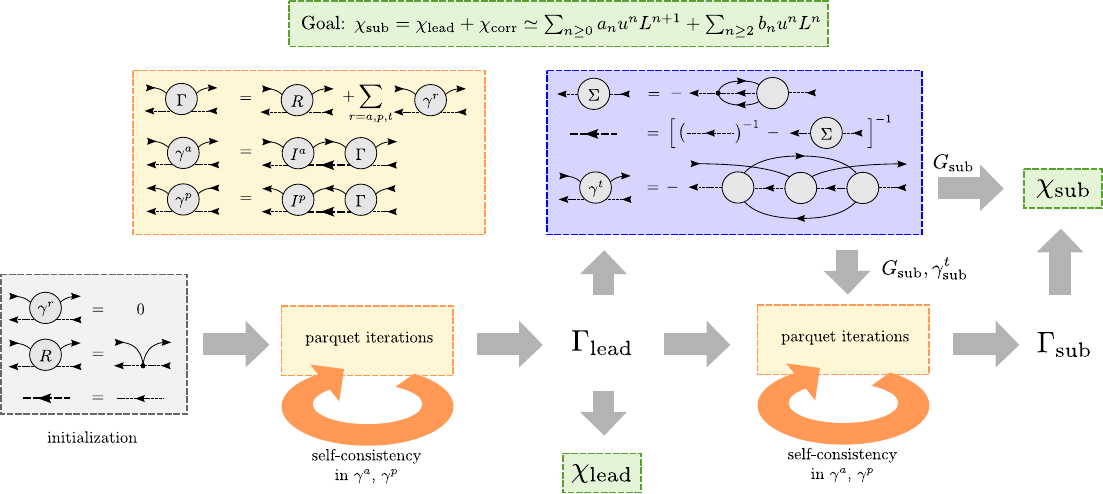}
    \caption{
    Overview of the computed quantities and used self-consistency schemes: The parquet iteration (yellow box) involves the Bethe--Salpeter equations \eqref{eq:parquet-leading-log} and \eqref{eq:Gamma_parquet_extended}. First, these are solved self-consistently (initialized using the gray box) without $\Sigma$ and $\gamma^t$ and give the vertex $\Gamma_\mathrm{lead}$ to leading-log accuracy. This output is used to compute $\Sigma$, \Eq{eq:SDE}, and $\gamma^t$, \Eq{eq:gamma_t} (blue box). These results in turn serve as an input to a second self-consistency loop, which yields $\Gamma$ up to subleading-log accuracy.
    }
    \label{fig:Parquet-overview}
\end{figure*}

In \Sec{sec:Perturbation-theory}, we introduced various quantities which appear in a diagrammatic description of the model's characteristic power laws, \Eqs{eq:exact_power-laws}. The perturbative expansion of the susceptibility $\chi$, \Eq{eq:chi_diagram}, involves terms scaling as $\chi^{(n)}\sim u^n L^{n+1-p}$, where $p=0$ encompasses the leading-log terms, while $p=1$ accounts for the subleading-log ones [$L$ is defined in \Eq{eq:definition_L}]. The full four-point vertex $\Gamma$ has an expansion $\Gamma^{(n)}\sim u^n L^{n-1-p}$ and is decomposed into two-particle reducible vertices $\gamma^r$ in the channels $r=a,p,t$ and a totally irreducible part $R$. (Note that what Refs.~\cite{roulet1969singularities,nozieres1969singularities2} call $R$ corresponds to $\gamma^t + R$ for us.) We argued that the lowest-order contributions of $\gamma^a$ and $\gamma^p$ are leading log while those of $\gamma^t$ are subleading log and those of $R$ are subsubleading log. Finally, the self-energy $\Sigma$ has an expansion $\Sigma^{(n)}\sim u^n \ii\tilde\nu \bar{L}^{n-1-p}$ and yields subleading-log contributions to $\chi$.
An overview of how the different objects are computed self-consistently is given in \Fig{fig:Parquet-overview}. Its details are explained throughout this section.

All leading-log diagrams can be summed within a parquet approach containing only the $a$- and $p$-reducible vertices $\gamma^{r=a,p}$~\cite{roulet1969singularities}. These fulfill Bethe--Salpeter equations involving the full vertex $\Gamma$ and the $a$- and $p$-irreducible vertices $I^{r=a,p}$. Since no fully irreducible diagram contributes to the leading-log behavior except for the bare vertex $\Gamma^{(1)}=-u$, we set $R=-u$ (often called \emph{parquet approximation}). The relevant equations are (cf.\ yellow box in \Fig{fig:Parquet-overview}):
\begin{subequations}
    \label{eq:parquet-leading-log}
    \begin{align}
        \Gamma^{r}_{\omega,\nu,\nu'}
        &=
        -u + \gamma^{r}_{\omega,\nu,\nu'} + \gamma^{\bar{r}}_{\omega-\nu-\nu',-\nu',-\nu}
        ,
        \label{eq:Gamma_leading-log}\\
        \gamma^{r}_{\omega,\nu,\nu'}
        &=
        \int_{\nu''}I^r_{\omega,\nu,\nu''}\Pi^r_{\omega,\nu''}\Gamma^r_{\omega,\nu'',\nu'}
        ,
        \label{eq:gamma_r_leading-log}\\
        I^r_{\omega,\nu,\nu'}
        &=
        \Gamma^r_{\omega,\nu,\nu'}-\gamma^r_{\omega,\nu,\nu'},
        \\
        \nonumber
        \begin{gathered}
            \includegraphics[width=45pt]{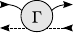}
        \end{gathered}
        &= ~
        \begin{gathered}
            \includegraphics[width=35pt]{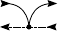}
        \end{gathered}
        ~ + ~
        \begin{gathered}
            \includegraphics[width=85pt]{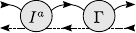}
        \end{gathered}
        \\
        &\phantom{=} ~ + ~
        \begin{gathered}
            \includegraphics[width=85pt]{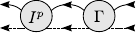}~.
        \end{gathered}
    \end{align}
\end{subequations}
Here, we use the channel indices $r=a,p$ and $\bar{r}=p,a$ and the bubbles $\Pi^r_{\omega,\nu}$, \Eq{eq:bubbles-definition}. The frequency arguments follow from the parametrization of the vertex, \App{sec:Vertex-conventions}. 

Equations~\eqref{eq:parquet-leading-log} must be solved self-consistently. In this process, one can apply the reasoning from \Sec{sec:logarithmic_behavior} to show that two leading-log vertices $\Gamma_\mathrm{lead}^{(n_{i=1,2})}\sim u^{n_i} L^{{n_i}-1}$ in the Bethe--Salpeter equation \eqref{eq:gamma_r_leading-log} yield again a leading-log vertex:
\begin{align}
    \nonumber \int_\nu \Gamma_\mathrm{lead}^{(n_1)}\Pi^r \Gamma_\mathrm{lead}^{(n_2)} &\sim \int_\nu u^{n_1} L^{n_1-1} \frac{\sgn(\nu)}{\ii\nu} u^{n_2} L^{n_2-1}\\
    &\sim u^{n_1+n_2}L^{n_1+n_2-1}=\Gamma_\mathrm{lead}^{(n_1+n_2)}.
\end{align}

In \Sec{sec:Perturbation-theory}, we showed that the self-energy $\Sigma$ and the $t$-reducible vertex $\gamma^t$ contribute to the subleading logarithm. We calculate $\Sigma$ from the Schwinger--Dyson equation. The ($d$-electron) Hartree term $u\xi_0$ yields a frequency-independent shift of the threshold frequency and is therefore irrelevant for the power-law exponent. Beyond the Hartree term, we have (cf.\ blue box in \Fig{fig:Parquet-overview})
\begin{align}
    \nonumber\Sigma_\nu
    &= \quad 
    \begin{gathered}
        \includegraphics[width=90pt]{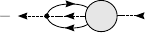}
    \end{gathered}
    \\
    \nonumber &=
    u\int_{\nu_1,\nu_2}\Pi^a_{\nu_1-\nu,\nu_2}g_{\nu_1}\Gamma^a_{\nu_1-\nu,\nu_2,\nu_1}
    \\
    &= u \int_{\nu_1,\nu_2}\Pi^p_{\nu_1-\nu,\nu_2}g_{-\nu_1}\Gamma^p_{\nu_1-\nu,\nu_2,\nu_1}.
    \label{eq:SDE}
\end{align}
Reference~\cite{nozieres1969singularities2} argued against using the Schwinger--Dyson equation \eqref{eq:SDE} as their analytic evaluation was limited to logarithmic accuracy of individual diagrams. By contrast, our numerics produce the full frequency dependence of vertex functions, allowing us to straightforwardly use \Eq{eq:SDE}.

Following \Sec{sec:logarithmic_behavior}, inserting a leading-log vertex $\Gamma_\mathrm{lead}^{(n)}\sim u^n L^{n-1}$ into the Schwinger--Dyson equation~\eqref{eq:SDE} yields a subleading-log self-energy:
\begin{align}
    \nonumber
    &
    \int_{\nu_1,\nu_2} u \Pi^r g \Gamma_\mathrm{lead}^{(n)} \sim \int_{\nu_1,\nu_2} u \frac{\sgn(\nu_2)}{\ii\nu_2}\sgn(\nu_1)u^n L^{n-1}
    \\
    &
    \sim \int_\nu u\, \sgn(\nu) u^n L^{n}
    \sim u^{n+1} \ii\nu L^{n} = \Sigma^{(n+1)}_\mathrm{sub}
    .
    \label{eq:SDE_leading-behavior}
\end{align}
Here, we first integrated over the $dc$ bubble and then over the last $c$ propagator.
It follows that self-energy corrections yield subleading-log contributions to $\chi$. Furthermore, inserting $\Sigma$ again into the $d$ propagator or inserting a subleading-log vertex $\Gamma^{(n)}\sim u^n L^{n-2}$ into \Eq{eq:SDE} would go beyond our approximation.

To include the $t$-reducible vertex $\gamma^t$, we use the subleading-log expression at third order, $[\gamma^t]^{(3)}$ in \Eq{eq:gamma_t_3}, and replace all bare vertices $\Gamma^{(1)}=-u$ by full vertices $\Gamma$. As we show in \App{sec:gamma_t_subleading}, this ansatz takes into account all subleading-log contributions starting from the most general parquet approach. The resulting expression for 
$\gamma^t$ (corresponding to Fig.~4(b) in Ref.~\cite{nozieres1969singularities2}) is (cf.\ blue box in \Fig{fig:Parquet-overview}):
\begin{align}
    \gamma^t_{\omega,\nu,\nu'}
    &= \quad 
    \begin{gathered}
        \includegraphics[width=150pt]{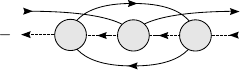}
    \end{gathered}
    \label{eq:gamma_t}
    \\
    \nonumber &= -\int_{\nu_1,\nu_2}\Gamma^t_{\omega,\nu,\nu_1}G_{\nu_1-\omega}G_{\nu_1}\Gamma^a_{\nu_2-\nu_1,\nu'-\nu_1+\nu_2,\nu_2}\\
    \nonumber &\phantom{= -\int_{\nu_1,\nu_2}}\times g_{\nu_2-\nu_1+\nu'}g_{\nu_2}\Gamma^a_{\omega-\nu_1+\nu_2,\nu_2,\nu'+\nu_2-\nu_1}.
\end{align}

If all vertices in \Eq{eq:gamma_t} are leading log, i.e., $\Gamma_\mathrm{lead}^{(n_{i=1,2,3})}\sim u^{n_i} L^{n_i-1}$, integrating first over the loop including the $c$ propagators and then over the loop with the two $d$ propagators yields
\begin{align}
    \nonumber&\int_{\nu_1,\nu_2}\!\!\!\!\!\!\!\Gamma_\mathrm{lead}^{(n_1)}GG\Gamma_\mathrm{lead}^{(n_2)}gg\Gamma_\mathrm{lead}^{(n_3)} \\
    \nonumber&\sim \int_{\nu_1}\Gamma_\mathrm{lead}^{(n_1)}GG\int_{\nu_2}u^{n_2}L^{n_2-1}\sgn^2(\nu_2)u^{n_3}L^{n_3-1}\\
    \nonumber&\sim \int_{\nu} u^{n_1} L^{n_1-1}\frac{1}{(\ii\nu)^2}u^{n_2+n_3}\ii\nu L^{n_2+n_3-2}\\
    &\sim u^{n_1+n_2+n_3}L^{n_1+n_2+n_3-2}=\Gamma_\mathrm{sub}^{n_1+n_2+n_3},
    \label{eq:gamma-t_leading-behavior}
\end{align}
which is subleading log. Including the self-energy in the $d$ propagators or subleading-log vertices in \Eq{eq:gamma_t} would go beyond the subleading approximation. In the expansion of the susceptibility \Eq{eq:Taylor-expansion}, such terms would contribute similarly as the totally irreducible diagrams $R$.

From the logarithmic behavior in \Eqs{eq:SDE_leading-behavior} and \eqref{eq:gamma-t_leading-behavior}, we conclude that the leading contributions to the self-energy $\Sigma \sim u^n \ii\tilde\nu \bar{L}^{n-1}$ and the $t$-reducible vertex $\gamma^t \sim u^n L^{n-2}$ are already fully recovered by inserting the leading-log vertex $\Gamma_\mathrm{lead}\sim u^n L^{n-1}$ into \Eqs{eq:SDE} and \eqref{eq:gamma_t}. Our strategy (depicted in \Fig{fig:Parquet-overview}) is thus to first compute the leading-log vertex $\Gamma$ by iteratively solving \Eqs{eq:parquet-leading-log}. In the next step, $\Sigma$ and $\gamma^t$ are determined from \Eqs{eq:SDE} and \eqref{eq:gamma_t}. These then form an input to a second iterative solution of the parquet equations, but now the $d$ lines are dressed through the Dyson equation $G_\nu = (1/G^{(0)}_\nu-\Sigma_\nu)^{-1}$ and the full vertex $\Gamma$ includes $\gamma^t$ from \Eq{eq:gamma_t}:
\begin{subequations}
    \label{eq:Gamma_parquet_extended}
    \begin{align}
        \nonumber\Gamma^a_{\omega,\nu,\nu'}
        &=
        -u + \gamma^a_{\omega,\nu,\nu'} +\gamma^p_{\omega-\nu-\nu',-\nu',-\nu}
        \\
        &\phantom{=}
        + \gamma^t_{\nu-\nu',\nu,\nu-\omega}
        ,\\
        \nonumber\Gamma^p_{\omega,\nu,\nu'}
        &=
        -u + \gamma^p_{\omega,\nu,\nu'} +\gamma^a_{\omega-\nu-\nu',-\nu',-\nu}
        \\
        &\phantom{=}
        + \gamma^t_{\nu-\nu',-\nu',\nu-\omega}
        .
    \end{align}
\end{subequations}
Thereby, Eqs.~\eqref{eq:Gamma_parquet_extended} replace \Eq{eq:Gamma_leading-log}. Finally, with the inclusion of $\Sigma$ and $\gamma^t$, we obtain $\gamma^a$ and $\gamma^p$ self-consistently up to subleading-log order and may altogether compute $\chi$ up to subleading-log order.

We note that further iterations over \Eqs{eq:SDE} and \eqref{eq:gamma_t} would yield subsubleading-log diagrams, but not in a complete and systematic manner since totally irreducible diagrams like the envelope diagram [cf.\ \Fig{fig:logarithmic-order}(b)] would not be taken into account. Moreover, also the $t$-reducible vertex from the full parquet solution includes further sub-subleading contributions, as discussed in \App{sec:gamma_t_subleading}, which exceed the scope of this work.

\subsection{Numerical results}
\label{sec:numerical-results}

In this section, we present our numerical results obtained from the self-consistency schemes discussed in \Sec{sec:Parquet-schemes} and compare them to the analytical power law in \Eq{eq:exact_power-law_chi}. Although, strictly speaking, these power laws hold very close to the threshold frequency and at $T=0$, they adequately describe the physical results in a much wider range (cf.\ \App{sec:FDA} and also Ref.~\cite{schmidt2018universal}). In our numerical implementations, the frequency dependence of the vertex is handled by a decomposition into single- and multi-boson exchange vertices~\cite{Krien2019,Krien2019a,Krien2019b,Krien2020a,Krien2020b,Krien2021,Harkov2021,Harkov2021a,krien2022plain,Bonetti2022SBE,fraboulet2022single,gievers2022multiloop} using the recently developed Julia library \texttt{MatsubaraFunctions.jl}~\cite{kiese2024matsubarafunctions}; further details are given in \App{sec:Numerics}.

\begin{figure}
    \centering
    \includegraphics[width=1.0\linewidth]{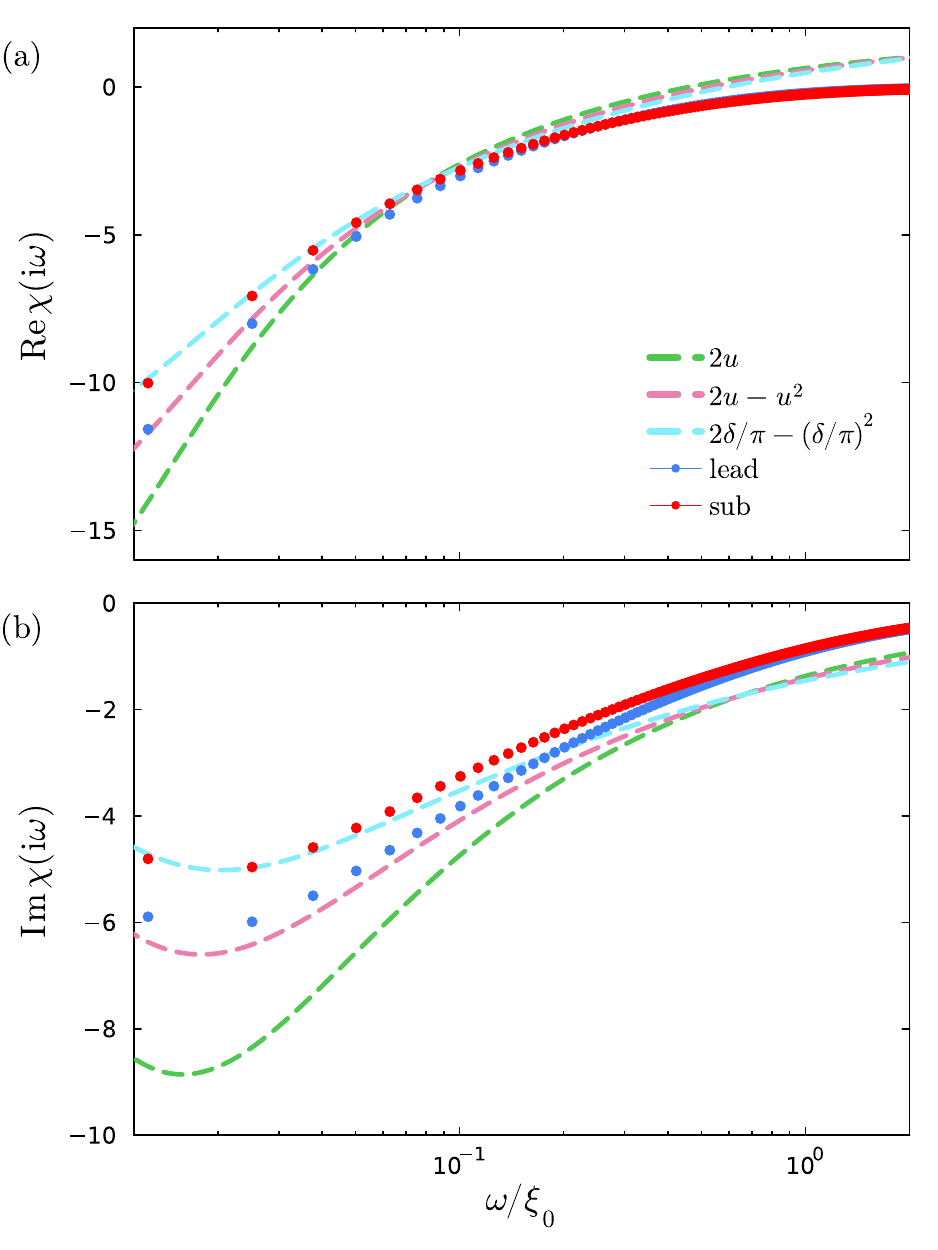}
    \caption{Frequency dependence of $\chi$ from self-consistent summations with $g^\mathrm{sm}$ at $u=0.28$, $T/\xi_0=0.002$, and $\xi_d/\xi_0=-0.01$. We compare numerical results from the leading-log scheme (blue dots) and subleading-log scheme (red dots) to the analytical power law~\eqref{eq:exact_power-law_chi} with exponents $\alpha_X= 2u$ (leading log, green), $\alpha_X = 2u-u^2$ (subleading log, pink), and $\alpha_X = 2\delta/\pi -(\delta/\pi)^2$ (exact, light blue).
    }
    \label{fig:chi_w_parquet}
\end{figure}

To start with, \Fig{fig:chi_w_parquet} shows how $\chi$ depends on imaginary frequencies at fixed $u$. We compare the result of the leading-log scheme (blue dots), \Eqs{eq:parquet-leading-log}, and the subleading-log scheme (red dots), \Eqs{eq:Gamma_parquet_extended}, to the power law~\eqref{eq:exact_power-law_chi} with exponents $\alpha_X = 2u$ (leading log, green), $\alpha_X = 2u-u^2$ (subleading log, pink) and $\alpha_X = 2\delta/\pi-(\delta/\pi)^2$ (exact, light blue). The analytical power laws describe the behavior at small frequencies, but of course do not capture the correct large-frequency behavior, including $\lim_{\omega\to\infty}\chi(\ii\omega)=0$. For the present choice of parameters, the results from our subleading-log parquet scheme are closest to the exact power law (red dots match light blue curve) while those from the leading-log parquet scheme are closest to the subleading-log power law (blue dots lie near pink curve). However, this strongly depends on the value of $\xi_d$, as elaborated below.

\begin{figure}
    \centering
        \includegraphics[width=1.0\linewidth]{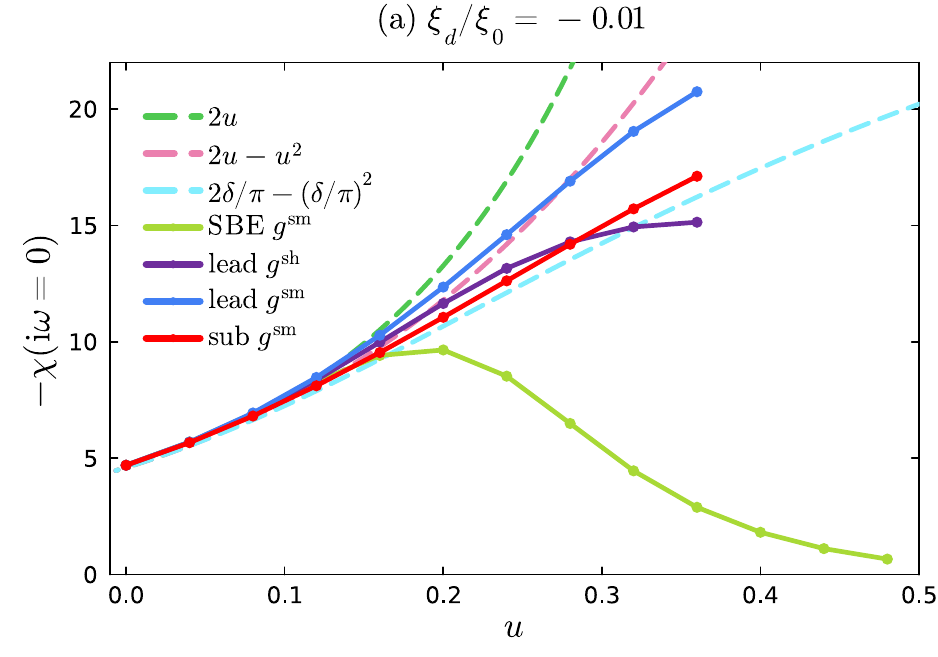}
    \vspace{-3mm}
    \includegraphics[width=1.0\linewidth]{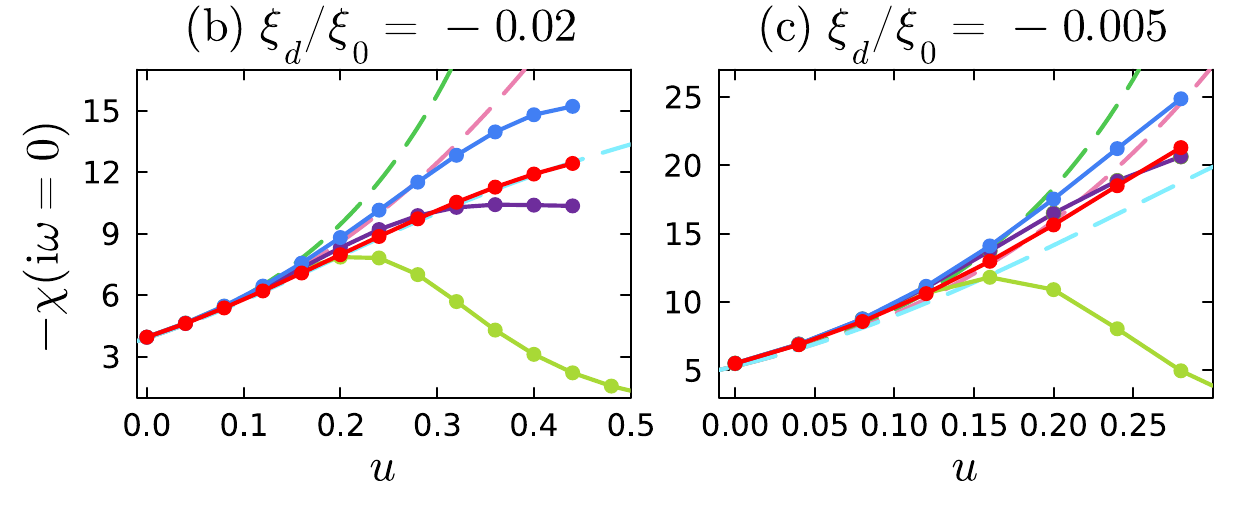}
    \caption{
        Interaction dependence of $\chi(\ii\omega=0)$ from self-consistent summations at $T/\xi_0=0.002$ and different choices of $\xi_d/\xi_0$. We compare results from the leading-log scheme, \Eqs{eq:parquet-leading-log}, in the SBE approximation with $g^\mathrm{sm}$ (light green dots) and including MBE diagrams with $g^\mathrm{sm}$ (blue dots) and $g^\mathrm{sh}$ (purple dots), as well as from the subleading scheme, \Eqs{eq:SDE}--\eqref{eq:Gamma_parquet_extended}, with $g^\mathrm{sm}$ including MBE diagrams (red dots). The numerical results are compared to the analytical power laws for leading-log $\alpha_X = 2u$ (green, dashed), subleading-log $\alpha_X = 2u-u^2$ (pink, dashed), and exact exponent $\alpha_X = \delta/\pi-(\delta/\pi)^2$ (light blue, dashed).
    }
    \label{fig:chi_w=0_U}
\end{figure}

To get an overview on the results for different parameters and obtained from the various self-consistent methods, we present $\chi$ at zero Matsubara frequency as a function of $u$ in \Fig{fig:chi_w=0_U}.
Here, we compare again our numerical results with the power law, \Eq{eq:exact_power-law_chi}, including different exponents $\alpha_X$. The power law with $\alpha_X = \delta/\pi-(\delta/\pi)^2$ matches the numerically exact results from the functional determinant approach, when evaluated at $\omega_0+\xi_d$ (cf.\ \App{sec:FDA}). We draw the following conclusions from \Fig{fig:chi_w=0_U}.

Dropping multi-boson exchange diagrams, which is known as the \emph{single-boson exchange approximation}~\cite{Bonetti2022SBE,gievers2022multiloop,fraboulet2022single}, clearly fails already at intermediate values of the interaction [cf.\ light green dots in \Fig{fig:chi_w=0_U}]. We anticipated that from our perturbative analysis since multi-boson diagrams contribute to leading-log diagrams and are therefore essential to obtain a power law at all (cf.\ \Sec{sec:MBE}).

The leading-log parquet solution using the sharp $c$ propagator $g^\mathrm{sh}$, \Eq{eq:Gc_sharp}, [cf.\ purple dots in \Fig{fig:chi_w=0_U}] bends down from the leading-log power law (green) at intermediate values of the interaction, similarly as in previous studies (see Fig.~4(c) in Ref.~\cite{KuglerPRL2018b}). This might originate from the artifacts around $|\omega|\simeq\xi_0$, encountered already in \Sec{sec:Perturbation-theory}.

Our main focus is on the results of the leading-log (\Eqs{eq:parquet-leading-log}, blue dots in \Fig{fig:chi_w=0_U}) and subleading-log parquet schemes (\Eqs{eq:Gamma_parquet_extended}, red dots), both using the exact propagator $g^\mathrm{sm}$, \Eq{eq:Gc_smooth}. The results start to deviate from one another and from the power-law curves with $\alpha_X = 2u$ (green) and $\alpha_X = 2u - u^2$ (pink), respectively, already at intermediate values of $u\gtrsim 0.2$. The results from the subleading-log parquet scheme are systematically improved in powers of the logarithmic factor $\ln(-\xi_d/\xi_0)$ [cf.\ \Eq{eq:definition_L}] in the region where the expansion of the power law in $u$, \Eq{eq:Taylor-expansion}, is valid.

The value of $\chi(\ii\omega=0)$ is strongly affected by the parameter $\xi_d/\xi_0$. At larger values of $|\xi_d|/\xi_0$, where different powers of the logarithm are hardly distinguishable, the results move down in magnitude [cf.\ \Fig{fig:chi_w=0_U}(b)]. The resulting apparent agreement between the self-consistent calculation to leading- and subleading-log accuracy with the power-law curves $\alpha_X = 2u-u^2$ (pink) and $\alpha_X = 2\delta/\pi-(\delta/\pi)^2$ (light blue), respectively, is likely coincidental. 

Smaller values of $|\xi_d|/\xi_0$ [larger values of $\ln(-\xi_d/\xi_0)$] yield a clearer separation between different powers of the logarithm. For $\xi_d/\xi_0 = -0.005$ [cf.\ \Fig{fig:chi_w=0_U}(c)], the numerical results come much closer to the expected behavior: the leading-log parquet results (blue dots) follow the $2u$ power law (green) and the subleading-log parquet (red dots) results the $2u-u^2$ power law (pink) up to $u \approx 0.2$. The value $\ln(-\xi_d/\xi_0) \approx -5.3$ is still relatively small. However, reducing $|\xi_d|/\xi_0$ further goes beyond our current numerical limitations, since this would require lower $T$ and thus more frequencies to resolve the vertex functions. It also becomes harder to converge the parquet equations 
at small $|\xi_d|/\xi_0$ and at large $u$ (hence, we computed less data points in that regime).

\begin{figure}
    \centering
    \includegraphics[width=1.0\linewidth]{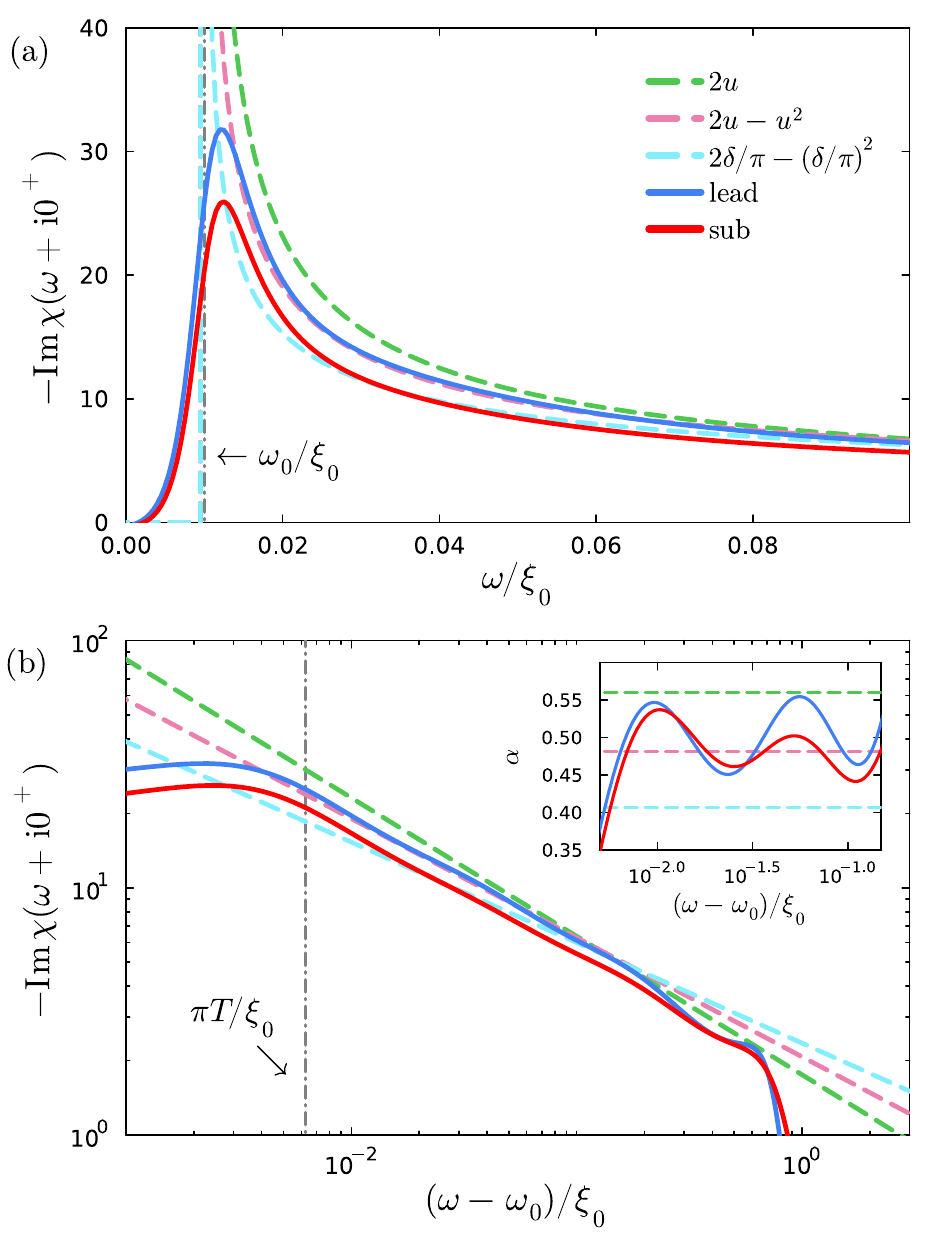}
    \caption{
    Absorption spectrum $-\Im\,\chi(\omega+\ii 0^+)$ obtained by analytical continuation of the imaginary-frequency data at $T/\xi_0=0.002$, $\xi_d/\xi_0=-0.01$, and $u=0.28$. We compare data from the leading-log (blue) and subleading-log scheme (red) to the respective analytical power laws. The upper plot has linear scales; here, the vertical dash-dotted line marks the threshold frequency $\omega_0=-\xi_0$. The lower plot has logarithmic scales and frequencies shifted by $\omega_0$; here, the vertical line marks the lowest fermionic Matsubara frequency $\pi T$ to indicate where $T$ cuts off the logarithmic behavior. The inset shows the negative logarithmic derivative.
    }
    \label{fig:chi_AC}
\end{figure}

Ideally, one would like to analyze the numerical results in real frequencies. To this end, we use analytical continuation via the recently developed minimal pole representation~\cite{Zhang2024Minimal,zhang2024minimalpolerepresentationanalytic}. Here, the susceptibilities are analytically continued as sums over a small number of complex poles, i.e., $\chi(z)=\sum_i A_i/(z-x_i), x_i\in\mathbb{C}$, by use of Prony's approximation method. The results, shown in \Fig{fig:chi_AC}, agree with the previous statements [cf.\ \Figs{fig:chi_w_parquet} and \ref{fig:chi_w=0_U}(a)]: for these parameters ($\xi_d/\xi_0=-0.01$ and $u=0.28$), the leading-log (blue dots) and subleading-log (red dots) numerical results are close to the curves of the analytical subleading-log (pink) and exact (light blue) power laws, respectively. As one would expect, the singularity at the threshold is cut off by $T$ at the corresponding energy scale $\pi T$ [cf.\ \Fig{fig:chi_AC}(b)]. The analytical continuation also confirms that correcting the self-energy by $\Sigma_\nu \to\Delta\tilde\Sigma_\nu$, \Eq{eq:Self-energy_subtraction_remedy}, does not lead to a detectable renormalization of the threshold frequency $\omega_0=-\xi_d$ [cf.\ \Fig{fig:chi_AC}(a)]. For a more rigorous analysis of the power law, one can extract the power-law exponents from the logarithmic derivative (see \App{sec:Analytical-continuation} for details). The result, shown in the inset of \Fig{fig:chi_AC}(b), exhibits strong oscillations somewhat centered around the subleading-log exponent $\alpha = 2u-u^2$ (dashed pink). We attribute the oscillations to the fact that the analytically continued data stem from only a small number of complex poles~\cite{Zhang2024Minimal,zhang2024minimalpolerepresentationanalytic}.

\section{Conclusion}
\label{sec:Conclusion}

In this work, we elucidated a conceptual aspect of a diagrammatic technique widely used in condensed-matter physics and beyond, the parquet formalism. The parquet formalism is well known for providing a way to sum all leading-log diagrams in a logarithmically divergent perturbation theory. On the example of the x-ray edge singularity, we showed that the parquet formalism actually allows for capturing all next-to-leading-log diagrams, too. To this end, one extends the first-order parquet approach~\cite{roulet1969singularities}, which involves only two two-particle channels and no self-energy, to the full parquet approach~\cite{Bickers1991Conserving}, often used for Hubbard-like models, involving all three two-particle channels and the self-energy determined from the Schwinger--Dyson equation.

We first examined the problem at low orders in perturbation theory. Thereby, we also provided exact results for the bare particle-hole susceptibility and the second-order self-energy, which to our knowledge had not been given before. We illustrated the vertex and self-energy contributions and formulated a general recipe for deducing the highest logarithmic power in a given diagram. This allowed us to formulate a self-consistent scheme, within the full parquet approach mentioned above, summing all leading- and subleading-log diagrams.

For all our results, we presented numerical data obtained in the finite-temperature Matsubara formalism. In doing so, we use the exact expressions for the bare propagators and resolve the full frequency dependence of any diagram, including four-point vertices, thus going beyond logarithmic accuracy used in previous works~\cite{roulet1969singularities,nozieres1969singularities2,Lange_2015,kugler2018fermi,KuglerPRL2018b,Diekmann2021,Diekmann2024Leading}. Our implementation exploits the recently introduced single- and multi-boson exchange decomposition, for which we showed that multi-boson exchange diagrams contribute already at the leading-log level.

In future work, our code could be used to treat other models with two distinct particle types. Examples are Fermi polarons with heavy impurities~\cite{knap2012time,schmidt2018universal,Gievers2024Probing} or Hubbard-like models without SU(2) spin symmetry. It would be interesting to lift the flat-band approximation of the $d$ electron, which however requires including momenta, significantly raising the computational costs.

On the technical level, it would be desirable to numerically resolve the power laws in an even cleaner fashion. One direction in this pursuit would be to lower the temperature $T/\xi_0$ and the excitation energy $|\xi_d|/\xi_0$. In the current implementation, using dense grids, this is not feasible since the memory scales as $(T/\xi_0)^{-3}$. However, techniques for using sparse grids or compression have recently been suggested~\cite{Shinaoka2020Sparse,Wallerberger2021Solving,Kaye2022Discrete,kiese2024discrete,Ritter2024Quantics,fernández2024learning,Rohshap2024TwoParticle}. Another direction is to directly work in real frequencies, thus circumventing the analytical continuation. The model can be numerically implemented using the zero-temperature formalism~\cite{Diekmann2021}. Beyond that, recent work has shown the viability of working in the Keldysh formalism with full frequency resolution of four-point functions~\cite{Kugler2021,Lee2021,Lihm2024Symmetric,ge2024real,Ritz_2024}.

\section*{Acknowledgments}

We thank Anxiang Ge, Dominik Kiese, Severin Jakobs, Benedikt Schneider, Nepomuk Ritz, Xin Chen, and Eugen Dizer for insightful discussions. Moreover, we thank Lei Zhang and Emanuel Gull for support with the analytical continuation of our data.
This research is part of the Munich Quantum Valley, which is supported by the Bavarian state government with funds from the Hightech Agenda Bayern Plus. 
We acknowledge funding for M.G.\ from the International Max Planck Research School for Quantum Science and Technology (IMPRS-QST),
for R.S.\ funding from the Deutsche Forschungsgemeinschaft under Germany’s Excellence Strategy EXC 2181/1–390900948 (the Heidelberg STRUCTURES Excellence Cluster),
and for J.v.D.\ funding from the Deutsche Forschungsgemeinschaft under Germany's Excellence Strategy EXC-2111 (Project No.\ 390814868) and the grant LE 3883/2-2.
The Flatiron Institute is a division of the Simons Foundation.
The numerical simulations were performed on the Linux clusters at the Leibniz Supercomputing Center with support by grant INST 86/1885-1 FUGG of the Deutsche Forschungsgemeinschaft (DFG).

\appendix

\section{Functional determinant approach}
\label{sec:FDA}

In this appendix, we discuss how we obtain numerically exact results for the particle-hole susceptibility $X(t)$, \Eq{eq:Susceptibility-definition}, and the $d$ propagator $G(t)$, \Eq{eq:Green_d-definition}, using the functional determinant approach~\cite{levitov1996electron,klich2003elementary,schmidt2018universal}.

\subsection{Exact computation of the spectra}

For large enough $|\varepsilon_d|$, the ground state is given by the occupied core level, $|1\rangle = \hat d^\dagger|0\rangle$, and the Fermi sea of conduction electrons $|\mathrm{FS}\rangle$ in the sense of Fermi-liquid theory, i.e., $|\Psi_0\rangle = |1\rangle\otimes|\mathrm{FS}\rangle$~\cite{Diekmann2024Leading}. Generally, the core level is either empty or occupied, so the full Hamiltonian, \Eq{eq:Hamiltonian}, can be brought into a form $\hat H = |0\rangle\langle 0|\otimes\hat H_1 + |1\rangle\langle 1|\otimes (\hat H_0+\varepsilon_d\hat{\doubleI})$, where $\hat H_0$ and $\hat H_1$ only act on the subspace of the conduction electrons:
\begin{align}
    \hat H_0
    =
    \sum_{\bm{k}}\varepsilon_{\bm{k}}\hat c_{\bm{k}}^\dagger \hat c^{\phantom{\dagger}}_{\bm{k}}
    , \quad
    \hat H_1
    =
    \sum_{\bm{k},\bm{k}'}\left(\varepsilon_{\bm{k}}\delta_{\bm{kk}'}-\frac{U}{V}\right)\hat c_{\bm{k}}^\dagger \hat c^{\phantom{\dagger}}_{\bm{k}'}
    .
    \label{eq:H0_H1}
\end{align}
Since these are quadratic, the system is exactly solvable. The time evolution of the many-body state with an empty $|0\rangle$ or occupied $|1\rangle$ core level is then determined by
\begin{subequations}
    \label{eq:time-evolutions}
    \begin{align}
        \ee^{-\ii\hat\Xi t}(|0\rangle\otimes|\Psi\rangle)
        &=
        |0\rangle\otimes\ee^{-\ii(\hat H_1-\mu\hat N_0)t}|\Psi\rangle
        ,
        \\
        \ee^{-\ii\hat\Xi t}(|1\rangle\otimes|\Psi\rangle)
        &=
        \ee^{-\ii(\varepsilon_d-\mu)t}|1\rangle \otimes \ee^{-\ii(\hat H_0-\mu\hat N_0)t}|\Psi\rangle
        .
    \end{align}
\end{subequations}
Here, we use the number operator of the conduction electrons $\hat N_0 = \sum_{\bm{k}}\hat c^\dagger_{\bm{k}}\hat c_{\bm{k}}$ and $\hat\Xi = \hat H-\mu\hat N$, with the total number operator $\hat N = \hat N_0 + \hat d^\dagger\hat d$.

As mentioned in the main text, we use the ground state $|\Psi_0\rangle = |1\rangle\otimes|\mathrm{FS}\rangle$ as reference state for the expectation value. The time-ordering operator $\mathcal{T}$ generates two terms. In the expressions for $X(t)$ and $G(t)$, however, only one remains according to the occupancy of the $d$ electron.

Inserting the time evolution $\hat d^{(\dagger)}(t)=\ee^{\ii\hat\Xi t}\hat d^{(\dagger)}\ee^{-\ii\hat\Xi t}$ into the definition of $G(t)$, \Eq{eq:Green_d-definition}, yields
\begin{align}
    \nonumber G(t)
    &=
    \ii\,\Theta(-t)(\langle 1|\otimes\langle\mathrm{FS}|)\hat d^\dagger\ee^{\ii \hat \Xi t}\hat d\ee^{-\ii \hat \Xi t}(|1\rangle\otimes|\mathrm{FS}\rangle)
    \\
    \nonumber &=
    \ii\,\Theta(-t)(\langle 0|\otimes\langle\mathrm{FS}|\ee^{\ii (\hat H_1-\mu \hat N_0)t})\\
    &\phantom{=}\quad\times (\ee^{-(\ii\varepsilon_d-\mu)t}|0\rangle\otimes\ee^{-\ii (\hat H_0-\mu\hat N_0) t}|\mathrm{FS}\rangle)
    ,
\end{align}
which, after evaluating the effect of the $d$ electron, gives
\begin{align}
    G(t)
        &=
        \ii\Theta(-t)\ee^{-\ii\xi_d t}\langle\mathrm{FS}|\ee^{\ii\hat H_1t}\ee^{-\ii\hat H_0t}|\mathrm{FS}\rangle
        .
        \label{eq:Gd-advanced}
\end{align}
Note that the terms including the number operators $\hat N_0$ cancel each other by conservation of particle number, i.e., $[\hat H_{0,1},\hat N_0]=0$. With $|\mathrm{FS}\rangle\langle\mathrm{FS}|=\ee^{-\beta(\hat H_0-\mu\hat N_0)}/Z_0$, where $Z_0=\mathrm{tr}\,\ee^{-\beta(\hat H_0-\mu\hat N_0)}$, the expectation value is expressed as a trace:
\begin{align}
    G(t) = \ii\Theta(-t)\ee^{-\ii(\varepsilon_d-\mu)t}\frac{1}{Z_0}\mathrm{tr}\left[\ee^{-\beta(\hat H_0-\mu\hat N_0)}\ee^{\ii\hat H_1 t}\ee^{-\ii\hat H_0 t}\right].
    \label{eq:G_d_intermediate}
\end{align}
As the Hamiltonians are bilinear $\hat H_{0,1} = \sum_{\bm{k},\bm{k}'}[\hat h_{0,1}]_{\bm{k}\bm{k}'}\hat c_{\bm{k}}^\dagger\hat c_{\bm{k}'}$, we use Klich's formula~\cite{klich2003elementary,schmidt2018universal} to express the many-particle trace as a determinant over single-particle operators $\hat h_{0,1}$:
\begin{align}
    \nonumber&\mathrm{tr}\left[\ee^{-\beta(\hat H_0-\mu\hat N_0)}\ee^{\ii\hat H_1 t}\ee^{-\ii\hat H_0 t}\right]\\
    &= \det\left[\hat\doubleI+\ee^{-\beta(\hat h_0-\mu)}\ee^{\ii\hat h_1t}\ee^{-\ii\hat h_0t}\right].
\end{align}
Due to the fermionic properties, the Green's function is finally written in terms of the Fermi--Dirac distribution:
\begin{align}
    G(t)
    &= \ii\Theta(-t)\ee^{-\ii(\varepsilon_d-\mu)t}\det\left[\hat\doubleI\!-\! f(\hat h_0)\! +\! f(\hat h_0)\ee^{\ii\hat h_1t}\ee^{-\ii\hat h_0t}\right].
    \label{eq:G(t)_FDA}
\end{align}
In frequency space, the expression for the advanced Green's function is obtained from fast Fourier transformation $G(\nu) = \int_t \ee^{\ii\nu t} G(t)$ after exact diagonalization of the single-particle Hamiltonians $[\hat h_1]_{\bm{kk}'}=\varepsilon_{\bm{k}}\delta_{\bm{kk}'}-U/V$ (for details, see Supplemental Material of Ref.~\cite{Gievers2024Probing}).

The susceptibility, \Eq{eq:Susceptibility-definition}, is computed in a similar fashion. Due to the occupancy of the $d$ electron, here only the retarded term of the time ordering survives:
\begin{align}
    \nonumber X(t)
    &=
    -\ii\,\Theta(t)\frac{1}{V}\sum_{\bm{k},\bm{k}'}(\langle 1|\otimes\langle\mathrm{FS}|)\ee^{\ii \hat\Xi t}\hat d^\dagger\ee^{-\ii\hat\Xi t}
    \\
    &\phantom{=-\ii\,\Theta(t)\frac{1}{V}} \quad
    \times \ee^{\ii \hat\Xi t}\hat c_{\bm{k}}\ee^{-\ii\hat\Xi t}\hat c^\dagger_{\bm{k}'}\hat d(|1 \rangle\otimes|\mathrm{FS}\rangle)
    ,
\end{align}
which according to the time evolutions, \Eq{eq:time-evolutions}, yields:
\begin{align}
    \nonumber X(t)
    &=
    -\ii\,\Theta(t)\frac{1}{V}\sum_{\bm{k},\bm{k}'}(\langle 0|\ee^{\ii(\varepsilon_d-\mu)t}\otimes\langle\mathrm{FS}|\ee^{\ii(\hat H_0-\mu\hat N_0)t}\hat c_{\bm{k}})
    \\
    &\phantom{=-\ii\,\Theta(t)\frac{1}{V}}\quad
    \times(|0\rangle\otimes\ee^{-\ii (\hat H_1-\mu \hat N_0) t}\hat c_{\bm{k}'}^\dagger|\mathrm{FS}\rangle)
    .
\end{align}
Again, the $d$ degree of freedom is evaluated straightforwardly. By particle-number conservation, we can write $\ee^{-\ii\mu\hat N_0t}\hat c_{\bm{k}}\ee^{\ii\mu\hat N_0t}=\ee^{\ii\mu t}\hat c_{\bm{k}}$, so the terms with the chemical potentials cancel, and we get the following expression:
\begin{align}
    X(t)
        &=
        -\ii\Theta(t)\ee^{\ii\varepsilon_dt}\frac{1}{V}\sum_{\bm{k},\bm{k}'}\langle\mathrm{FS}|\ee^{\ii\hat H_0t}\hat c^{\phantom{\dagger}}_{\bm{k}}\ee^{-\ii\hat H_1t}\hat c_{\bm{k}'}^\dagger|\mathrm{FS}\rangle
        .
        \label{eq:chi-dc-retarded}
\end{align}
By using $[\hat H_1,\hat c_{\bm{k}}]=\sum_{\bm{k}'}[\hat h_1]_{\bm{k}\bm{k}'}\hat c_{\bm{k}'}$ with the single-particle operator $\hat h_1$, we can write
\begin{align}
    \hat c_{\bm{k}}\ee^{-\ii\hat H_1t}=\ee^{-\ii\hat H_1t}\sum_{\bm{k}'}[\ee^{-\ii\hat h_1t}]_{\bm{k}\bm{k}'}\hat c_{\bm{k}'}.
\end{align}
The term including the single-particle operator $\hat h_1$ can be pulled out of the expectation value and the susceptibility yields
\begin{align}
    \nonumber X(t) &= -\ii\Theta(t)\ee^{\ii\varepsilon_dt}\frac{1}{V}\sum_{\bm{k},\bm{k}',\bm{k}''}[\ee^{-\ii\hat h_1t}]_{\bm{k}\bm{k}''}\\
    &\phantom{=}\quad\quad\quad\quad\quad\times\langle\mathrm{FS}|\ee^{\ii\hat H_0t}\ee^{-\ii\hat H_1t}\hat c_{\bm{k}''}\hat c^\dagger_{\bm{k}'}|\mathrm{FS}\rangle.
    \label{eq:X(t)_intermediate}
\end{align}
Applying the anti-commutation relation $\hat c_{\bm{k}''}\hat c^\dagger_{\bm{k}'} = \delta_{\bm{k}'\bm{k}''}-\hat c^\dagger_{\bm{k}'}\hat c_{\bm{k}''}$ generates two terms. The first term is analogous to the $d$ propagator. In the second term,
\begin{align}
    \nonumber&\langle\mathrm{FS}|\ee^{\ii\hat H_0t}\ee^{-\ii\hat H_1t}\hat c^\dagger_{\bm{k}'}\hat c_{\bm{k}''}|\mathrm{FS}\rangle\\
    &= \frac{1}{Z_0}\mathrm{tr}\left[\ee^{-\beta(\hat H_0-\mu\hat N_0)}\ee^{\ii\hat H_0t}\ee^{-\ii\hat H_1t}\hat c^\dagger_{\bm{k}'}\hat c_{\bm{k}''}\right],
\end{align}
the density operator can be treated as a derivative of a bilinear operator:
\begin{align}
    \hat c^\dagger_{\bm{k}'}\hat c_{\bm{k}''} = \left.\frac{\dd}{\dd a}\ee^{a\hat c^\dagger_{\bm{k}'}\hat c_{\bm{k}''}}\right\vert_{a=0} \equiv \left.\frac{\dd}{\dd a}\ee^{a\sum_{\bm{q},\bm{q}'}[\hat A_{\bm{k}'\bm{k}''}]_{\bm{q}\bm{q}'}\hat c^\dagger_{\bm{q}}\hat c_{\bm{q}'}}\right\vert_{a=0}.
\end{align}
Here, the single-particle operator $\hat A_{\bm{k}'\bm{k}''}$ just picks the mode with the corresponding momenta. Consequently, Klich's formula is applicable:
\begin{align}
    \nonumber&\langle\mathrm{FS}|\ee^{\ii\hat H_0t}\ee^{-\ii\hat H_1t}\hat c^\dagger_{\bm{k}'}\hat c_{\bm{k}''}|\mathrm{FS}\rangle\\
    &= \left.\frac{\dd}{\dd a}\det\left[\hat\doubleI-f(\hat h_0)+f(\hat h_0)\ee^{\ii\hat h_0t}\ee^{-\ii\hat h_1 t}\ee^{a \hat A_{\bm{k}'\bm{k}''}}\right]\right\vert_{a=0}.
\end{align}
Let us define $B(t) = \hat\doubleI-f(\hat h_0)+f(\hat h_0)\ee^{\ii\hat h_0t}\ee^{-\ii\hat h_1 t}$. Using Jacobi's formula for the derivative of a determinant,
\begin{align}
	\frac{\dd}{\dd a}\det A(a)=\det A(a)\,\mathrm{tr}\left( A^{-1}(a)\frac{\dd A(a)}{\dd a}\right),
\end{align}
yields
\begin{align}
    \nonumber&\langle\mathrm{FS}|\ee^{\ii\hat H_0t}\ee^{-\ii\hat H_1t}\hat c^\dagger_{\bm{k}''}\hat c_{\bm{k}'}|\mathrm{FS}\rangle\\
    \nonumber &= \det B(t)\,\mathrm{tr}\left[B^{-1}(t) f(\hat h_0)\ee^{\ii\hat h_0t}\ee^{-\ii\hat h_1 t}\hat A_{\bm{k}'\bm{k}''}\right]\\
    &= \det B(t) \left[B^{-1}(t) f(\hat h_0)\ee^{\ii\hat h_0t}\ee^{-\ii\hat h_1 t}\right]_{\bm{k}''\bm{k}'}.
\end{align}
In the last line, we used the property of $\hat A_{\bm{k}'\bm{k}''}$. It picks up only one mode and thus gives only one matrix element in the trace, i.e., $\mathrm{tr}(MA_{\bm{k}'\bm{k}''})=M_{\bm{k}''\bm{k}'}$. Combining everything, we get the final form for the susceptibility, \Eq{eq:X(t)_intermediate}:
\begin{align}
    \nonumber X(t) &= -\ii\Theta(t)\ee^{\ii\varepsilon_d t}\det B(t)\\
    \nonumber&\phantom{=}\times\frac{1}{V}\sum_{\bm{k},\bm{k}''}\left[\ee^{-\ii\hat h_1 t}(\hat\doubleI - B^{-1}(t)f(\hat h_0)\ee^{\ii\hat h_0t}\ee^{-\ii\hat h_1 t})\right]_{\bm{k}\bm{k}''}\\
    \nonumber &= -\ii\Theta(t)\ee^{\ii\varepsilon_d t}\det B(t)\\
    &\phantom{=}\quad\times\frac{1}{V}\sum_{\bm{k},\bm{k}''}\left[\ee^{-\ii\hat h_1 t} B^{-1}(t)\left(\hat\doubleI-f(\hat h_0)\right)\right]_{\bm{k}\bm{k}''}.
    \label{eq:X(t)_FDA}
\end{align}
Also here, the retarded correspondent is obtained via Fourier transformation $X(\omega) = \int_t \ee^{\ii\omega t}X(t)$.

Equations \eqref{eq:G(t)_FDA} and \eqref{eq:X(t)_FDA} are computed by exact diagonalization of the single-particle matrix $[\hat h_1]_{\bm{k}\bm{k}'} = \varepsilon\delta_{\bm{k}\bm{k}'}-U/V$. To be more precise, we discretize the single-particle states with non-interacting energies $\varepsilon_n = 2\xi_0 n/n_\mathrm{max}$ where the number of states is given by $n_\mathrm{max}=\lfloor 2\xi_0 \rho V \rfloor$~\cite{Diekmann2024Leading}. This way, we simulate a constant density of states $\rho$ for a finite volume $V$ [cf.\ \Eq{eq:local_dos}]. We observe convergence of our data with respect to the finite size at high enough volumes, i.e., $n_\mathrm{max} \gtrsim 1000$.

The finite size, however, discretizes the energy spectrum $\delta\varepsilon = 2\xi_0/n_\mathrm{max}$, so we limit our Fourier integral up to time scales $t_\mathrm{max}=\pi/\delta\varepsilon \simeq \pi \rho V$. Furthermore, we broaden the frequency-dependent data by applying an exponential decay $\ee^{\pm 2t/t_\mathrm{max}}$ in the Fourier transform. Here $+$ is used in the exponent for the advanced Green's function $G(t)$, \Eq{eq:G(t)_FDA}, and $-$ for the retarded susceptibility $X(t)$, \Eq{eq:X(t)_FDA}.

\begin{figure}
    \centering
    \includegraphics[width=1.0\linewidth]{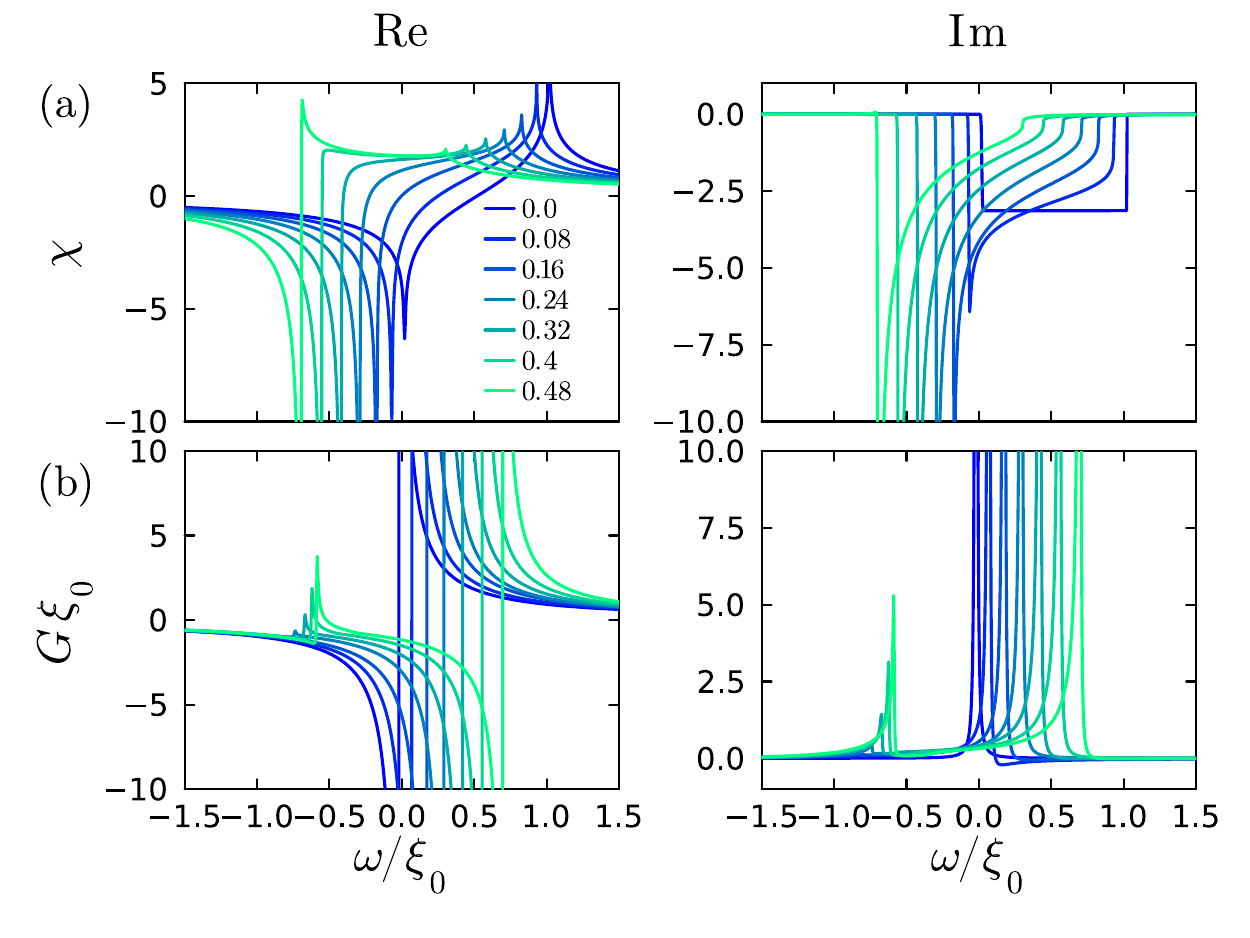}
    \caption{Results for the (a) the particle-hole susceptibility $\chi(\omega) = X(\omega)/\rho$ and (b) the $d$ propagator $G(\nu)$ from the functional determinant approach for $T/\xi_0 = 0.002$, $\xi_d=-0.02$, $n_\mathrm{max}=1000$, and different values of $u$ (marked by different colors). 
    }
    \label{fig:FDA_results}
\end{figure}

\begin{figure}
    \centering
    \includegraphics[width=1.0\linewidth]{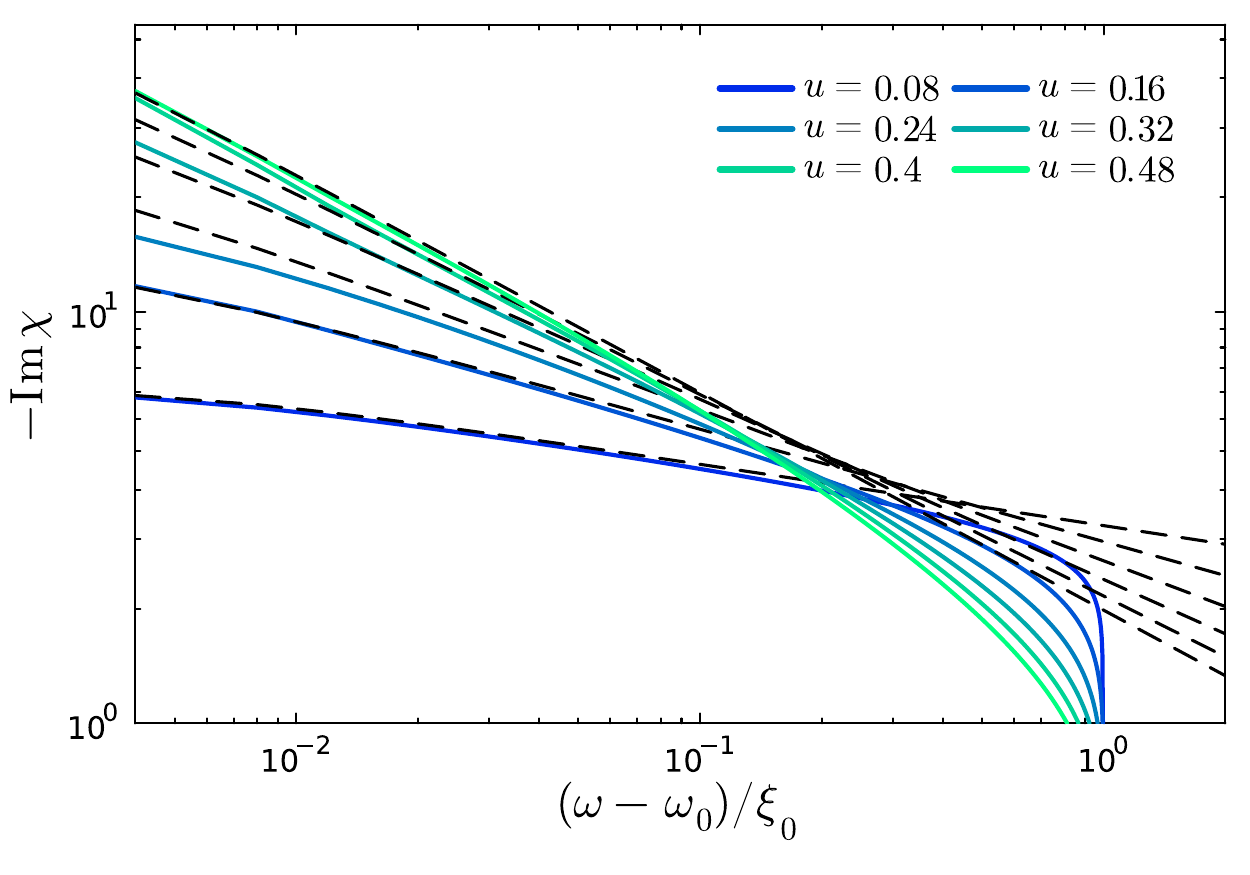}
    \caption{Susceptibility from the functional determinant approach with parameters as in \Fig{fig:FDA_results} compared to the analytical power-law behavior \Eq{eq:exact_power-law_chi} (black dashed lines).
    }
    \label{fig:FDA_power-law}
\end{figure}

We present our numerically exact data from the functional determinant approach in \Fig{fig:FDA_results}. The Fermi edge as well as the corresponding power laws are visible in both quantities. Note that the singularities are cut due to finite-size effects and the regularization we implement in the Fourier transforms. The additional peaks at negative frequencies in $G$ for large enough interactions $u$ [cf.\ \Fig{fig:FDA_results}(b)] mark the additional bound states~\cite{Mahan1967Excitons}. Close to the threshold frequency $\omega_0$, we can confirm that the analytical power-law behavior \Eq{eq:exact_power-law_chi} is well described by the functional determinant approach (cf.\ \Fig{fig:FDA_power-law}).

\subsection{Numerical results of the threshold}
\label{sec:FDA_Threshold}

\begin{figure}
    \centering
    \includegraphics[width=1.0\linewidth]{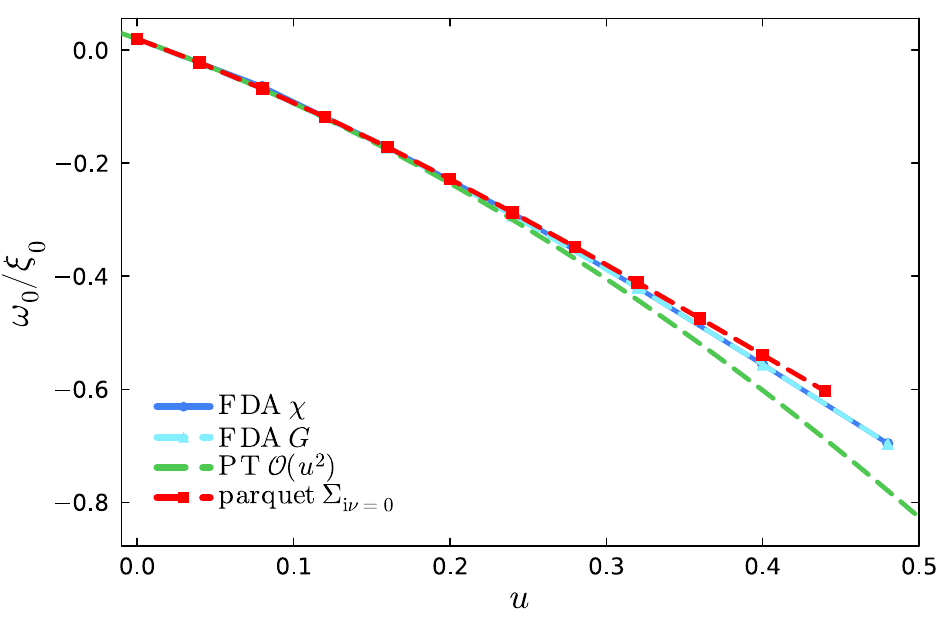}
    \caption{Threshold frequencies $\omega_0$ obtained from the functional determinant approach for $\chi$ (blue) and $G$ (light blue) [cf.\ \Fig{fig:FDA_results}] compared to the values obtained from second-order perturbation theory, \Eq{eq:Threshold_second-order}, (green, dashed) and the subleading-log self-energy $\Sigma_{\ii\nu=0}$ at zero imaginary frequency (red squares) [cf.\ \Eq{eq:Threshold_Sigma_sub}]. The data are evaluated at $T/\xi_0 = 0.002$ and $\xi_d/\xi_0 = -0.02$.}
    \label{fig:FDA_threshold}
\end{figure}

From the positions corresponding to the x-ray edge singularity in $\mathrm{Im}\,\chi$ [cf.\ \Fig{fig:FDA_results}(a)] and the peak due to Anderson's orthogonality catastrophe in $\mathrm{Im}\,G$ [cf.\ \Fig{fig:FDA_results}(b)], we extract the values for the threshold frequency $\omega_0$ (i.e., for $\chi$ the peaks are located at $\omega_0$ and for $G$ at $-\omega_0$). Figure \ref{fig:FDA_threshold} shows that the data for $\omega_0$ obtained from $\chi$ and $G$ lie on top (blue and light blue data points).

Furthermore, we compare these numerically exact data to the values form the diagrammatic approaches. First of all, we have the form in second-order perturbation theory,
\begin{align}
    \omega_0 = -\xi_d-u\xi_0-u^2\, 2\xi_0 \ln 2+\mathcal{O}(u^3),
    \label{eq:Threshold_second-order}
\end{align}
which describes the behavior quite accurately up to intermediate values $u\lesssim 0.3$ (cf.\ green dashed line in \Fig{fig:FDA_threshold}).

Moreover, \Fig{fig:FDA_threshold} also serves as a check that our empirical strategy for an adjustment of the renormalized threshold frequency in the parquet formalism (cf.\ \Sec{sec:Threshold}) is quite accurate. To this end, we determine values for the threshold as
\begin{align}
    \omega_0 = -\xi_d - u\xi_0 - [\Sigma_\mathrm{sub}]_{\ii\nu=0} - u^2 \xi_d L(0)
    .
    \label{eq:Threshold_Sigma_sub}
\end{align}
Here, $\Sigma_\mathrm{sub}$ is the imaginary-frequency self-energy obtained by an insertion of the leading-log vertex $\Gamma_\mathrm{lead}$, \Eqs{eq:parquet-leading-log}, into the Schwinger--Dyson equation~\eqref{eq:SDE}. The terms, which do not affect the threshold frequency are hereby compensated perturbatively, by adding the second-order logarithmic term $-u^2 \xi_d L(0)$ [cf.\ \Eq{eq:Sigma_2_exact_main_text_full}]. The data points extracted from \Eq{eq:Threshold_Sigma_sub} (red squares in \Fig{fig:FDA_threshold}) are closer to the actual values (blue) than the prediction from second-order perturbation theory (green dashed line). 

In a general treatment, the threshold frequency $\tilde\omega_0$ predicted by parquet results differs from the actual value $\omega_0$. To compare the power-law behavior predicted by parquet data with that predicted by the functional determinant approach, we need to adjust the threshold frequencies accordingly. Analytical continuation of parquet data to real frequencies provides a behavior $\chi^\mathrm{R}(\omega) = \chi(\omega+\ii 0^+) \sim (\omega+\ii 0^+-\tilde\omega_0)^{-\alpha_X}$ near the presumed threshold $\tilde\omega_0$ [cf.\ \Eq{eq:exact_power-law_chi}]. The value at zero Matsubara frequency $\chi_\mathrm{parq}(\ii\omega=0) \sim (-\tilde\omega_0)^{-\alpha_X}$ is approximately reproduced by the exact data at a shifted frequency $\chi^\mathrm{R}(\omega=\omega_0-\tilde\omega_0)$. Following this reasoning, the data $\chi_\mathrm{parq}(\ii\omega=0)$ presented in \Fig{fig:chi_w=0_U} that are computed with $\tilde{\omega}_0\simeq -\xi_d$ (cf.\ \Sec{sec:Threshold}) correspond to the values $\chi_\mathrm{FDA}(\omega=\omega_0+\xi_d)$ in \Fig{fig:FDA_results}.

\section{Further power-law expansions}
\label{sec:Taylor-expansions_supplement}

For completeness, let us give the Taylor expansion of \Eqs{eq:exact_power-laws} using the full exponents depending on the phase shift $\delta=\arctan(\pi u)$:
\begin{subequations}
    \begin{align}
    \nonumber\chi(\ii\omega)
        &=
        L-uL^2+u^2[\tfrac{2}{3}L^3+\tfrac{1}{2}L^2]\\
        \nonumber&\phantom{=}-u^3[\tfrac{1}{3}L^4+\tfrac{2}{3}L^3-\tfrac{1}{3}\pi^2L^2]
        \\
        \nonumber &\phantom{=}
        + u^4[\tfrac{2}{15}L^5+\tfrac{1}{2}L^4+(\tfrac{1}{6}-\tfrac{4}{9}\pi^2)L^3-\tfrac{1}{3}\pi^2L^2]
        \\
        &\phantom{=}
        +\mathcal{O}(u^5)
        ,\\
        G(\ii\nu)
        &=
        \frac{1}{\ii\nu-\xi_d}\left[1+u^2 \bar{L}+u^4(\tfrac{1}{2}\bar{L}^2-\tfrac{2}{3}\pi^2\bar{L})+\mathcal{O}(u^6)\right]
        \label{eq:full_power_law_expansion_G}
        .
\end{align}
\end{subequations}
As $2\delta/\pi-(\delta/\pi)^2=2u-u^2-2\pi^2u^3/3+\mathcal{O}(u^4)$, the leading- and subleading-log terms in $\chi$ are not changed compared to \Eqs{eq:Taylor-expansion}. Only the subsubleading terms are changed by the addends including $\pi$. Similarly, in the expansion of $G$, the highest power of logarithms at each order remains unchanged compared to \Eq{eq:Taylor-expansion_G}, since $(\delta/\pi)^2 = u^2 -2\pi^2 u^4/3 + \mathcal{O}(u^6)$.

It is worth to mention that the power law for the self-energy $\Sigma$ is completely analogous to that of the Green's function $G$. Applying the Dyson equation to \Eq{eq:exact_power-law_G} yields
\begin{align}
    \Sigma(\nu-\ii 0^+) &= (\nu-\ii 0^+\! +\omega_0)\left[1-\left(\frac{\nu-\ii 0^+\! +\omega_0}{\xi_0}\right)^{-\alpha_G}\right].
    \label{eq:power_law_Sigma}
\end{align}

Let us come back to the subleading-log power law $\alpha_X=2u-u^2$. Strictly speaking, the threshold frequency in the exact power law, \Eq{eq:exact_power-law_chi}, also depends on the interaction, i.e., $\omega_0=\omega_0(u)=-\xi_d + \mathcal{O}(u)$. This causes additional terms in the Taylor series of \Eq{eq:Taylor-expansion} (recall $\ii\tilde\omega = \ii\omega+\xi_d$):
\begin{align}
    \nonumber\chi(\ii\omega) &= \frac{1}{2u-u^2}\left[1-\left(\frac{\ii\omega-\omega_0(u)}{-\xi_0}\right)^{-2u+u^2}\right]\\
    \nonumber &= L - u \left[L^2 + \frac{\omega_0^\prime(0)}{\ii\tilde\omega}\right] + u^2 \left[\tfrac{2}{3}L^3+\tfrac{1}{2}L^2\right]\\
    &\phantom{=} +u^2\left[2L\frac{\omega_0^\prime(0)}{\ii\tilde\omega}-\tfrac{1}{2}\frac{[\omega_0^\prime(0)]^2}{(\ii\tilde\omega)^2}-\tfrac{1}{2}\frac{\omega_0^{\prime\prime}(0)}{\ii\tilde\omega}\right].
    \label{eq:Taylor-expansion_chi_w_threshold}
\end{align}
The terms involving $\omega_0^\prime(0)=-\xi_0$ originate from the $d$ Hartree self-energy $\Sigma_\mathrm{H}=u\xi_0$, \Eq{eq:Hartree}. [(i) $-u\,\omega_0^\prime(0)/(\ii\tilde\omega)$ is generated by a bubble including one $d$ Hartree term, (ii) $2u^2L\,\omega_0^\prime(0)/(\ii\tilde\omega)$ is generated by two connected bubbles where one includes a $d$ Hartree term, (iii) $-\tfrac{1}{2}u^2[\omega_0^\prime(0)]^2/(\ii\tilde\omega)^2$ is generated by one bubble including two $d$ Hartree terms.] Further, $\omega_0^{\prime\prime}(0)$ comes from the frequency-independent part of $\Sigma^{(2)}$, \Eq{eq:Sigma_2_approximated}, inserted into $\chi_{\Sigma}^{(2)}$, \Eq{eq:chi_2_Sigma}. We thus identify $\omega_0^{\prime\prime}(0) = -4 \xi_0 \ln 2$, which corresponds to \Eq{eq:Threshold_second-order}.

This interpretation is confirmed in the expansion of the self-energy $\Sigma = [G_0]^{-1}-G^{-1}$ with $G$ given by an extended form of \Eq{eq:Taylor-expansion_G} (recall $\ii\tilde\nu = \ii\nu-\xi_d$):
\begin{align}
    \nonumber\Sigma(\ii\nu) &= \frac{1}{G_0(\ii\nu)}-\frac{1}{G(\ii\nu)}\\
    \nonumber &= \ii\tilde\nu - (\ii\nu+\omega_0(u))\left(\frac{\ii\nu+\omega_0(u)}{\xi_0}\right)^{-u^2}\\
    &= -u\,\omega_0'(0) + u^2 \left[\ii\tilde\nu\bar L -\tfrac{1}{2}\omega_0''(0)\right] +\mathcal{O}(u^3).
    \label{eq:Taylor-expansion_Sigma_w_threshold}
\end{align}
Clearly, the first term corresponds to the $d$ Hartree self-energy, \Eq{eq:Hartree}, and the second term coincides with $\Sigma^{(2)}$, \Eq{eq:Sigma_2_approximated}.

\section{Frequency parametrization of the vertex}
\label{sec:Vertex-conventions}

\begin{figure}
    \centering
    \includegraphics[width=1.0\linewidth]{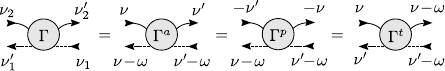}
    \caption{Channel-specific frequency conventions of the four-point vertices $\Gamma$.}
    \label{fig:Frequency-conventions}
\end{figure}

As a four-leg object, in general, the full vertex has four entries each depending on a frequency and particle type ($c$ vs.\ $d$). Our analysis in the main text is conducted exclusively in one realization of particle types with four distinguishable legs: one ingoing and outgoing $d$ leg and one ingoing and outgoing $c$ leg. We denote the four entries of the four-point vertex $\Gamma_{1'2'|12}$ where $1'$ refers to outgoing $d$, $2'$ outgoing $c$, $1$ ingoing $d$ and $2$ outgoing $c$. By energy conservation, $\Gamma$ only depends on three frequencies~\cite{gievers2022multiloop}. In this work, we use the following conventions:
\begin{subequations}
    \label{eq:Frequency-conventions}
	\begin{align}
		\Gamma^a_{\omega,\nu,\nu'}
        &=
        \Gamma_{\nu-\omega,\nu'|\nu'-\omega,\nu}
        , \\ 
		\Gamma^p_{\omega,\nu,\nu'}
        &=
        \Gamma_{\nu-\omega,-\nu|\nu'-\omega,-\nu'}
        , \\
		\Gamma^t_{\omega,\nu,\nu'}
        &=
        \Gamma_{\nu',\nu-\omega|\nu'-\omega,\nu}
        .
	\end{align}
\end{subequations}
These are illustrated in \Fig{fig:Frequency-conventions}. Note that in contrast to $\gamma^r$ and $I^r$, the channel index $r$ in $\Gamma^r$ does not indicate any reducibility property but just marks the channel taken for the frequency dependence. Due to \Eqs{eq:Frequency-conventions}, the Bethe--Salpeter equations are performed using the following summation/integration
\begin{align}
    \int_{\nu''}\Gamma^r_{\omega,\nu,\nu''}\Pi^r_{\omega,\nu''}\tilde\Gamma^r_{\omega,\nu'',\nu'}.
\end{align}
If the vertices $\Gamma$ or $\tilde{\Gamma}$ are replaced by ${r}'$-reducible vertices, one has to transform the arguments from channel $r$ to ${r}'$ according to the parametrization \Eq{eq:Frequency-conventions}. The transformations are given by the following linear maps:
\begin{subequations}
    \begin{alignat}{3}
        &\left(\begin{matrix}
            \omega \\ \nu \\ \nu'
        \end{matrix}\right)^a
        &&=
        \left(\begin{matrix}
            1 & -1 & -1 \\
            0 & \phantom{-}0 & -1 \\
            0 & -1 & \phantom{-}0
        \end{matrix}\right)
        \left(\begin{matrix}
            \omega \\ \nu \\ \nu'
        \end{matrix}\right)^p
        &&=
        \left(\begin{matrix}
            \phantom{-}0 & 1 & -1 \\
            \phantom{-}0 & 1 & \phantom{-}0 \\
            -1 & 1 & \phantom{-}0
        \end{matrix}\right)
        \left(\begin{matrix}
            \omega \\ \nu \\ \nu'
        \end{matrix}\right)^t
        , \\
        &\left(\begin{matrix}
            \omega \\ \nu \\ \nu'
        \end{matrix}\right)^p
        &&=
        \left(\begin{matrix}
            1 & -1 & -1 \\
            0 & \phantom{-}0 & -1 \\
            0 & -1 & \phantom{-}0
        \end{matrix}\right)
        \left(\begin{matrix}
            \omega \\ \nu \\ \nu'
        \end{matrix}\right)^a
        &&=
        \left(\begin{matrix}
            1 & -1 & -1 \\
            1 & -1 & \phantom{-}0 \\
            0 & -1 & \phantom{-}0
        \end{matrix}\right)
        \left(\begin{matrix}
            \omega \\ \nu \\ \nu'
        \end{matrix}\right)^t
        , \\
        &\left(\begin{matrix}
            \omega \\ \nu \\ \nu'
        \end{matrix}\right)^t
        &&=
        \left(\begin{matrix}
            \phantom{-}0 & 1 & -1 \\
            \phantom{-}0 & 1 & \phantom{-}0 \\
            -1 & 1 & \phantom{-}0
        \end{matrix}\right)
        \left(\begin{matrix}
            \omega \\ \nu \\ \nu'
        \end{matrix}\right)^a
        &&=
        \left(\begin{matrix}
            \phantom{-}0 & 1 & -1 \\
            \phantom{-}0 & 0 & -1 \\
            -1 & 1 & \phantom{-}0
        \end{matrix}\right)
        \left(\begin{matrix}
            \omega \\ \nu \\ \nu'
        \end{matrix}\right)^p
        .
    \end{alignat}
\end{subequations}

\section{Details of the perturbation series}
\label{sec:Perturbation-theory_details}

Individual diagrams in perturbation theory are obtained from successive integration over logarithmic terms. In the following, we give the most important integral expressions.

The integral over a logarithm to some power $n$ multiplied with the $d$ propagator $G\sim 1/(\ii\nu)$ raises the power of the logarithm according to
\begin{align}
    \int_a^b\dd\nu \frac{1}{\ii\nu+\ii\omega}\ln^n\frac{\ii\nu+\ii\omega}{\xi_0}
    &=
    \left[\frac{-\ii}{n+1}\ln^{n+1}\frac{\ii\nu+\ii\omega}{\xi_0}\right]_a^b
    ,
    \label{eq:integral-identity}
\end{align}
where $n \in \mathbb{N}_0$.

The power of the logarithm is not raised when there is no additional $d$ propagator. This becomes obvious from the simple integral
\begin{align}
    \int_a^b\dd\nu\ln\frac{\ii\nu+\ii\omega}{\xi_0}
    &=
    \left[(\nu+\omega)\ln\frac{\ii\nu+\ii\omega}{\xi_0}-\nu\right]_a^b
    ,
    \label{eq:integral_log}
\end{align}
which inductively can be generalized to higher powers of the logarithm 
\begin{align}
    \nonumber&\int_a^b\dd\nu\ln^n\frac{\ii\nu+\ii\omega}{\xi_0}\\
    &= \left[(\nu+\omega)\ln^{n}\frac{\ii\nu+\ii\omega}{\xi_0}\right]_a^b - n\int_a^b\dd\nu\ln^{n-1}\frac{\ii\nu+\ii\omega}{\xi_0}.
    \label{eq:integral_log^n}
\end{align}

Furthermore, if there are multiple $d$ propagators combined with a logarithm, the power of the logarithm is also not raised. This is shown by
\begin{align}
    \nonumber&\int_a^b \dd\nu\frac{1}{(\ii\nu+\ii\omega)^n}\ln\frac{\ii\nu+\ii\omega}{\xi_0}\\
    &= \left[\frac{\ii}{n-1}\frac{1}{(\ii\nu+\ii\omega)^{n-1}}\left(\ln\frac{\ii\nu+\ii\omega}{\xi_0}+\frac{1}{n-1}\right)\right]_a^b,
    \label{eq:integral_log/iv^n}
\end{align}
where $n>1$. For higher powers of the logarithm $m>1$, this has the following recursive generalization:
\begin{align}
    \nonumber&\int_a^b \dd\nu \frac{1}{(\ii\nu+\ii\omega)^n}\ln^m\frac{\ii\nu+\ii\omega}{\xi_0}\\
    \nonumber&= \left[\frac{\ii}{n-1}\frac{1}{(\ii\nu+\ii\omega)^{n-1}}\right.\\
    \nonumber&\phantom{=}\quad\times\left.\left(\ln^m\frac{\ii\nu+\ii\omega}{\xi_0}+\frac{1}{n-1}\ln^{m-1}\frac{\ii\nu+\ii\omega}{\xi_0}\right)\right]_a^b\\
    \nonumber&\phantom{=}+\frac{m-1}{n-1}\int_a^b\dd\nu\frac{1}{(\ii\nu+\ii\omega)^n}\\
    &\phantom{=}\quad\times\left(\ln^{m-1}\frac{\ii\nu+\ii\omega}{\xi_0}+\frac{1}{n-1}\ln^{m-2}\frac{\ii\nu+\ii\omega}{\xi_0}\right).
    \label{eq:integral_log^m/iv^n}
\end{align}

\subsection{Leading-log diagrams}
\label{sec:leading-logarithms_details}

The integral over the bare bubble, \Eq{eq:simple-bubble}, including the smooth propagator $g^\mathrm{sm}$, \Eq{eq:Gc_smooth}, is exactly solvable for $T=0$. For this, we keep the original energy integral, which comes from the density of states, \Eq{eq:local_dos} (we use $\ii\tilde\omega=\ii\omega + \xi_d$):
\begin{align}
    \int_\nu \Pi^a_{\omega,\nu}
    = 
    \int_\nu G_{\omega-\nu}g^\mathrm{sm}_\nu
    = 
    \int_{-\infty}^\infty\!\frac{\dd\nu}{2\pi}\frac{1}{\ii\nu-\ii\tilde\omega}
    \int_{-\xi_0}^{\xi_0}\!\frac{\dd\xi}{\ii\nu-\xi}.
\end{align}
First, the $\nu$ integral is solved via the residue theorem
\begin{align}
    \int_{-\infty}^\infty\!\frac{\dd\nu}{2\pi}\frac{1}{\ii\nu-\ii\tilde\omega}\frac{1}{\ii\nu-\xi} = \frac{\Theta(\xi)}{-\xi+\ii\tilde\omega}.
\end{align}
Next, the $\xi$ integral gives the logarithmic behavior:
\begin{align}
    \int_\nu\Pi^a_{\omega,\nu} &= -\int_{0}^{\xi_0}\frac{\dd\xi}{\xi-\ii\tilde\omega}=\ln\frac{\ii\tilde\omega}{\ii\tilde\omega-\xi_0}.
    \label{eq:simple-bubble_exact}
\end{align}
We are interested in the power-law behavior near the threshold frequency $\omega\simeq-\xi_d$. So, after analytic continuation, i.e., $\ii\omega\to\omega+\ii 0^+$, $|\tilde\omega|\ll\xi_0$, and we may use [cf.\ \Eq{eq:simple-bubble}]~\cite{roulet1969singularities,nozieres1969singularities2},
\begin{align}
    \ln\frac{\ii\tilde\omega}{\ii\tilde\omega-\xi_0} = \ln\frac{\ii\tilde\omega}{-\xi_0}+\mathcal{O}\bigg(\frac{\ii\tilde\omega}{-\xi_0}\bigg).
    \label{eq:simple-bubble_exact_close_to_threshold}
\end{align}

For general diagrams, the exact treatment of $g^\mathrm{sm}$ becomes difficult. Therefore, we now use the approximation $g^\mathrm{sh}$, \Eq{eq:Gc_sharp}, which holds close to the threshold frequency.

We first compute a general integral over a product of the bubble $\Pi^a$, involving the sharp Green's function $g^\mathrm{sh}$ \Eq{eq:Gc_sharp}, and an arbitrary function $f(\nu)$:
\begin{align}
    \nonumber \int_\nu \Pi^a_{\omega,\nu}f(\nu)
    &=
    -\frac{\ii}{2}\int_{-\xi_0}^{\xi_0}\dd\nu\frac{\sgn(\nu)}{\ii\nu-\ii\tilde\omega}f(\nu)
    \\
    &
    =
    -\frac{\ii}{2}\int_0^{\xi_0}\dd\nu\sum_{\sigma=\pm}\frac{f(\sigma\nu)}{\ii\nu-\sigma\ii\tilde\omega}
    .
    \label{eq:integral_bubble-f}
\end{align}
Setting $f(\nu)=1$, we find the 
integrated bubble
\begin{align}
    \chi^{(0)}(\omega)
    =
    \int_\nu\Pi^a_{\omega,\nu}
    &=
    \frac{1}{2}\sum_{\sigma=\pm}\ln\frac{\ii\tilde\omega}{\ii\tilde\omega-\sigma\ii\xi_0}
    .
    \label{eq:simple-bubble_details}
\end{align}
This resembles the exact result, \Eq{eq:simple-bubble_exact}, yet the usage of $g^\mathrm{sh}$ generates some artifacts at the UV cutoff $|\omega|\simeq\xi_0$. Using $\ln\ii + \ln-\ii = 0$, we may rewrite this as
\begin{align}
    \nonumber
    \chi^{(0)}(\omega)
    &=
    \tfrac{1}{2}\sum_{\sigma=\pm}
    \ln\frac{\ii\tilde\omega}{\sigma\tilde\omega-\xi_0}
    =
    \ln\frac{\ii\tilde\omega}{-\xi_0}
    - \tfrac{1}{2}\sum_{\sigma=\pm}
    \ln \big( 1 + \frac{\tilde\omega}{\xi_0} \big)
    \\
    & = \ln\frac{\ii\tilde\omega}{-\xi_0} + \mathcal{O} \bigg[\bigg(\frac{\tilde{\omega}}{\xi_0}\bigg)^2\bigg]
    ,
    \label{eq:simple-bubble_details2}
\end{align}
consistent with \Eq{eq:simple-bubble_exact_close_to_threshold}, \Eq{eq:simple-bubble} in the main text and Eq.~(16) in Ref.~\cite{roulet1969singularities}.

Logarithms with more complicated arguments are simplified up to logarithmic accuracy in order to apply the integral \Eq{eq:integral-identity}:
\begin{align}
    \nonumber \ln\frac{\ii\nu+\ii\nu'+\ii\tilde\omega}{-\xi_0}
    &\simeq
    \Theta(|\nu|-|\nu'|)\ln\frac{\ii\nu+\ii\tilde\omega}{-\xi_0}
    \\
    &\phantom{=}
    + \Theta(|\nu'|-|\nu|)\ln\frac{\ii\nu'+\ii\tilde\omega}{-\xi_0}
    ,
	\label{eq:logarithm-max}
\end{align}
which was first used before Eq.~(29) in Ref.~\cite{roulet1969singularities}.

Turning to the crossed diagram \Eq{eq:cross-diagram}, the integral over $\nu'$ is solved by inserting $[\gamma^p]^{(2)}=-u^2 L$ into \Eq{eq:integral_bubble-f}:
\begin{align}
    \nonumber &\int_{\nu'}[\gamma^p]^{(2)}_{\omega-\nu-\nu'}\Pi^a_{\omega,\nu'}
    \\ 
    &=
    \frac{\ii u^2}{2}\sum_{\sigma'}\int_0^{\xi_0}\!\!\dd\nu'\frac{1}{\ii\nu'-\sigma'\ii\tilde\omega}\ln\frac{\ii\nu'+\sigma'\ii\nu-\sigma'\ii\tilde\omega}{\sigma'\xi_0}
    .
    \label{eq:crossed-diagram-intermediate1}
\end{align}
Via \Eqs{eq:integral-identity} and \eqref{eq:logarithm-max}, the integral is evaluated as
\begin{align}
    \nonumber &
    \int_0^{\xi_0} \dd\nu' 
    \frac{1}{\ii\nu'-\sigma'\ii\tilde\omega}
    \ln\frac{\ii\nu'+\ii\sigma'\nu-\ii\sigma'\omega}{\sigma'\xi_0}
    \\
    \nonumber&\simeq 
    \int_0^{|\nu|} \dd\nu' 
    \frac{1}{\ii\nu'-\sigma'\ii\tilde\omega}
    \ln\frac{\ii\sigma'\nu-\ii\sigma'\omega}{\sigma'\xi_0}
    \\
    \nonumber& \quad +
    \int_{|\nu|}^{\xi_0} \dd\nu' 
    \frac{1}{\ii\nu'-\sigma'\ii\tilde\omega}
    \ln\frac{\ii\nu'-\ii\sigma'\omega}{\sigma'\xi_0}
    \\
    &= 
    -\ii \ln\frac{\ii\nu-\ii\tilde\omega}{\xi_0}
    \ln\frac{\ii|\nu|-\sigma'\ii\tilde\omega}{-\sigma'\ii\tilde\omega}
    -
    \frac{\ii}{2}\ln^2\frac{\ii\nu'-\ii\sigma'\omega}{\sigma'\xi_0}
    \bigg|_{|\nu|}^{\xi_0}
    .
    \label{eq:crossed-diagram-intermediate2}
\end{align}
We write the result as
\begin{align}
    \ii\ln\frac{\ii\nu-\ii\tilde\omega}{\xi_0}
    \left[
    \ln\frac{-\ii\tilde\omega}{\xi_0}
    -\ln\frac{\ii|\nu|-\sigma'\ii\tilde\omega}{\sigma'\xi_0}
    \right]
    +
    \frac{\ii}{2}
    \ln^2\frac{\ii|\nu|-\ii\sigma'\tilde\omega}{\sigma'\xi_0}
    ,
    \label{eq:crossed-diagram-intermediate3}
\end{align}
where, in the second term, we neglected terms arising from the upper limit $\nu'=\xi_0$ as up to logarithmic accuracy we can set $|\ii\xi_0-\ii\sigma'\tilde\omega|\simeq \xi_0$. Eventually, we combine the terms in \Eq{eq:crossed-diagram-intermediate3} by neglecting differences in the signs, as this is also correct up to logarithmic accuracy:
\begin{align}
    \left\vert\ln\frac{-\ii\nu+\ii\tilde\omega}{\xi_0}-\ln\frac{\ii\nu+\ii\tilde\omega}{\xi_0}\right\vert
    \simeq
    \left\vert\ln\frac{-\ii\nu + \ii\tilde\omega}{\ii\nu + \ii\tilde\omega}\right\vert\ll\left\vert\ln\frac{\ii\nu+\ii\tilde\omega}{\xi_0}\right\vert.
	\label{eq:neglect_signs_log}
\end{align}
This way, we obtain
\begin{align}
    \nonumber &\int_{\nu'}[\gamma^p]^{(2)}_{\omega-\nu-\nu'}\Pi^a_{\omega,\nu'}
    \\
	&\simeq
    u^2\left[\frac{1}{2}\ln^2\frac{\ii\nu-\ii\tilde\omega}{\xi_0}-\ln\frac{\ii\nu-\ii\tilde\omega}{\xi_0}\ln\frac{-\ii\tilde\omega}{\xi_0}\right]
    .
	\label{eq:result-gamma_p-Pi_a-1_a}
\end{align}
The integration over the second frequency $\nu$ in the crossed diagram, \Eq{eq:cross-diagram}, is then performed straightforwardly by \Eq{eq:integral-identity} and yields the logarithmic behavior $-\tfrac{1}{3}u^2 L^3$ given in the main text.

\subsection{Self-energy}
\label{sec:Self-energy_details}

The first-order Hartree diagram for the $d$ electron $\Sigma$ can be calculated exactly using $g^\mathrm{sm}$:
\begin{align}
    \nonumber\Sigma_\mathrm{H}^{(1)} 
    &= 
    u \int_\nu g_\nu^\mathrm{sm} \ee^{\ii \nu 0^+}
    = 
    u\int_{-\xi_0}^{\xi_0}\dd\xi\int_{-\infty}^\infty\frac{\dd\nu}{2\pi}\frac{\ee^{\ii \nu 0^+}}{\ii\nu-\xi}
    \\
    &= 
    u \int_{-\xi_0}^{\xi_0}\dd\xi\,\Theta(-\xi) = u\xi_0.
    \label{eq:Hartree}
\end{align}
Also the self-energy to second order, \Eq{eq:Sigma-2}, can be integrated exactly using $g^\mathrm{sm}$, \Eq{eq:Gc_smooth} (recall $\ii\tilde\nu=\ii\nu-\xi_d$):
\begin{align}
    \nonumber \tfrac{1}{u^2} \Sigma^{(2)}_\nu &= -\int_{-\infty}^\infty\frac{\dd\omega}{2\pi}\frac{1}{\ii\tilde\nu-\ii\omega}\int_{-\infty}^\infty\frac{\dd\nu'}{2\pi}\\
    &\phantom{=}\quad \times \int_{-\xi_0}^{\xi_0}\frac{\dd\xi_1}{\ii\nu'-\ii\omega-\xi_1}\int_{-\xi_0}^{\xi_0} \frac{\dd\xi_2}{\ii\nu'-\xi_2}.
\end{align}
The integrals over $\omega$ and $\nu$ are performed by the residue theorem, and we are left with
\begin{align}
    \nonumber \tfrac{1}{u^2} \Sigma^{(2)}_\nu
    &=
    \int_{-\xi_0}^{\xi_0}\!\dd\xi_1\int_{-\xi_0}^{\xi_0}\!\dd\xi_2\frac{\Theta(\xi_1)\Theta(-\xi_2)}{\xi_1-\xi_2+\ii\tilde\nu}\\
    &= \ii\tilde\nu\ln\frac{\ii\tilde\nu}{\ii\tilde\nu+\xi_0}
    +
    (\ii\tilde\nu+2\xi_0)\ln\frac{\ii\tilde\nu+2\xi_0}{\ii\tilde\nu+\xi_0}
    .
    \label{eq:Sigma_2_exact}
\end{align}
The leading behavior [cf.\ \Eq{eq:Sigma_2_approximated}] is extracted as 
\begin{align}
    \tfrac{1}{u^2} \Sigma^{(2)}_\nu
    &=
    \ii\tilde\nu \Big( \ln\frac{\ii\tilde\nu}{\xi_0} + \ln 2 - 1 \Big) +  2 \xi_0 \ln 2
    + \mathcal{O}\bigg( \frac{(\ii\tilde\nu)^2}{\xi_0} \bigg)
    .
    \label{eq:Sigma_2_exact3}
\end{align}
Setting $\ii\tilde\nu=0\Rightarrow \ii\nu=\xi_d$ in \Eq{eq:Sigma_2_exact3} yields the term $u^2 2\xi_0 \ln 2$, which appears in the expansions \Eqs{eq:Taylor-expansion_chi_w_threshold}--\eqref{eq:Taylor-expansion_Sigma_w_threshold} as threshold renormalization.

We now turn back to the approximated version using $g^\mathrm{sh}$, \Eq{eq:Gc_sharp}. Let us start by first integrating the bubble $\int\Pi^a\simeq L$, \Eq{eq:simple-bubble}, inside \Eq{eq:Sigma-2} for the self-energy $\Sigma^{(2)}$. Using the following integral expression including $g^\mathrm{sh}$, \Eq{eq:Gc_sharp}, and an arbitrary function $f(\nu)$:
\begin{align}
    \int_\nu f(\nu)g_\nu
    =
    -\frac{\ii}{2}\sum_{\sigma=\pm}\sigma\int_0^{\xi_0}\dd\nu\,f(\sigma\nu)
    ,
    \label{eq:integral_Gc-f}
\end{align}
the self-energy term yields
\begin{align}
    \nonumber \tfrac{1}{u^2} \Sigma^{(2)}_\nu
    &=
    -\int_{\nu',\nu''}\Pi^a_{\nu'-\nu,\nu''}g_{\nu'}
    \simeq
    -\int_{\nu'}\ln\frac{\ii\nu'-\ii\tilde\nu}{-\xi_0}g_{\nu'}
    \\
    &=
    \frac{\ii}{2}\sum_{\sigma'}\sigma'\int_0^{\xi_0}\dd\nu'\ln\frac{\sigma'\ii\nu'-\ii\tilde\nu}{-\xi_0}
    ,
    \label{eq:Sigma^(2)_cd-bubble}
\end{align}
which is solved by \Eq{eq:integral_log}. So, we obtain
\begin{align}
    \nonumber \tfrac{1}{u^2} \Sigma^{(2)}_\nu
    &=
    \frac{\ii}{2}\sum_{\sigma'}\sigma'\left[(\xi_0-\sigma'\tilde\nu)\ln\frac{\ii\xi_0-\sigma'\ii\tilde\nu}{-\sigma'\xi_0}-\xi_0\right.
    \\
    &\phantom{=\frac{\ii}{2}\sum_\sigma\sigma}
    \quad\quad\left.+\sigma'\tilde\nu\ln\frac{\ii\tilde\nu}{-\xi_0}\right]
    .
    \label{eq:Sigma-2_gamma_first}
\end{align}
To extract the logarithmic behavior in the first term, we approximate $\xi_0-\sigma'\tilde\nu\simeq\xi_0$ such that the sum over $\sigma'$ yields $-\xi_0\ln(\ii)+\xi_0\ln(-\ii) = -\ii\pi\xi_0$. Together with the other terms, we obtain $u^2(\ii\tilde\nu \bar{L}+\pi\xi_0/2)$. Here, the first term yields the correct logarithmic behavior given in the main text [cf.\ \Eq{eq:Sigma_2_approximated}], but the constant term beyond logarithmic accuracy is incorrect compared to the exact result, \Eq{eq:Sigma_2_exact}. We conclude that using $g^\mathrm{sh}$ instead of $g^\mathrm{sm}$ causes inconsistencies. That is why, in our numerical evaluation, we refrain from using $g^\mathrm{sh}$. 

The logarithmic behavior of the corresponding susceptibility, \Eq{eq:chi_2_Sigma}, is calculated by \Eqs{eq:integral_bubble-f} (recall $\ii\tilde\omega=\ii\omega+\xi_d$):
\begin{align}
    \nonumber\chi^{(2)}_{\Sigma}(\omega)
    &=
    \int_{\nu}\Pi^a_{\omega,\nu}u^2(\ii\nu-\ii\tilde\omega)\ln\frac{-\ii\nu+\ii\tilde\omega}{-\xi_0}\frac{1}{\ii\nu-\ii\tilde\omega}
    \\
    &=
    -u^2\frac{\ii}{2}\int_0^{\xi_0}\dd\nu\sum_\sigma\frac{1}{\ii\nu-\sigma\ii\tilde\omega}\ln\frac{\ii\nu-\sigma\ii\tilde\omega}{\sigma\xi_0}
    .
\end{align}
The expression is directly applicable to \Eq{eq:integral-identity} and yields the subleading-log term $\frac{1}{2}u^2L^2$ [cf.\ \Eq{eq:chi_2_Sigma}].

Let us briefly comment how the self-energy diagrams of third order cancel each other. In the Schwinger--Dyson equation~\eqref{eq:SDE}, one can replace $\Gamma$ by the second-order diagrams $\gamma_a^{(2)}$ and $\gamma_p^{(2)}$. This yields the following expression for the self-energy:
\begin{align}
    \nonumber \Sigma^{(3)}_\nu\!
    &= \,
    \begin{gathered}
        \includegraphics[width=105px]{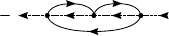}~
        \includegraphics[width=105px]{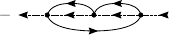}
    \end{gathered}
    \\
    &=
    - \int_{\nu''}\left([\gamma^a_\mathrm{lad}]^{(3)}_{\nu''-\nu}g_{\nu''}+[\gamma^p_\mathrm{lad}]^{(3)}_{\nu''-\nu}g_{-\nu''}\right)
    =
    0.
    \label{eq:Sigma_3}
\end{align}
The terms cancel each other as $[\gamma^a_\mathrm{lad}]_\omega^{(3)}=[\gamma^p_\mathrm{lad}]_\omega^{(3)}$ [cf.\ \Eqs{eq:ladder-diagrams_gamma}] and $g_{-\nu''}=-g_{\nu''}$.

\subsection{$t$-reducible diagram}
\label{sec:gamma_t_details}

The $t$-reducible diagram $[\gamma_t]^{(3)}$, \Eq{eq:gamma_t_3}, contains the integrated bubble of two conduction-electron propagators, $\int_{\nu_2}g_{\nu_2-(\nu'-\nu_1)}g_{\nu_2}$ similar to the self-energy $\Sigma^{(2)}$. We expand the product $G_{\nu_1-\omega}G_{\nu_1}$ into partial fractions (recall $\ii\tilde\nu=\ii\nu-\xi_d$ and $\ii\tilde\omega=\ii\omega+\xi_d$):
\begin{align}
    \nonumber G_{\nu_1-\omega}G_{\nu_1}
    &=
    \frac{1}{\ii\tilde\nu_1-\ii\omega}\frac{1}{\ii\tilde\nu_1} = \frac{1}{\ii\omega}\left[\frac{1}{\ii\tilde\nu_1-\ii\omega}-\frac{1}{\ii\tilde\nu_1}\right]
    \\
    &=
    \frac{1}{\ii\omega}\left[G_{\nu_1-\omega}-G_{\nu_1}\right]
    ,
\end{align}
and then we manipulate \Eq{eq:gamma_t_3} as
\begin{align}
     [\gamma^t]^{(3)}_{\omega,\nu'}
     &=
     \frac{u^3}{\ii\omega}\int_{\nu_1,\nu_2}[G_{\nu_1-\omega}-G_{\nu_1}]g_{\nu_1+\nu_2-\nu'}g_{\nu_2}
     .
\end{align}
By substituting $\nu_1\to\nu'-\omega'$ and $\nu_2\to\nu''$, we identify terms from the self-energy $\Sigma^{(2)}$ \Eq{eq:Sigma-2}:
\begin{align}
    [\gamma^t]^{(3)}_{\omega,\nu'}
    &=
    -\frac{u}{\ii\omega}(-u^2)\!\int_{\nu',\omega'}\![G_{\nu'-\omega-\omega'}-G_{\nu'-\omega'}]g_{\nu''-\omega'}g_{\nu''}
    ,
\end{align}
which yields the final expression, \Eq{eq:gamma_t_3}, given in the main text.

Inserting the logarithmic terms for the self-energies, \Eq{eq:Sigma_2_exact_main_text_full}, yields
\begin{align}
    [\gamma^t]^{(3)}_{\omega,\nu'}
    &=
    \frac{u^3}{\ii\omega}\left[\ii\tilde\nu'\ln\frac{\ii\tilde\nu'}{\xi_0}-(\ii\tilde\nu'-\ii\omega)\ln\frac{\ii\tilde\nu'-\ii\omega}{\xi_0}\right]
    .
    \label{eq:gamma_t_3_details}
\end{align}

In numerical calculations, 
a discontinuity appears at $\omega=0$, which reflects that the analytical behavior is critical there. Actually, the limit $\lim_{\omega\to 0}[\gamma^t]^{(3)}_{\omega,\nu}u^3L(-\nu')+\mathcal{O}(1)$ is well behaved. To regularize our numerical results, we linearly interpolate $\gamma^t_{\omega,\nu}$ at $\omega=0$ using the values for the first bosonic Matsubara frequencies $\omega=\pm\pi T$.

Finally, we derive the corresponding third-order term for the susceptibility, \Eq{eq:chi_gamma_t_3}. First, we use \Eq{eq:integral_bubble-f} and get
\begin{align}
    \nonumber \chi^{(3)}_{\gamma^t}(\omega)
    &=
    -\frac{1}{4}\sum_{\sigma,\sigma'}\int_0^{\xi_0}\dd\nu\int_0^{\xi_0}\dd\nu'\frac{\sigma}{\sigma\ii\nu-\ii\tilde\omega}\frac{\sigma'}{\sigma'\ii\nu'-\ii\tilde\omega}
    \\
    \nonumber &\phantom{=}\quad\quad\quad
    \times [\gamma^t]^{(3)}_{\sigma\nu-\sigma'\nu',\sigma\nu-\omega}
    \\
    \nonumber &\simeq
    \frac{u^3}{4}\sum_{\sigma,\sigma'}\int_0^{\xi_0}\dd\nu\int_0^{\xi_0}\dd\nu'
    \\
    \nonumber&\phantom{=}
    \times\left[\frac{\sigma\sigma'}{\sigma'\ii\nu'-\ii\tilde\omega}\frac{1}{\ii\sigma\nu-\sigma'\ii\nu'}\ln\frac{\ii\nu-\sigma\ii\tilde\omega}{\sigma\xi_0}\right.
    \\
	&\phantom{=}\quad
    \left.-\frac{\sigma\sigma'}{\sigma\ii\nu-\ii\tilde\omega}\frac{1}{\sigma\ii\nu-\sigma'\ii\nu'}\ln\frac{\ii\nu'-\sigma'\ii\tilde\omega}{\sigma'\xi_0}\right]
    .
    \label{eq:gamma_t_3_intermediate}
\end{align}
By exchanging the integration and summation variables $\nu\leftrightarrow\nu'$ and $\sigma\leftrightarrow\sigma'$, the two terms are the same. The integral over $\nu'$ can be performed by use of
\begin{align}
        &\nonumber\int_a^b\dd\nu\frac{1}{(\ii\nu+\ii\nu_1)(\ii\nu+\ii\nu_2)}
        \\
        \nonumber &=
        \frac{-\ii}{\ii\nu_1-\ii\nu_2}\left[-\ln(\nu+\nu_1)+\ln(\nu+\nu_2)\right]_a^b
        \\
		  &=
        \frac{-\ii}{\ii\nu_1-\ii\nu_2}\left[-\ln\frac{\ii\nu+\ii\nu_1}{\xi_0}+\ln\frac{\ii\nu+\ii\nu_2}{\xi_0}\right]_a^b
        ,
        \label{eq:integral_1/nu^2}
\end{align}
which is a special case of a product of $d$ propagators (including different frequencies $\nu+\nu_i$) without a logarithmic term: 
\begin{align}
    \int_a^b\!\dd\nu\prod_{i=1}^n\frac{1}{\ii\nu\!+\!\ii\nu_i}\!=\!\left[\ii(-1)^n\sum_{i=1}^n\ln\frac{\ii\nu\!+\!\ii\nu_i}{\xi_0}\prod_{j\neq i}\frac{1}{\ii\nu_j\!-\!\ii\nu_i}\right]_a^b\! .
    \label{eq:integral_1/iv^n}
\end{align}
We have
\begin{align}
    \nonumber&\int_0^{\xi_0}\dd\nu'\frac{-1}{(\ii\nu'-\sigma'\ii\tilde\omega)(\ii\nu'-\sigma\sigma'\ii\nu)}
    \\
    &\simeq
    \frac{\ii\sigma'}{\ii\tilde\omega-\sigma\ii\nu}\left[\ln\frac{\ii\tilde\omega}{-\xi_0}-\ln\frac{\ii\nu}{\sigma\xi_0}\right],
\end{align}
where we restricted the terms to the lower boundary of the integral $\nu'=0$. Equation~\eqref{eq:gamma_t_3_intermediate} then gives
\begin{align}
    \nonumber \chi^{(3)}_{\gamma^t}(\omega)
    &=
    \frac{u^3}{2}\sum_{\sigma,\sigma'}\int_0^{\xi_0}\dd\nu\,\ln\frac{\ii\nu-\sigma\ii\tilde\omega}{\sigma\xi_0}
    \\
    &\phantom{=}\quad
    \times\frac{-\ii}{\ii\nu-\sigma\ii\tilde\omega}\left[\ln\frac{\ii\tilde\omega}{-\xi_0}-\ln\frac{\ii\nu}{\sigma\xi_0}\right]
    .
\end{align}
The remaining expression is computed by \Eq{eq:integral-identity} using $\ln[\ii\nu/(\sigma\xi_0)]\simeq\ln[(\ii\nu-\sigma\ii\tilde\omega)/(\sigma\xi_0)]$ and again only the lower boundary $\nu=0$ is evaluated. This way we find the subleading-log behavior $\tfrac{1}{3}u^3 L^3$, given in the main text. Interpolating $\gamma^t$ at $\omega=0$ slightly improves the results for $\chi_{\gamma^t}^{(3)}$ shown in \Fig{fig:selfenergy-2}(c).

In analogy to our perturbative analysis, we interpolate the full $t$-reducible vertex $\gamma^t_{\omega,\nu,\nu'}$, \Eq{eq:gamma_t}, linearly around $\omega=0$ to avoid further numerical instabilities during the self-consistency loop of the parquet equations.

\subsection{Perturbative $U(1)$ Ward identity}
\label{sec:Ward-identity}

The $U(1)$ Ward identity in the Matsubara formalism yields~\cite{Kopietz_2010,krien2018conserving,Ritz2024Ward}
\begin{align}
    \nonumber&\Sigma^\sigma\!(\nu)\!-\!\Sigma^\sigma\!(\nu\!+\!\omega)\!=\!\int_{\nu'}\sum_{\sigma'}\left([G_0^{-1}]^{\sigma'}\!(\nu')\!-\![G_0^{-1}]^{\sigma'}\!(\nu'\!+\!\omega)\right)\\
    &\quad\quad\quad\times\Gamma_{\sigma\sigma'|\sigma'\sigma}(\nu+\omega,\nu'|\nu'+\omega,\nu)G^{\sigma'}(\nu')G^{\sigma'}(\nu'+\omega).
    \label{eq:Ward-identity}
\end{align}
In this section, we use the general notation for the vertex indices with particle types $\sigma = c, d$ introduced in \App{sec:Vertex-conventions-general}. For the x-ray edge singularity model with $G_0^d(\nu) \equiv G_\nu = 1/(\ii\nu-\xi_d)$, we get (using the notation from the main text where possible)
\begin{align}
    \nonumber \Sigma_\nu\!-\!\Sigma_{\nu+\omega} &= -\ii\omega\int_{\nu'}[\Gamma_{dd}]_{\nu+\omega,\nu'|\nu'+\omega,\nu}G_{\nu'}G_{\nu'+\omega}\\
    &+ \int_{\nu'}\left(g^{-1}_{\nu'}-g^{-1}_{\nu'+\omega}\right)[\Gamma_{\hat{dc}}]_{\nu+\omega,\nu'|\nu'+\omega,\nu}g_{\nu'}g_{\nu'+\omega}.
    \label{eq:Ward-identity-XES}
\end{align}
In second order of the interaction, the second term exactly vanishes, which can be seen as follows. First, the second-order contributions to $\Gamma_{\hat{dc}}$ are $[\gamma^t_{\hat{dc}}]^{(2)}$ and $[\gamma^p_{\hat{dc}}]^{(2)}$, which are related to the well studied terms by crossing symmetry:
\begin{align}
    \nonumber [\Gamma_{\hat{dc}}]^{(2)}_{\nu+\omega,\nu'|\nu'+\omega,\nu} &= -[\Gamma_{{dc}}]^{(2)}_{\nu+\omega,\nu'|\nu,\nu'+\omega}\\
    &= -[\gamma^a]^{(2)}_{\nu'-\nu}-[\gamma^p]^{(2)}_{-\nu'-\nu-\omega}.
\end{align}
From \Eqs{eq:ladder-diagrams_gamma}, we know that those are related by a minus sign, i.e., $[\gamma^a]_\omega^{(2)}=-[\gamma^p]_\omega^{(2)}$. The clue is now that the $c$ propagator is odd, i.e., $g_{-\nu}=-g_\nu$. By performing the substitution $\nu'\to\nu'-\omega/2$ of the integral variable $\nu'$ and then subdividing the integral for negative and positive $\nu'$, it becomes clear that the second term in \Eq{eq:Ward-identity-XES} vanishes:
\begin{align}
    \nonumber&\int_{\nu'}\left(g^{-1}_{\nu'-\tfrac{\omega}{2}}-g^{-1}_{\nu'+\tfrac{\omega}{2}}\right)\left(-[\gamma^a]^{(2)}_{\nu'-\tfrac{\omega}{2}-\nu}+[\gamma^a]^{(2)}_{-\nu'-\tfrac{\omega}{2}-\nu}\right)\\
    \nonumber&\phantom{\int_{\nu'}}\times g_{\nu'-\tfrac{\omega}{2}}g_{\nu'+\tfrac{\omega}{2}}\\
    \nonumber &= \int_{\nu'>0}\sum_{\sigma'=\pm}\left(g^{-1}_{\sigma'\nu'-\tfrac{\omega}{2}}-g^{-1}_{\sigma'\nu'+\tfrac{\omega}{2}}\right)\\
    \nonumber&\phantom{=} \times \left(-[\gamma^a]^{(2)}_{\sigma'\nu'-\tfrac{\omega}{2}-\nu}+[\gamma^a]^{(2)}_{-\sigma'\nu'-\tfrac{\omega}{2}-\nu}\right)g_{\sigma'\nu'-\tfrac{\omega}{2}}g_{\sigma'\nu'+\tfrac{\omega}{2}}\\
    &= 0.
\end{align}
The remaining first term in \Eq{eq:Ward-identity-XES} is simplified in second order. Here, $\Gamma_{dd}$ only gives a contribution in the $a$ channel [the $p$-channel diagram $[\gamma^p_{dd}]^{(2)}$ vanishes directly while $[\gamma^t_{dd}]^{(2)}$ yields a closed $d$ bubble when integrated multiplied by the two $d$ propagators in \Eq{eq:Ward-identity-XES}]. We now multiply both sides of \Eq{eq:Ward-identity-XES} by $u$ and get
\begin{align}
    \frac{u}{\ii\omega}\left(\Sigma^{(2)}_{\nu+\omega}-\Sigma^{(2)}_\nu\right) &= u \int_{\nu'}[\gamma^a_{dd}]^{(2)}_{\nu'-\nu}G_{\nu'}G_{\nu'+\omega}=[\gamma^t]^{(3)}_{\omega,\nu+\omega}.
\end{align}
Diagrammatically, we could identify the right-hand side with the third-order diagram in the $t$ channel in \App{sec:gamma_t_details}. A frequency shift $\nu\to\nu'-\omega$ yields the expression \eqref{eq:gamma_t_3} in the main text.

\subsection{Multi-boson exchange diagram}
\label{sec:MBE4_details}

Inserting the $a$-reducible multi-boson diagram $[M^a]^{(4)}$, \Eq{eq:MBE4}, into the susceptibility $\chi$ yields \Eq{eq:MBE-diagram} (recall $\ii\tilde\omega=\ii\omega+\xi_d$):
\begin{align}
    \nonumber \chi^{(4)}_{M^a}(\omega)
    &=
    \int_{\nu,\nu',\nu''}\!\!\!\!\!\Pi^a_{\omega,\nu}[\gamma^p]^{(2)}_{\omega-\nu-\nu''}\Pi^a_{\omega,\nu''}[\gamma^p]^{(2)}_{\omega-\nu''-\nu'}\Pi^a_{\omega,\nu'}
    \\
    &=
    \int_{\nu''}\Pi^a_{\omega,\nu''}\left[\int_{\nu}\Pi^a_{\omega,\nu}[\gamma^p]^{(2)}_{\omega-\nu-\nu''}\right]^2
    .
\end{align}
The integrals over $\nu$ and $\nu'$, which consist each of one bubble $\Pi^a$ and one vertex $[\gamma^p]^{(2)}$, could be integrated independently. They yield the same and, in fact, they coincide with the integral, \Eq{eq:result-gamma_p-Pi_a-1_a}, which has already been performed in the context of the crossed diagram,
\begin{align}
    \nonumber&\left[\tfrac{1}{u^2}\int_{\nu}\Pi^a_{\omega,\nu}[\gamma^p]^{(2)}_{\omega-\nu-\nu''}\right]^2\\
    &\simeq
    \frac{1}{4}\!\ln^4\!\frac{\ii\nu''\!-\!\ii\tilde\omega}{\xi_0}\!-\!\ln^3\!\frac{\ii\nu''\!-\!\ii\tilde\omega}{\xi_0}\ln\!\frac{-\!\ii\tilde\omega}{\xi_0}\!+\!\ln^2\!\frac{\ii\nu''\!-\!\ii\tilde\omega}{\xi_0}\ln^2\!\frac{-\!\ii\tilde\omega}{\xi_0}
    .
\end{align}
So in the end, one only needs to perform the integration over $\nu''$, which can be recast according to \Eq{eq:integral_bubble-f}:
\begin{align}
    \nonumber \chi^{(4)}_{M^a}(\omega)
    &\simeq
    -\frac{\ii u^4}{2}\int_0^{\xi_0}\dd\nu''\sum_\sigma\frac{1}{\ii\nu''\!-\!\sigma\ii\tilde\omega}\left[\frac{1}{4}\ln^4\frac{\ii\nu''\!-\!\sigma\ii\tilde\omega}{\sigma\xi_0}\right.
    \\
    &\phantom{=}
    \left. -\!\ln\frac{-\ii\tilde\omega}{\xi_0}\ln^3\frac{\ii\nu''\!-\!\sigma\ii\tilde\omega}{\sigma\xi_0}\!+\!\ln^2\frac{-\ii\tilde\omega}{\xi_0}\ln^2\frac{\ii\nu''\!-\!\sigma\ii\tilde\omega}{\sigma\xi_0}\right]\!
    .
\end{align}
This expression is solvable by \Eq{eq:integral-identity},
\begin{align}
    \nonumber \chi^{(4)}_{M^a}
    &=
    -\frac{\ii u^4}{2}\sum_\sigma\left[\frac{-\ii}{20}\ln^5\frac{\ii\nu''-\sigma\ii\tilde\omega}{\sigma\xi_0}\right.\\
    \nonumber &\phantom{= -\frac{\ii u^4}{2}\sum_\sigma} \quad+\frac{\ii}{4}\ln\frac{-\ii\tilde\omega}{\xi_0}\ln^4\frac{\ii\nu''-\sigma\ii\tilde\omega}{\sigma\xi_0}
    \\
    \nonumber &\phantom{= -\frac{\ii u^4}{2}\sum_\sigma} \quad
    \left.-\frac{\ii}{3}\ln^2\frac{-\ii\tilde\omega}{\xi_0}\ln^3\frac{\ii\nu''-\sigma\ii\tilde\omega}{\sigma\xi_0}\right]
    \\
    &\simeq
    u^4 \left[-\frac{1}{20}+\frac{1}{4}-\frac{1}{3}\right]\ln^5\frac{\ii\tilde\omega}{-\xi_0}
    ,
\end{align}
and yields the leading-log result $\tfrac{2}{15}u^4 L^5$ (cf.\ \Eq{eq:MBE-diagram} in the main text).

\section{Numerical results for the self-energy}

\label{sec:AOC_Self-energy}

From \Eq{eq:SDE_leading-behavior}, we concluded that inserting a leading-log vertex $\Gamma_\mathrm{lead}^{(n)} \sim u^n L^{n-1}$ into the Schwinger--Dyson equation~\eqref{eq:SDE} yields a subleading-log contribution to the self-energy $\Sigma_\mathrm{sub}^{(n+1)}\sim u^{n+1}\ii\tilde\nu \bar{L}^n$. Analogously, inserting a subleading-log vertex $\Gamma_\mathrm{sub}\sim u^n L^{n-2}$ yields a subsubleading-log term $\Sigma_\mathrm{subsub}^{(n+1)}\sim u^{n+1}\ii\tilde\nu\bar{L}^{n-1}$. Consequently, inserting the full leading-log vertex $\Gamma_\mathrm{lead}$, solved by \Eqs{eq:parquet-leading-log}, into the Schwinger--Dyson equation~\eqref{eq:SDE} reproduces the full subleading logarithm of the self-energy, while inserting the subleading-log $\Gamma_\mathrm{sub}$, solved by \Eqs{eq:SDE}--\eqref{eq:Gamma_parquet_extended} and the full $d$ propagator, reproduces the full subsubleading logarithm of the self-energy.

As mentioned at the end of \Sec{sec:Model}, the expansion of Anderson's orthogonality power law, \Eq{eq:Taylor-expansion_G}, involves powers of $u^2\bar{L}$. The second-order term $u^2\ii\tilde\nu \bar{L}$ for the self-energy [cf.\ \Eq{eq:Sigma_2_approximated}] involves the subleading logarithm and is thus correctly reproduced by inserting $\Gamma_\mathrm{lead}$ into the Schwinger--Dyson equation~\eqref{eq:SDE}. The fourth-order term $u^4\ii\tilde\nu\bar{L}^2$, in contrast, already goes beyond the subleading logarithm and can only be correctly reproduced when also the subleading contributions to $\Gamma_\mathrm{sub}$ and the renormalized propagator $G_\mathrm{sub}$ are included in \Eq{eq:SDE}. Furthermore, a complete computation of the sixth-order term $u^6\ii\tilde\nu\bar{L}^3$ would already require inclusion of the envelope diagram $R^{(4)}_\boxtimes$.

We see that terms involving the subleading logarithm $u^n \ii\tilde\nu \bar{L}^{n-1}$ have to exactly cancel in higher orders of perturbation theory $\mathcal{O}(u^{n\geq 3})$ in order to reproduce Anderson's orthogonality power law. Using our scheme, we cannot guarantee the complete reproduction of the first non-vanishing logarithmic terms at arbitrary orders of perturbation theory for the self-energy without the inclusion of totally irreducible diagrams (beyond the bare vertex). These are beyond the scope of this work. Nonetheless, an insertion of $\Gamma_\mathrm{lead}$ into \Eq{eq:SDE} already generates a lot of terms beyond the subleading logarithm $u^n \ii\tilde\nu \bar{L}^{n-1-p}$ with $p>0$. We evaluate them numerically, being aware that their summation is incomplete.

\begin{figure}
    \centering
    \includegraphics[width=1.0\linewidth] {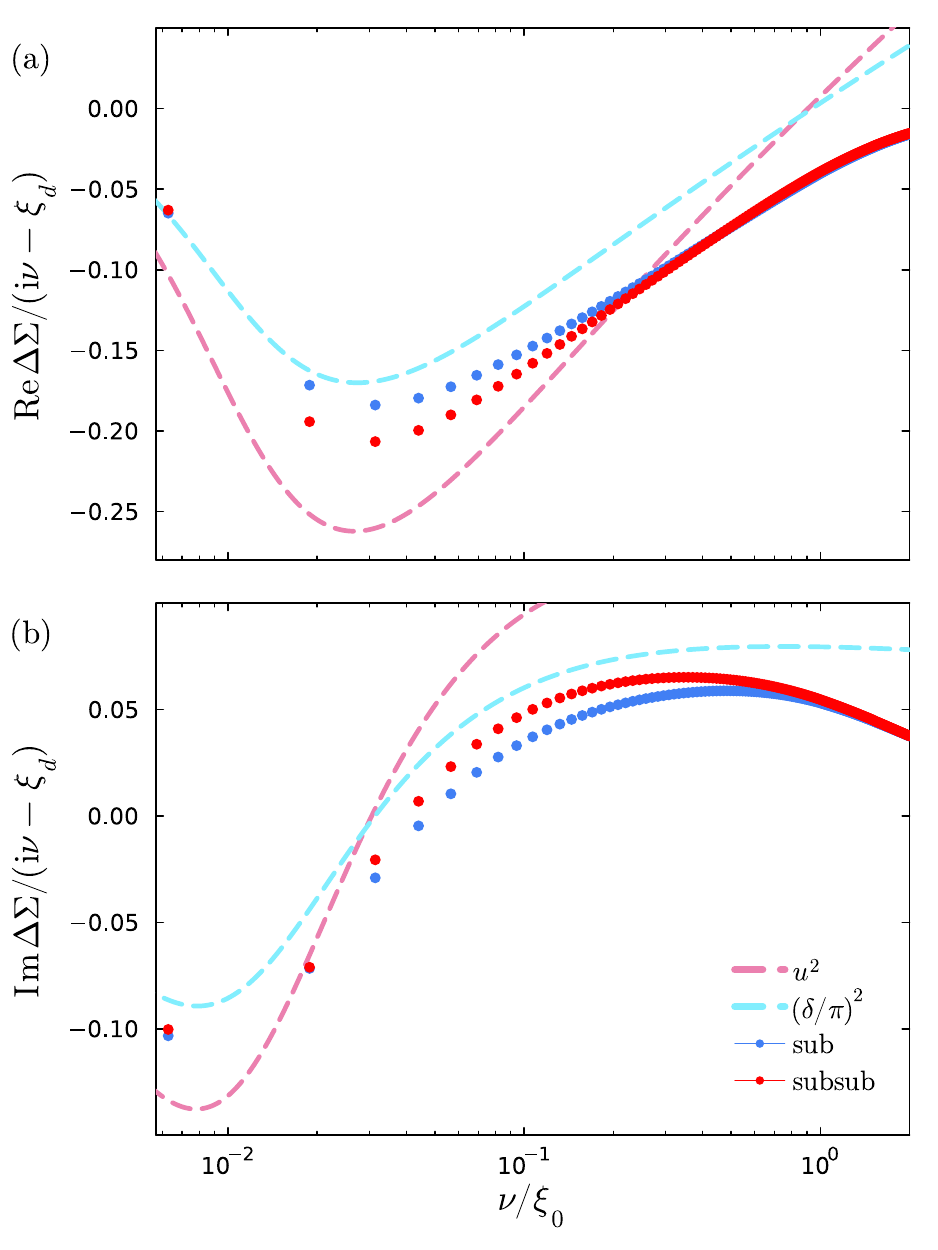}
    \caption{
    Frequency dependence of $\Sigma$ from self-consistent summations at $u=0.28$, $T/\xi_0=0.002$, and $\xi_d/\xi_0=-0.01$. The self-energy differences $\Delta\Sigma$, \Eq{eq:Self-energy_subtraction}, are divided by $\ii\tilde\nu=\ii\nu-\xi_d$. We compare the numerical results after inserting the leading-log vertex into the Schwinger--Dyson equation (blue dots) and after inserting the subleading-log vertex and the renormalized $d$ propagator into the Schwinger--Dyson equation (red dots) to the analytically determined $T=0$ power law with the subleading-log exponent $\alpha_G = u^2$ (pink, dashed) and the exact exponent $\alpha_G = (\delta/\pi)^2$ (light blue, dashed).}
    \label{fig:Sigma_v_parquet}
\end{figure}

Although within the parquet approximation, it is impossible to capture the $u^2$ power law at all orders of perturbation theory, we may still compare our numerical data to the power law, \Eq{eq:power_law_Sigma}. Figure~\ref{fig:Sigma_v_parquet} shows the results when inserting the leading-log vertex $\Gamma_\mathrm{lead}$, \Eq{eq:parquet-leading-log}, into the Schwinger--Dyson equation~\eqref{eq:SDE} (blue dots) and when additionally including the subleading-log vertex $\Gamma_\mathrm{sub}$, \Eq{eq:Gamma_parquet_extended}, and the full $d$ propagator (red dots). The analytic power laws with $\alpha_G = u^2$ and $\alpha_G = (\delta/\pi)^2$ (pink and light blue, dashed), applicable in a rather small frequency regime, are not too far from the numerical results. Moreover, it is remarkable that the quantitative difference between inserting $\Gamma_\mathrm{lead}$ or $\Gamma_\mathrm{sub}$ into the Schwinger--Dyson equation is rather small.

\section{Details on the $t$-reducible vertex}
\label{sec:gamma_t-full}

Here, we give details on how our expression for the $t$-reducible vertex, $\gamma^t$ \Eq{eq:gamma_t}, can be motivated from the full parquet formalism and how we have implemented its numerical computation.

\subsection{More general vertex conventions}
\label{sec:Vertex-conventions-general}

\begin{figure}
    \centering
    \includegraphics[width=0.98\linewidth]{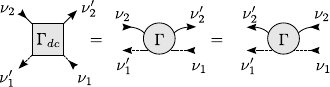}
    \caption{Translation between the two diagrammatic conventions: The square vertex is used in Hubbard-like models. Here, the positions of the legs are fixed. This more general notation is exclusively used in this section. The round vertex on the other hand is used in the main text. There, the frequencies are defined according to the respective leg type, not its leg position.}
    \label{fig:translation_diagrams}
\end{figure}

The bare interaction $u$ appearing in the action $S$, \Eq{eq:action_rescaled}, describes only a single scattering event between the $d$ electron with one $c$ electron. In a general diagrammatic treatment, however, the full vertex $\Gamma$ describes all scattering events between $d$ electrons and $c$ electrons involving two particles, in particular, also scattering events within one particle type. For this, it naturally comes with four indices $\Gamma_{1'2'|12}$ representing the four different particle types of the legs (cf.\ \App{sec:Vertex-conventions} and Ref.~\cite{gievers2022multiloop}). While in the main text only one component is needed, namely that with four distinguishable legs, in the general treatment, we have to include the particle-type in the notation. Figure~\ref{fig:translation_diagrams} points out the difference between the two conventions: In Hubbard-like models, we represent the full vertex by a square, where the indices of its four legs are identified by their position ($\Gamma_{1'2'|12}$: $1'$ bottom-left, $2'$ top-right, $1$ bottom-right, $2$ top-left). When using only one component with four distinguishable legs, this notation becomes superfluous; so, in the main text, we always take $\Gamma \equiv \Gamma_{1'2'|12}$ (with $1'$ outgoing $d$, $2'$ outgoing $c$, $1$ ingoing $d$, $2$ ingoing $c$). There, the indices are identified by its particle types and the position is not decisive. To mark the difference, we represent $\Gamma$ by a circle.

\begin{figure}
    \centering
    \includegraphics[width=0.85\linewidth]{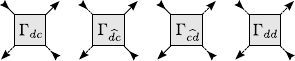}
    \caption{Convention for the particle-type components of the full vertex $\Gamma$.}
    \label{fig:Gamma_spin}
\end{figure}

Following the convention with four indices of the full vertex $\Gamma_{1'2'|12}$ introduced in Ref.~\cite{gievers2022multiloop}, we denote the spin components of vertices in the following way:
\begin{align}
    \Gamma_{dc} = \Gamma_{dc|dc}, \quad \Gamma_{\hat{dc}} = \Gamma_{dc|cd}, \quad \Gamma_{dd} = \Gamma_{dd|dd}.
    \label{eq:Spin-conventions}
\end{align}
The other components $\Gamma_{cd}$, $\Gamma_{\hat{cd}}$ and $\Gamma_{cc}$ are obtained by exchanging $c\leftrightarrow d$. The corresponding diagrams are illustrated in \Fig{fig:Gamma_spin}. In contrast to Hubbard-like models obeying SU($2$) spin symmetry, $\Gamma_{dd}$ cannot be retrieved from $\Gamma_{dc}$ and $\Gamma_{\hat{dc}}$. Moreover, due to the advanced property of the $d$ propagator, closed loops of dashed $d$ lines are suppressed, so $\Gamma_{cc}$ is negligible as it must contain closed dashed loops. In the main text, we exclusively use the component $\Gamma_{dc}$ and drop the corresponding indices.

Exchanging two fermionic legs of the vertex yields an additional minus sign:
\begin{align}
    \Gamma_{1'2'|12} = -\Gamma_{2'1'|12} = -\Gamma_{1'2'|21} = \Gamma_{2'1'|21}.
    \label{eq:Crossing-symmetries}
\end{align}
An insertion of the frequencies \Eqs{eq:Frequency-conventions} and spin indices \Eqs{eq:Spin-conventions} yields the so-called \emph{crossing symmetries}~\cite{Bickers2004,Rohringer2013}. Applying symmetries interrelates the different vertex components and heavily simplifies the numerical effort. Note that under exchange of two legs, the $a$ and $t$ channels are translated into each other while the $p$ channel translates into itself.

In the main text, only bubbles including one $d$ line $G$ and one $c$ line $g$ were used [cf.\ \Eq{eq:bubbles-definition}]. In general, however, the products of two propagators can appear in all possible combinations of particle-type indices and diagrammatic channels:
\begin{subequations}
    \label{eq:bubbles-general}
    \begin{align}
        [\Pi^a_{ij}]_{\omega,\nu} &= G^i_{\nu}G^j_{\nu+\omega},\\
        [\Pi^p_{ij}]_{\omega,\nu} &= \tfrac{1}{2}G^i_{-\nu}G^j_{\nu+\omega},\\
        [\Pi^t_{ij}]_{\omega,\nu} &= -G^i_{\nu}G^j_{\nu+\omega}.
    \end{align}
\end{subequations}
Note that in contrast to \Eq{eq:bubbles-definition} in the main text, we inserted an additional factor $1/2$ in the definition of $\Pi^p_{ij}$ to compensate overcounting, which appears as the additional sum over particle types in the Bethe--Salpeter equations includes both $\Gamma_{dc}$ and $\Gamma_{cd}$~\cite{gievers2022multiloop}. Moreover, in the general framework, we denote $G^c_\nu=g_\nu$ and $G^d_{\nu}=G_{\nu}$.

\subsection{Subleading-log parts of the $t$-reducible vertex}
\label{sec:gamma_t_subleading}

\begin{figure}
    \centering
    \includegraphics[width=1.0\linewidth]{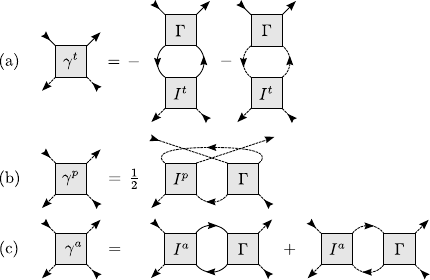}
    \caption{Vertices appearing in the full parquet expression for the $t$-reducible vertex $\gamma^t$.}
    \label{fig:gamma_t_full}
\end{figure}

Here, we motivate that our expression for $\gamma^t$, \Eq{eq:gamma_t}, can be derived from the full parquet formalism~\cite{KuglerNJP2018e,gievers2022multiloop} by taking into account only subleading-log diagrams. In the full parquet formalism, the $dc$ component of the $t$-reducible vertex is given by [cf.\ \Fig{fig:gamma_t_full}(a)]
\begin{align}
    \nonumber [\gamma^t_{dc}]_{\omega,\nu,\nu'} &= \int_{\nu''}[\Gamma^t_{cc}]_{\omega,\nu,\nu''} [\Pi^t_{cc}]_{\omega,\nu''}[I^t_{dc}]_{\omega,\nu'',\nu'}\\
    &\phantom{=} + \int_{\nu''}[\Gamma^t_{dc}]_{\omega,\nu,\nu''} [\Pi^t_{dd}]_{\omega,\nu''}[I^t_{dd}]_{\omega,\nu'',\nu'}.
    \label{eq:full_gamma_t}
\end{align}
The first term does not contribute as $\Gamma_{cc}$ contains closed dashed loops. According to \Sec{sec:logarithmic_behavior}, $\gamma^t_{dc}$ at the most contains subleading-log terms if both the inserted vertices are leading log. Insertions of the self-energy into $\Pi^t_{dd}$ and subleading contributions of the two vertices $\Gamma_{dc}$ and $I^t_{dd}$ are beyond our subleading-log scheme. Thus, they are on the same footing as the higher-order totally irreducible vertices, which are anyways dropped according to the parquet approximation.

So, we focus on $\Gamma_{dc}$ and $I^t_{dd}$ in leading-log order. The irreducible vertex is given by $I^t_{dd} = \gamma^a_{dd} + \gamma^p_{dd}$. There is only one contribution in the $p$-reducible channel $\gamma^p_{dd}$ [cf.\  \Fig{fig:gamma_t_full}(b)]:
\begin{align}
    [\gamma^p_{dd}]_{\omega,\nu,\nu'} = \int_{\nu''}[\Gamma^p_{dd}]_{\omega,\nu,\nu''} [\Pi^p_{dd}]_{\omega,\nu''}[I^p_{dd}]_{\omega,\nu'',\nu'}.
\end{align}
This eventually leads to closed dashed loops when inserted into \Eq{eq:full_gamma_t} and thus leads to a vanishing contribution. From the two contributions in the $a$-reducible channel $\gamma^a_{dd}$ [cf.\ \Fig{fig:gamma_t_full}(c)],
\begin{align}
    \nonumber [\gamma^a_{dd}]_{\omega,\nu,\nu'} &= \int_{\nu''}[\Gamma^a_{\hat{dc}}]_{\omega,\nu,\nu''} [\Pi^a_{cc}]_{\omega,\nu''}[I^a_{\hat{cd}}]_{\omega,\nu'',\nu'}\\
    &\phantom{=} + \int_{\nu''}[\Gamma^a_{dd}]_{\omega,\nu,\nu''} [\Pi^a_{dd}]_{\omega,\nu''}[I^a_{dd}]_{\omega,\nu'',\nu'},
\end{align}
the second one also leads to closed dashed loops and is therefore negligible. Hence, the remaining term includes $I^a_{\hat{dc}}$ and $\Gamma_{\hat{cd}}$. By crossing symmetry, the $a$-irreducible vertex is related to the $t$-irreducible one, i.e., $I^a_{\hat{cd}}=-I^t_{cd}=u-\gamma^a_{cd}-\gamma^p_{cd}$, which, to leading-log order coincides with the full vertex $-\Gamma_{cd}$ (there are no leading-log contributions in the transversal channel). So, only the term
\begin{align}
    \int_{\nu''}[\Gamma^a_{\hat{dc}}]_{\omega,\nu,\nu''} [\Pi^a_{cc}]_{\omega,\nu''}[\Gamma^a_{\hat{cd}}]_{\omega,\nu'',\nu'} \subseteq [I^t_{dd}]_{\omega,\nu,\nu'}
\end{align}
leads to the subleading logarithm of the full parquet $t$-reducible vertex $\gamma^t_{dc}$, \Eq{eq:full_gamma_t}. Thus, the remaining term in \Eq{eq:full_gamma_t} [cf.\ \Eq{eq:gamma_t_in_components}] reproduces the expression used in the main text [cf.\ \Eq{eq:gamma_t}].

\subsection{Numerical implementation of the $t$-reducible vertex}
\label{sec:gamma_t-numerics}

In the subleading-log parquet scheme, \Eq{eq:Gamma_parquet_extended}, the $dc$ component of the $t$-reducible vertex $\gamma^t$ is taken additionally, which, on the other hand, includes the $dd$ component of the $a$-reducible vertex $\gamma^a$. Using the conventions introduced in \Sec{sec:Vertex-conventions-general}, the $t$-reducible vertex $\gamma^t$, \Eq{eq:gamma_t}, from the main text is equivalent to
\begin{subequations}
    \label{eq:gamma_t_in_components}
    \begin{align}
        [\gamma^t_{dc}]_{\omega,\nu,\nu'}
        &=
        \int_{\nu''}[\Gamma^t_{dc}]_{\omega,\nu,\nu''}[\Pi^t_{dd}]_{\omega,\nu''}[\gamma^a_{dd}]_{\nu''-\nu',\omega+\nu',\nu'}
        ,
        \label{eq:gamma_t^dc}\\
        [\gamma^a_{dd}]_{\omega,\nu,\nu'}
        &=
        \int_{\nu''}[\Gamma^a_{\hat{dc}}]_{\omega,\nu,\nu''}[\Pi^a_{cc}]_{\omega,\nu''}[\Gamma^a_{\hat{cd}}]_{\omega,\nu'',\nu'}
        ,
        \label{eq:gamma_a^dd}
        \\
        \begin{gathered}
            \includegraphics[width=40pt]{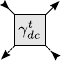}
        \end{gathered}
        &=~ -
        \begin{gathered}
            \includegraphics[width=40pt]{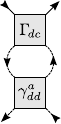}
         \end{gathered}
         ~=~ -
         \begin{gathered}
            \includegraphics[width=80pt]{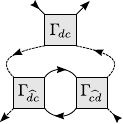}~.
         \end{gathered}
    \end{align}
\end{subequations}
To minimize the effort in numerical computations, the additional vertex components $\Gamma^t_{dc}$, $\Gamma^a_{\hat{dc}}$ and $\Gamma^a_{\hat{cd}}$ are expressed in terms of $\Gamma^t_{dc}$ and $\Gamma^a_{dc}$ by using the crossing symmetries, \Eqs{eq:Crossing-symmetries}. Explicitly we have
\begin{subequations}
    \label{eq:Crossing-symmetries_for_dcx_cdx}
    \begin{align}
        \nonumber[\Gamma^t_{dc}]_{\omega,\nu,\nu'}
        &=
        -u + [\gamma^t_{dc}]_{\omega,\nu,\nu'} + [\gamma^a_{dc}]_{\nu-\nu',\omega+\nu',\nu'}
        \\
        &\phantom{=}
        + [\gamma^p_{dc}]_{\omega+\nu+\nu',-\omega-\nu',-\nu'}
        ,
        \\
        \nonumber[\Gamma^a_{\hat{dc}}]_{\omega,\nu,\nu'}
        &=
        -[\Gamma^a_{dc}]_{\nu'-\nu,\nu,\nu+\omega}
        \\
        \nonumber&=
        u - [\gamma^a_{dc}]_{\nu'-\nu,\nu,\nu+\omega} - [\gamma^p_{dc}]_{\omega+\nu+\nu',-\nu,-\nu-\omega}
        \\
        &\phantom{=}
        - [\gamma^t_{dc}]_{-\omega,\nu',\nu+\omega}
        ,
        \label{eq:Gamma_a^dcx}
        \\
        \nonumber[\Gamma^a_{\hat{cd}}]_{\omega,\nu,\nu'}
        &=
        -[\Gamma^a_{dc}]_{\nu-\nu',\nu'+\omega,\nu'}
        \\
        \nonumber&=
        u - [\gamma^a_{dc}]_{\nu-\nu',\nu'+\omega,\nu'} - [\gamma^p_{dc}]_{\omega+\nu+\nu',-\nu'-\omega,-\nu'}
        \\
        &\phantom{=}
        - [\gamma^t_{dc}]_{\omega,\nu,\nu'}
        .
        \label{eq:Gamma_a^cdx}
    \end{align}
\end{subequations}
Note that for the first-order contribution we have $\Gamma^{(1)}_{dc}=-u=-\Gamma^{(1)}_{\hat{dc}}=-\Gamma^{(1)}_{\hat{cd}}$. By inserting \Eqs{eq:Crossing-symmetries_for_dcx_cdx} into \Eqs{eq:gamma_t_in_components}, we receive the equations given in the main text, were everything is expressed in the $dc$ component. For clarity, let us elaborate the derivation. In a first step, \Eqs{eq:Gamma_a^dcx}--\eqref{eq:Gamma_a^cdx} are inserted into \Eq{eq:gamma_a^dd}:
\begin{align}
    [\gamma^a_{dd}]_{\omega,\nu,\nu'}
    &=
    \int_{\nu''}[\Gamma^a_{dc}]_{\nu''-\nu,\nu,\nu+\omega}[\Pi^a_{cc}]_{\omega,\nu''}[\Gamma^a_{dc}]_{\nu''-\nu',\nu'+\omega,\nu'}
    ,
\end{align}
which is then inserted into \Eq{eq:gamma_t^dc}:
\begin{align}
    \nonumber [\gamma^t_{dc}]_{\omega,\nu,\nu'}\!\!
    &=\!\!
    \int_{\nu_1,\nu_2}\!\!\!\!\!\![\Gamma^t_{dc}]_{\omega,\nu,\nu_1}[\Pi^t_{dd}]_{\omega,\nu_1}[\Gamma^a_{dc}]_{\nu_2-\omega-\nu',\omega+\nu',\nu+\nu_1-\nu'}
    \\
    &\phantom{=}\quad
    \times [\Pi^a_{cc}]_{\nu_1-\nu',\nu_2}[\Gamma^a_{dc}]_{\nu_2-\nu',\nu_1,\nu'}
    .
\end{align}
Inserting the bubbles, \Eqs{eq:bubbles-general}, and dropping the $dc$ indices gives \Eq{eq:gamma_t}.

In our code, we save the three vertices $\Gamma^t_{dc}$, $\Gamma^a_{dc}$ and $\Gamma^a_{dc}$ (minus their constant first-order contributions, i.e., $\tilde\Gamma = \Gamma-\Gamma^{(1)}$) on three-dimensional frequency grids.

To calculate $\gamma^a_{dd}$, \Eq{eq:gamma_a^dd}, we subdivide the equation into contributions of different asymptotic classes~\cite{Wentzell2020,gievers2022multiloop}, i.e., $\gamma^a_{dd}=[\K1a]^{dd}+[\Kb2a]^{dd}+[\K2a]^{dd}+[\K3a]^{dd}$, which are given by
\begin{subequations}
    \begin{align}
		[\K1a]^{dd}_\omega
        &=
        \Gamma^{(1)}_{\hat{dc}}\int_{\nu''}[\Pi^a_{cc}]_{\omega,\nu''}\Gamma^{(1)}_{\hat{cd}}
        ,
        \\
		[\Kb2a]^{dd}_{\omega,\nu'}
        &=
        \int_{\nu''} \Gamma^{(1)}_{\hat{dc}}[\Pi^a_{cc}]_{\omega,\nu''}[\tilde{\Gamma}^a_{\hat{cd}}]_{\omega,\nu'',\nu'}
        ,
        \\
		[\K2a]^{dd}_{\omega,\nu}
        &=
        \int_{\nu''}[\tilde{\Gamma}^a_{\hat{dc}}]_{\omega,\nu,\nu''}[\Pi^a_{cc}]_{\omega,\nu''}\Gamma^{(1)}_{\hat{cd}}
        ,
        \\
		[\K3a]^{dd}_{\omega,\nu,\nu'}
        &=
        \int_{\nu''}[\tilde{\Gamma}^a_{\hat{dc}}]_{\omega,\nu,\nu''}[\Pi^a_{cc}]_{\omega,\nu''}[\tilde{\Gamma}^a_{\hat{cd}}]_{\omega,\nu'',\nu'}
        .
	\end{align}
\end{subequations}
Also the numerical result for $\gamma^t_{dc}$, \Eq{eq:gamma_t}, is subdivided into asymptotic classes $\gamma^t_{dc} = [\Kb2t]^{dc}+[\K3t]^{dc}$, which are given by
\begin{subequations}
    \label{eq:K2t_K3t}
    \begin{align}
        \nonumber [\Kb2t]^{dc}_{\omega,\nu'}
        &=
        \int_{\nu''}\Gamma^{(1)}_{dc}[\Pi^t_{cc}]_{\omega,\nu''}[\gamma^a_{dd}]_{\nu''-\nu',\nu'+\omega,\nu'}
        \\
        &=~ -
        \begin{gathered}
            \includegraphics[width=80pt]{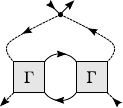}~,
         \end{gathered}
        \\
		  \nonumber [\K3t]^{dc}_{\omega,\nu,\nu'}
        &=
        \int_{\nu''}[\tilde{\Gamma}^t_{dc}]_{\omega,\nu,\nu''}[\Pi^t_{dd}]_{\omega,\nu''}[\gamma^a_{dd}]_{\nu''-\nu',\nu'+\omega,\nu'}
        \\
        &=~ -
        \begin{gathered}
            \includegraphics[width=80pt]{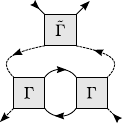}~.
         \end{gathered}
    \end{align}
\end{subequations}

\section{Details on the numerical implementation}
\label{sec:Numerics}

The self-consistent schemes presented in \Sec{sec:Parquet-schemes} are implemented using the recently developed Julia library \texttt{MatsubaraFunctions.jl}~\cite{kiese2024matsubarafunctions}. To efficiently handle the frequency dependence, the two-particle reducible vertices $\gamma^r$ are parametrized in single-boson exchange vertices~\cite{Krien2019,Krien2019a,Krien2019b,Krien2020a,Krien2020b,Krien2021,Harkov2021,Harkov2021a,krien2022plain,Bonetti2022SBE,fraboulet2022single,gievers2022multiloop}:
\begin{align}
    \gamma^r_{\omega,\nu,\nu'} &= \bar\lambda^r_{\omega,\nu}\eta^r_{\omega}\lambda^r_{\omega,\nu'}+u + M^r_{\omega,\nu,\nu'}.
\end{align}
(Note that the bare vertex is defined with an additional minus sign, i.e., $\Gamma^{(1)}=-u$.)
Here the $U$-reducible contribution is a product of one bosonic propagator $\eta^r_\omega$ and two Hedin vertices $\bar\lambda^r_{\omega,\nu}$ and $\lambda^r_{\omega,\nu'}$ coupling fermionic degrees of freedom with exchange bosons. The remaining term is incorporated in the multi-boson vertex $M^r_{\omega,\nu,\nu'}$.

The parquet equations \eqref{eq:parquet-leading-log} and \eqref{eq:Gamma_parquet_extended} are then solved in terms of the single-boson vertices using the following set of self-consistent equations (cf.\ Eqs.~(41) in Ref.~\cite{gievers2022multiloop}):
\begin{subequations}
    \label{eq:SBE-equations}
    \begin{align}
        P^r_\omega &= \int_{\nu''}\lambda^r_{\omega,\nu''}\Pi^r_{\omega,\nu''},\\
        \eta^r_\omega &= -u -u P^r_\omega \eta^r_\omega,\\
        \bar\lambda^r_{\omega,\nu} &= 1+\int_{\nu''}T^r_{\omega,\nu,\nu''}\Pi^r_{\omega,\nu''},\\
        \lambda^r_{\omega,\nu'} &= 1+\int_{\nu''}\Pi^r_{\omega,\nu''}T^r_{\omega,\nu'',\nu'},\\
        T^r_{\omega,\nu,\nu'} &= \Gamma^r_{\omega,\nu,\nu'} - \bar\lambda^r_{\omega,\nu}\eta^r_{\omega}\lambda^r_{\omega,\nu'},\\
        \nonumber M^r_{\omega,\nu,\nu'} &= \int_{\nu''}(T^r_{\omega,\nu,\nu''}-M^r_{\omega,\nu,\nu''})\Pi^r_{\omega,\nu''}T^r_{\omega,\nu'',\nu'}\\
        &= \int_{\nu''}T^r_{\omega,\nu,\nu''}\Pi^r_{\omega,\nu''}(T^r_{\omega,\nu'',\nu'}-M^r_{\omega,\nu'',\nu'}).
        \label{eq:MBE-equation}
    \end{align}
\end{subequations}
Here, the polarization $P^r$ is the bosonic self-energy. $T^r$ represent the $U$-irreducible vertices in a respective channel. In practice, we use a symmetrized form of the two expressions for the multi-boson vertex, \Eq{eq:MBE-equation}. We solve \Eqs{eq:SBE-equations} self-consistently by using the Anderson acceleration method, which leads to a faster convergence involving adaptive mixing of prior solutions.

The susceptibility, \Eq{eq:chi_diagram}, is directly obtained from the bosonic propagator,
\begin{align}
    \chi(\omega) = \frac{1}{u^2}(\eta^a_\omega+u).
\end{align}
Using $-u-u\int_{\nu''}\Pi^r_{\omega,\nu''}\Gamma_{\omega,\nu'',\nu'}=\eta^r_\omega\lambda^r_{\omega,\nu'}$ (cf.\ Eq.~(42b) in Ref.~\cite{gievers2022multiloop,Patricolo2024Single}), the Schwinger--Dyson equation~\eqref{eq:SDE} represented in single-boson exchange vertices yields
\begin{align}
    \nonumber\Sigma_\nu &= -\int_{\nu''}\eta^a_{\nu''-\nu}\lambda^a_{\nu''-\nu,\nu''}g_{\nu''}\\
    &= -\int_{\nu''}\eta^p_{-\nu''-\nu}(2\lambda^p_{-\nu''-\nu,-\nu''}-1)g_{\nu''}.
\end{align}
Here, the Hartree term $\Sigma_\mathrm{H}=u\int_\nu g_\nu e^{\ii \nu 0^+}$ is implicitly added.

\begin{figure}
    \centering
    \includegraphics[width=1.0\linewidth]{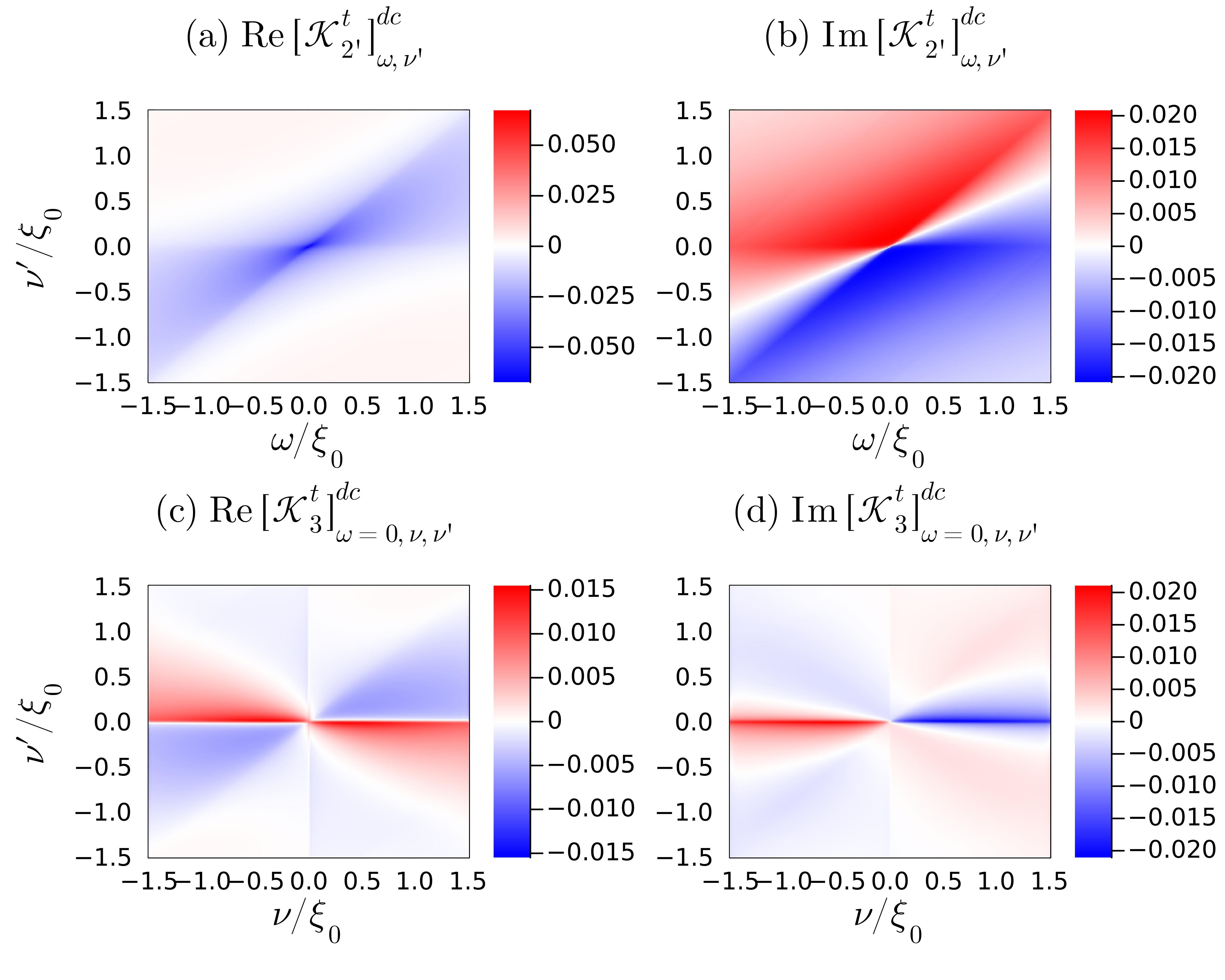}
    \caption{Frequency dependence of the $t$-reducible vertices $[\mathcal{K}^t_{2'}]^{dc}$ and $[\mathcal{K}^t_3]^{dc}$, \Eqs{eq:K2t_K3t}. These are obtained for $T/\xi_0 = 0.002$, $\xi_d/\xi_0=-0.01$, and $u=0.28$ from the subleading parquet scheme, \Eqs{eq:SDE}--\eqref{eq:Gamma_parquet_extended}.}
    \label{fig:gammat_t-vertices}
\end{figure}

\begin{figure}
    \centering
    \includegraphics[width=1.0\linewidth]{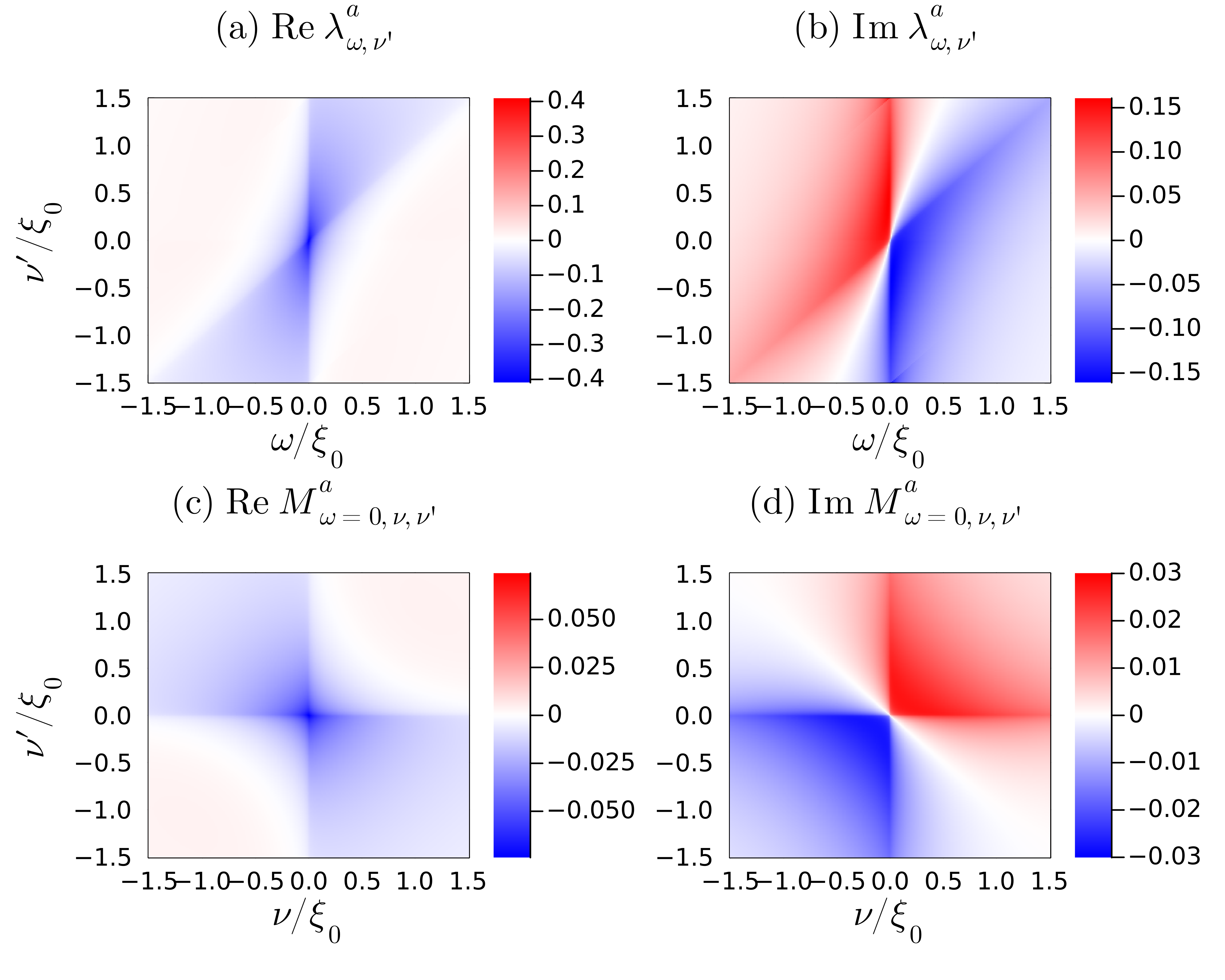}
    \caption{Frequency dependence of the $a$-reducible Hedin vertex $\lambda^a_{\omega,\nu'}$ and the multi-boson vertex $M^a_{\omega,\nu,\nu'}$, \Eqs{eq:SBE-equations}, obtained for $T/\xi_0 = 0.002$, $\xi_d/\xi_0=-0.01$, and $u=0.28$ from the subleading parquet scheme, \Eqs{eq:SDE}--\eqref{eq:Gamma_parquet_extended}.}
    \label{fig:MBE-vertices}
\end{figure}

We save the objects $P^r_\omega, \eta^r_\omega, \Sigma_\nu$ on one-dimensional frequency grids, $\bar\lambda^r_{\omega,\nu}, \lambda^r_{\omega,\nu'}, [\mathcal{K}^t_{2'}]^{dc}_{\omega,\nu'}$ on two-dimensional frequency grids and $M^r_{\omega,\nu,\nu'}, [\mathcal{K}_3^t]^{dc}_{\omega,\nu,\nu'}$ on three-dimensional frequency grids. In doing so, we ensure that the largest frequencies of the three-dimensional quantities exceed the bandwidth $\nu_\mathrm{max}\simeq 1.5\,\xi_0$. The frequency boxes corresponding to the lower-dimensional vertices are taken much larger. To represent the high-frequency asymptotics of the one-dimensional quantities $P^r_\omega$ and $\eta^r_\omega$ in a more sophisticated way, we approximate the Matsubara summation over bubbles outside the frequency box by an integral over the bare bubble:
\begin{align}
    \nonumber\frac{1}{\beta}\sum_{\nu''}\Pi^r_{\omega,\nu''}&\approx\frac{1}{\beta}\sum_{|\nu''|\leq\nu_\mathrm{max}}\Pi^r_{\omega,\nu''}\\
    &\quad +\beta\int_{|\nu''|>\nu_\mathrm{max}}\frac{\dd\nu''}{2\pi}[\Pi^r]^{(0)}_{\omega,\nu''}.
\end{align}
Figures~\ref{fig:gammat_t-vertices} and \ref{fig:MBE-vertices} show exemplary numerical results for vertex functions obtained from the subleading parquet scheme, \Eqs{eq:SDE}--\eqref{eq:Gamma_parquet_extended}.

\section{Analytical continuation}
\label{sec:Analytical-continuation}

To test how well the minimal pole representation for analytical continuation~\cite{Zhang2024Minimal,zhang2024minimalpolerepresentationanalytic} predicts the power-law behaviors, we start from the exact power law, \Eq{eq:exact_power-law_chi},
\begin{align}
    \chi(z) = \frac{1}{\alpha}\left[1-\left(\frac{z-\omega_0}{-\xi_0}\right)^{-\alpha}\right],
    \label{eq:Powerlaw-complex}
\end{align}
continued to complex variables $z$. The power law for $\chi(z)$ is an approximation near the threshold and does not decay to zero for $|z|\to\infty$. Nevertheless, after subtracting the offset $1/\alpha$, we find a spectral representation. By separating the power law into real and complex parts,
\begin{align}
    \nonumber (-x-\ii 0^+)^{-\alpha} &=
    |x|^{-\alpha}\ee^{-\ii\alpha\,\mathrm{arctan2}(-x,-0^+)}\\
    &=|x|^{-\alpha}\left[\Theta(-x)+\Theta(x)\ee^{\ii\pi\alpha}\right],
    \label{eq:power_real_complex}
\end{align}
with $x=(\omega-\omega_0)/\xi_0$, the spectral function $A(\omega)=-\Im\,\chi(\omega+\ii 0^+)/\pi$ yields
\begin{align}
    A(\omega) 
    &= 
    \frac{\sin\pi\alpha}{\pi\alpha}\Theta(\omega-\omega_0)\bigg(\frac{\omega-\omega_0}{\xi_0}\bigg)^{-\alpha}.
    \label{eq:spectral-function_power-law}
\end{align}
In the Matsubara formalism, this gives
\begin{align}
    \chi(\ii\omega)-\frac{1}{\alpha}=\int_{-\infty}^\infty{\dd\omega'}\frac{A(\omega')}{\ii\omega-\omega'} = -\frac{1}{\alpha}\left(\frac{\ii\omega-\omega_0}{-\xi_0}\right)^{-\alpha}.
\end{align}
The expression in imaginary frequencies is thus identical to a simple transformation $z\to\ii\omega$ in \Eq{eq:Powerlaw-complex}.

We generated imaginary-frequency data including different levels of artificial noise. This test showed that data at a temperature $T/\xi_0\simeq 0.002$ and with relative error $10^{-4}$ are sufficient to reproduce the exact power when only taking a few Matsubara frequencies, i.e., the lowest $50$. From this procedure, we can be confident about the validity of the analytically continued data presented in the main text (cf.\ \Fig{fig:chi_AC}).

The power-law exponent is extracted from the slope of the log-log plot in \Fig{fig:chi_AC}(b). For $\omega>\omega_0$, we have
\begin{align}
    \nonumber &\frac{\dd}{\dd\ln(\omega-\omega_0)}\ln[-\Im\,\chi(\omega+\ii 0^+)] \\
    &= \frac{\omega-\omega_0}{\Im\,\chi(\omega+\ii 0^+)}\Im\frac{\dd}{\dd\omega}\chi(\omega+\ii 0^+).
\end{align}
Inserting the analytical power law, \Eq{eq:Powerlaw-complex}, gives exactly $-\alpha$. For the logarithmic derivative of our numerical data, we use the minimal pole expansion,
\begin{align}
    \frac{\dd}{\dd z}\chi(z) = \frac{\dd}{\dd z}\sum_i \frac{A_i}{z-x_i} = \sum_i \frac{-A_i}{(z-x_i)^2},
\end{align}
defined in Refs.~\cite{Zhang2024Minimal,zhang2024minimalpolerepresentationanalytic}.

\bibliography{main}

\end{document}